\documentclass[a4paper,12pt]{article}
\usepackage[utf8]{inputenc}
\usepackage{amsmath,amsfonts,amssymb}
\usepackage{graphicx}
\usepackage{xcolor}
\usepackage{float}
\usepackage[section]{placeins}
\let\origsubsection\subsection
\renewcommand{\subsection}{\FloatBarrier\origsubsection}
\usepackage{caption}
\usepackage{subcaption}
\usepackage{booktabs}
\usepackage[left=2.3cm,right=2.3cm,top=2cm,bottom=2cm]{geometry}
\usepackage[skip=0.5\baselineskip]{parskip}

\usepackage[style=numeric-comp,sorting=none,sortcites=true,giveninits=true,uniquename=false]{biblatex}
\usepackage{hyperref}

\title{Thermally Conditioned Time-Series Forecasting of Grain Growth in Polycrystalline Materials}
\author{
    Eliane Younes, Elie Hachem, Marc Bernacki \\
    Mines Paris, PSL University\\
Centre for material forming (CEMEF), UMR CNRS\\
06904 Sophia Antipolis, France
}
\date{}

\begin{document}
\maketitle
\begin{abstract}
Grain growth during heat treatment strongly affects the microstructure of polycrystalline materials and, consequently, their mechanical properties. Although numerical methods such as front-tracking or front-capturing simulations can accurately describe grain growth, their high computational cost limits their use for long or non-isothermal thermal cycles. In previous work, grain growth was successfully predicted under isothermal conditions by treating grain size distributions as time series and forecasting their evolution using recurrent neural networks, allowing complete grain growth trajectories to be generated in seconds rather than through costly PDE-based simulations. However, such models cannot represent the strong influence of thermal history on grain boundary kinetics. In the present work, this forecasting framework is reformulated to handle non-isothermal conditions by explicitly incorporating thermal information into the model architecture. Grain size distributions are encoded using a convolutional neural network, thermal descriptors are embedded through a dedicated multilayer perceptron, and temporal evolution is predicted using a recurrent network. Instantaneous temperature and thermal rate are provided at each time step, allowing the model to adapt its predictions to changing thermal conditions. The framework is trained and evaluated on datasets generated from validated front-tracking simulations across a wide range of non-isothermal thermal paths. The results show that the thermally conditioned model accurately reproduces grain growth dynamics under complex thermal histories, achieving, in $k$-fold cross-validation over all simulations, mean relative errors of 2.99\% (one-hour cycles) and 4.51\% (four-hour cycles) with coefficients of determination exceeding 0.99, and mean relative errors between 1.91\% and 14.66\% on ten simulations following four unseen thermal histories of increasing complexity.
\end{abstract}


\section{Introduction}
Thermal processing leads to significant changes in the microstructure of polycrystalline materials, which in turn affect their mechanical behaviour.
In this context, grain growth plays an important role~\cite{smith1948,rohrer2023grain,burke1952,mullins1956,neumann1952, rollett2017}, as it governs grain size evolution during annealing, which strongly influences mechanical properties such as flow stress and creep resistance~\cite{ZHANG2022161710,lasalmonie1986influence}.
Accurate prediction of grain growth therefore remains a key objective in materials science, particularly in the context of heat treatment design \cite{tep2026predicting}.

At the mesoscale, grain growth in single-phase polycrystals is commonly described as a curvature-driven process, in which grain boundary migration results from the reduction of total interfacial energy~\cite{herring1999, burke1952, smith1948}.
In such kinetic descriptions, the normal velocity of a grain boundary can be reasonably approximated at high temperature and for polycrystal scale predictions as:
\begin{equation}
\mathbf{v} = - \mu \, \gamma \, \kappa \, \mathbf{n},
\end{equation}
where $\mu$ denotes the grain boundary mobility, $\gamma$ the grain boundary energy, $\kappa$ the mean curvature, and $\mathbf{n}$ the outward unit normal to the interface.

Although simplified kinetic formulations have demonstrated strong predictive capability at the polycrystalline scale, grain boundary mobility is known to depend strongly on temperature, most often through Arrhenius-type relations~\cite{chen2020temperature}.
It is therefore commonly expressed as:
\begin{equation}
\mu= \mu_0 \exp\left(-\frac{Q}{RT}\right),
\label{eq:arrhenius_mobility}
\end{equation}
where $\mu_0$ is the pre-exponential factor, $Q$ the activation energy, $R$ the gas constant, and $T$ the absolute temperature.
Consequently, the evolution of a given microstructure is governed not only by its initial state, but also by the thermal history imposed during processing.

Numerical approaches such as Monte Carlo Potts methods~\cite{wu1982potts,peczak1993monte}, cellular automata~\cite{liu1996simulation,hesselbarth1991simulation}, front-tracking~\cite{frost1988two,barrales2007novel,florez2020novel,florez2025,bernacki2024vertex}, phase-field~\cite{chen1997computer,krill2002computer}, and level-set methods~\cite{elsey2009diffusion,zhao1996variational,bernacki2008level,bernacki2024kinetic} have been extensively developed to simulate grain growth with high physical fidelity. These methods accurately capture the topology and kinetics of evolving grain boundary networks. However, because microstructural evolution must be resolved incrementally in time, such Partial Differential Equation (PDE)-based simulations incur substantial computational cost, particularly for large systems and long thermal histories.

To reduce these limitations, reduced or mean-field representations have been developed as an alternative to full-field descriptions by representing the evolving microstructure through statistical descriptors rather than explicitly resolving the full spatial topology~\cite{roth2023comparison}. Although these approaches significantly decrease data dimensionality and computational requirements, the temporal prediction of microstructural evolution over extended time scales remains challenging.

In our previous study~\cite{younes2026predicting}, we addressed these limitations by introducing a data-driven surrogate model for grain growth prediction under isothermal conditions. Grain size distributions extracted from validated front-tracking simulations ToRealMotion (TRM)~\cite{florez2020novel,florez2021parallelization} were represented as evolving time series and predicted using recurrent neural networks. Using this formulation, the full grain growth trajectory could be generated from a short temporal history, with prediction accuracies on the order of 90\% relative to reference simulations. Once trained, the surrogate produced complete grain growth evolutions in a few seconds, compared to simulation times on the order of tens of minutes for the underlying front-tracking model, corresponding to speedups exceeding two orders of magnitude. At the same time, physically consistent trends in the predicted grain-size distributions and mean grain size evolution were preserved.

Despite these results, the earlier surrogate formulation was developed and assessed only for constant temperature conditions.
In practice, industrial heat treatments rarely follow strictly isothermal paths and instead involve non-isothermal thermal schedules composed of successive heating, holding, and cooling stages, often with variable heating rates.
Since grain boundary mobility depends explicitly on temperature, identical initial microstructures subjected to different thermal histories may evolve along distinct trajectories.
Such effects cannot be captured by models trained solely on isothermal data without explicit incorporation of thermal information.

The present work extends the previously proposed time-series forecasting framework to non-isothermal grain growth prediction.
The proposed extension preserves the representation based on grain size distributions developed previously \cite{younes2026predicting}, while additional thermal descriptors are introduced to account for temperature-dependent grain growth kinetics.
Instantaneous temperature and thermal rate are provided as conditioning variables, enabling the surrogate to adapt its predictions to evolving thermo-kinetic conditions while preserving the computational efficiency observed under isothermal training.

The extended framework is trained and evaluated using datasets generated with the TRM framework, whose grain growth predictions have been validated against EBSD measurements for nickel-based superalloys and stainless steels, under a wide range of non-isothermal thermal cycles. Several recurrent architectures are examined under identical training conditions, allowing a controlled comparison of their ability to propagate grain growth dynamics over time.
In addition, different conditioning strategies are investigated to distinguish the respective influence of temperature, thermal rate, and their combined effect.
Beyond standard train--validation--test evaluation, thermal histories excluded from training are considered to assess the robustness and generalization capability of the proposed surrogate under increasingly complex conditions.

By extending data-driven grain growth prediction from isothermal to non-isothermal regimes, this work provides a practical step toward surrogate models applicable to realistic heat treatment scenarios.
The proposed approach offers a computationally efficient alternative to high-fidelity simulations while maintaining physically meaningful predictive behaviour, with direct relevance for large-scale studies and process-level applications.

The paper is organized as follows. In Section~\ref{sec:methodology}, we describe the dataset generation procedure under non-isothermal conditions, along with the dataset organization, preprocessing steps, and the architecture of the thermally conditioned surrogate model. In Section~\ref{sec:results}, we present the prediction results and analyze the effects of different conditioning strategies, recurrent formulations, and generalization capabilities. We discuss the predictive behavior of the model and the impact of thermal conditioning on grain growth forecasting in Section~\ref{sec:discussion}. We summarize the main findings and conclude in Section~\ref{sec:conclusion}.

\section{Methodology}
\label{sec:methodology}
\subsection{Dataset Generation}

The initial polycrystalline microstructure states are generated following the same procedure as in the previous isothermal study~\cite{younes2026predicting}, using the LaVoGen (Laguerre--Voronoï tessellation) algorithm \cite{hitti2012precise}. These microstructures are characterized by different domain sizes, mean equivalent circle radii, and grain size distribution standard deviations, representative of polycrystalline microstructures.

Grain growth evolutions are then simulated using the TRM front-tracking framework~\cite{florez2020novel,florez2021parallelization}, with each initial microstructure subjected to different non-isothermal thermal conditions during the TRM simulations.
Within this framework, grain boundary migration is driven by curvature under the assumption of isotropic grain boundary energy. The effect of temperature on grain boundary kinetics is accounted for through the Arrhenius-type mobility relation introduced in Eq.~(\ref{eq:arrhenius_mobility}).
The parameters $\mu_0$, $Q$, and $\gamma$ are chosen to be representative of a 304L stainless steel, ensuring realistic grain boundary behavior over the considered temperature range \cite{bernacki2024kinetic, maire2017modeling}.
Their numerical values are reported in Table~\ref{tab:combined_params_article2}.

\begin{figure}[htbp]
    \centering

    \begin{subfigure}[b]{0.95\textwidth}
        \centering
        \includegraphics[width=\textwidth]{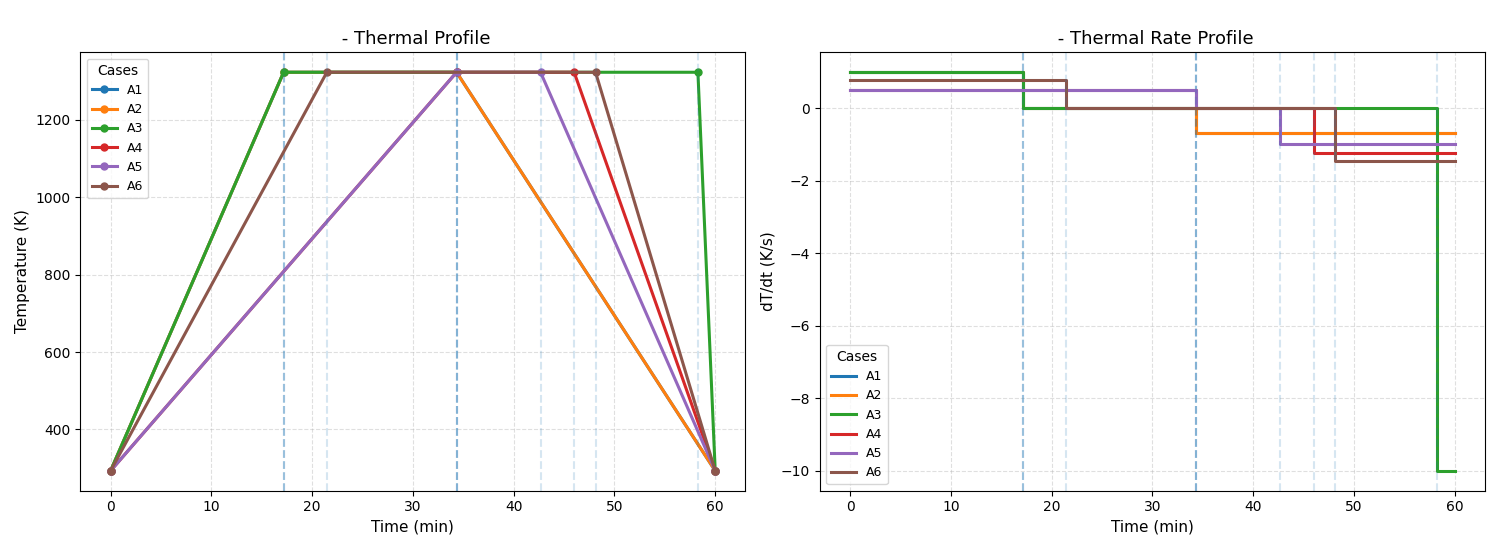}
        \caption{One-hour thermal cycles ($3600$ s): temperature $T(t)$ and thermal rate $\dot{T}(t)$ for cases A1--A6.}
        \label{fig:thermal_1h}
    \end{subfigure}

    \vspace{0.5cm}

    \begin{subfigure}[b]{0.95\textwidth}
        \centering
        \includegraphics[width=\textwidth]{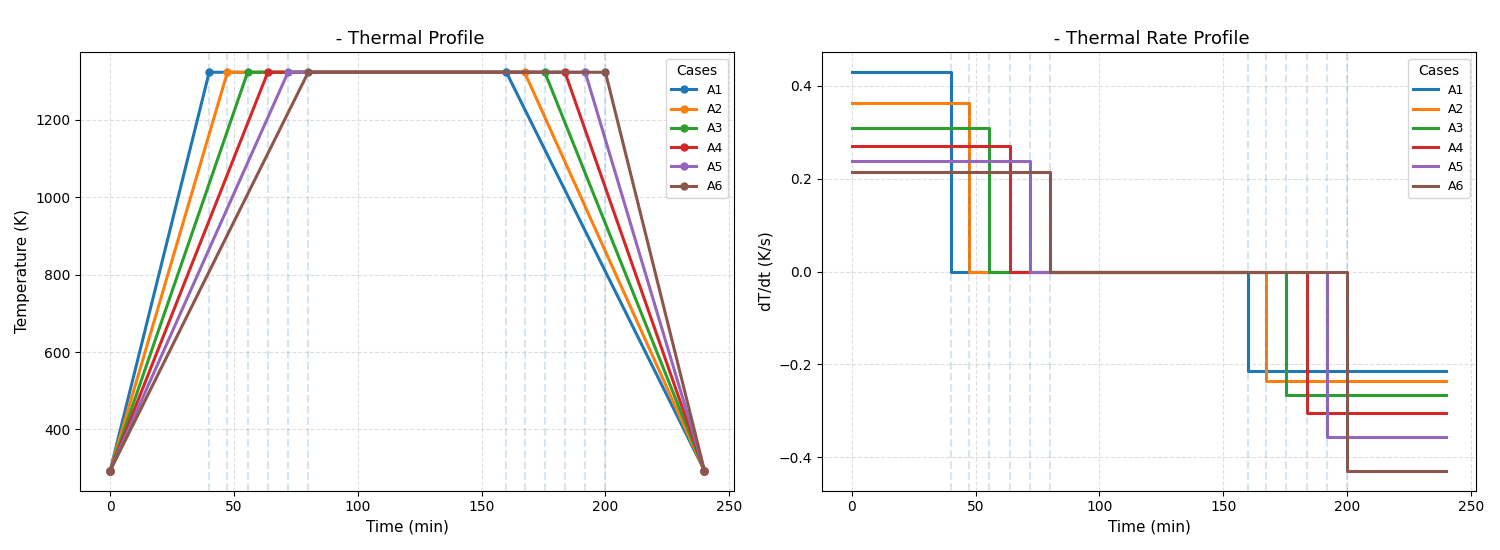}
        \caption{Four-hour thermal cycles ($14400$ s): temperature $T(t)$ and thermal rate $\dot{T}(t)$ for cases A1--A6.}
        \label{fig:thermal_4h}
    \end{subfigure}

    \caption{Non-isothermal thermal histories considered in this study. Short-duration cycles (1 h) are used to investigate fast thermo-kinetic regimes, whereas long-duration cycles (4 h) are introduced to examine slower grain growth conditions. In each case, the thermal evolution is characterized by the instantaneous temperature $T(t)$ and the corresponding thermal rate $\dot{T}(t)$.}

    \label{fig:thermal_cycles_combined}
\end{figure}

For each initial microstructure, two categories of thermal cycles are applied to investigate both fast and slow thermo-kinetic regimes. Short ($3600$ s) cycles correspond to rapidly evolving thermal histories, whereas longer ($14400$ s) cycles are associated with slower thermo-kinetic conditions through extended heating and cooling stages under identical peak temperature conditions. The complete set of thermal cycles is illustrated in Figs.~\ref{fig:thermal_1h} and~\ref{fig:thermal_4h}.

Each thermal cycle is described by a temperature profile composed of successive heating, holding, and cooling stages, with linear variations of temperature within each phase. The thermal evolution is characterized by the instantaneous temperature $T(t)$ and the corresponding thermal rate $\dot{T}(t)$, shown in Figs.~\ref{fig:thermal_1h} and~\ref{fig:thermal_4h}. The thermal rate is computed from the temperature values of the neighbouring stored time steps (central differences, one-sided at the first and last steps):
\begin{equation}
\dot{T}^{\,t}
=
\frac{T^{\,t+dt}-T^{\,t-dt}}{2\,dt}.
\end{equation}

where $dt$ is the time interval between two stored states (60~s after the temporal downsampling described in Section~2.2).

The temporal evolution is discretized using a constant time step, and microstructural states are stored at each step. Equivalent circle radius (ECR) are extracted and converted into normalized grain size histograms with a fixed number of bins (30 bins of equal width between 0 and a maximum size chosen just above the largest grain observed in each dataset: 0.11~mm for the one-hour cycles and 0.15~mm for the four-hour cycles, so that no grain is excluded), providing a compact statistical representation suitable for time-series forecasting. The thermal descriptors $T(t)$ and $\dot{T}(t)$ are used as conditioning variables for the learning framework.

The full set of parameters used for both LaVoGen microstructure generation and TRM grain growth simulations is summarized in Table~\ref{tab:combined_params_article2}.

\begin{table}[htbp]
\centering
\begin{tabular}{lll}
\toprule
\textbf{Parameter} & \textbf{Value(s)} & \textbf{Description} \\
\midrule
$a$ (mm)
& 2, 3, 4, 5
& Side length of the square domain \\

Distribution
& $\mathcal{N}(\bar{R}, \sigma^2)$
& Initial ECR distribution (LaVoGen) \\

$\bar{R}$ ($\mu$m)
& 20
& Mean equivalent circle radius (ECR) \\

$\sigma$ ($\mu$m)
& 2, 8, 16, 32
& Standard deviation of the ECR distribution \\

$T$ (K)
& 293.15--1323.15
& Temperature range \\
$dT/dt$ heating (K\,s$^{-1}$)
& 0.215--1.0
& Heating rate range \\

$|dT/dt|$ cooling (K\,s$^{-1}$)
& 0.05--10
& Cooling rate range \\

$dt$ (s)
& 10
& Simulation time step \\

$t_{\mathrm{sim}}$ (min)
& 60, 240
& Simulation duration \\

Storage frequency
& 1
& Microstructure storage frequency \\

$Q$ (J\,mol$^{-1}$)
& 280\,000
& Activation energy \\

$\gamma$ (J\,m$^{-2}$)
& 0.6
& Grain boundary energy \\

$\mu_0$ (m$^{4}$\,J$^{-1}$\,s$^{-1}$)
& 0.156
& Pre-exponential mobility constant \\

\bottomrule
\end{tabular}
\caption{Main parameters used for LaVoGen microstructure generation and TRM grain growth simulations under non-isothermal thermal cycles.}
\label{tab:combined_params_article2}
\end{table}

\subsection{Dataset Organization and Preprocessing}

Each simulation is represented as a temporal dataset stored through a pair of aligned files: an \texttt{Evolution} file and a \texttt{Property} file. The \texttt{Evolution} file contains the normalized grain size distributions arranged in a matrix of size $(30 \times N_t)$, where $N_t$ is the number of stored time steps and each column consists of 30 values defining the grain size distribution at a given time step. The sequence of columns therefore captures the temporal evolution of the grain size distribution over the course of the simulation.

Following our previous work~\cite{younes2026predicting}, the grain size distributions are normalized at each time step by dividing the number of grains in each bin by the total number of grains. Each bin therefore represents the relative frequency of grains within the corresponding size interval, and each distribution is described by 30 relative-frequency values.

The corresponding thermal information is stored in a \texttt{Property} file, which contains the instantaneous temperature $T(t)$ and the thermal rate $\dot{T}(t)$, forming a matrix of size $(2 \times N_t)$.

Although the TRM simulations are performed with a time step of $dt = 10$~s and microstructural states are stored at each step, the dataset used for training is temporally downsampled by retaining one state every six time steps.
This results in an effective temporal resolution of $1$~min for all learning experiments.

In this framework, the grain size distributions correspond to the quantities to be predicted, while the thermal descriptors serve as conditioning variables that provide the physical context of the non-isothermal evolution.

The one-hour and four-hour datasets, composed of 86 and 208 simulations respectively, are treated separately, and a dedicated model is trained for each of them. Model performance is evaluated by $k$-fold cross-validation at the sequence level, with $k = 10$ for the one-hour dataset and $k = 5$ for the four-hour dataset. In this way, each simulation is predicted by a model that did not use it during training, and all simulations contribute to the reported results. Since the partitioning is performed at the sequence level, all time steps of a given simulation remain in the same subset, which prevents information leakage between sequences.

Thermal descriptors are standardized using z-score normalization based exclusively on statistics computed from the training subset in order to improve numerical conditioning and training stability while avoiding information leakage between data partitions~\cite{goodfellow2016deep}:

\begin{equation}
T_{\mathrm{norm}}^{\,t}
=
\frac{T^{\,t}-\bar{T}}{s_T},
\qquad
\dot{T}_{\mathrm{norm}}^{\,t}
=
\frac{\dot{T}^{\,t}-\overline{\dot{T}}}{s_{\dot{T}}},
\end{equation}

where $\bar{T}$ and $\overline{\dot{T}}$ are the mean values of temperature and thermal rate computed from the training subset, and $s_T$ and $s_{\dot{T}}$ are the corresponding standard deviations. The resulting normalization parameters are subsequently applied unchanged to the validation and test subsets in order to ensure consistency and avoid data leakage~\cite{goodfellow2016deep}.

Samples are constructed using a sliding-window strategy with a fixed window size of $W=5$, corresponding to five consecutive time steps of the downsampled evolution.
Each sample consists of the grain size histograms and associated thermal descriptors contained within the temporal window.

All experiments are conducted with fixed random seeds (seed 42) to ensure reproducibility of data splitting and model training.

\subsection{Surrogate Model Architecture}

The present study extends the time-series forecasting framework introduced in our previous work, where grain growth under isothermal conditions was modeled using sequences of grain size distributions and recurrent neural networks.
In this earlier framework, the temporal evolution of the microstructure was learned exclusively from histogram-based representations obtained at constant temperature.
While this approach proved effective for isothermal annealing, it does not explicitly capture the changes in grain growth kinetics associated with heating and cooling stages.

To address this limitation, the forecasting architecture is modified so that the evolving microstructural state and the applied thermal history are considered jointly.
At each time step $t$, the model receives a temporal window of length $W$ containing the past grain size distributions $x(t-W), \dots, x(t-1)$ together with the corresponding thermal descriptors $T(t-W), \dots, T(t-1)$ and $\dot{T}(t-W), \dots, \dot{T}(t-1)$.
Rather than predicting the subsequent distribution $x(t)$ directly, the model predicts its variation with respect to the last known state, $x(t) = x(t-1) + \alpha\,\Delta(t)$, where $\Delta(t)$ is the output of the network and $\alpha = 0.1$ is a fixed scaling factor. This residual formulation keeps the increments small at each step and stabilizes the auto-regressive rollout. The value $\alpha = 0.1$ was retained after comparison with the unscaled formulation ($\alpha = 1$) using the same cross-validation: over the 86 one-hour simulations, the MRE of the LSTM decreases from 3.56\% to 2.99\% (from 7.68\% to 4.52\% for the GRU), and the only simulation whose error exceeded 30\% at some time step with $\alpha = 1$ no longer does. In addition, the distributions are processed in square-root space ($\sqrt{x}$), which reduces the weight of the dominant bins relative to the sparsely populated bins at the edges of the distribution. The prediction is then squared back to obtain the relative frequencies.
During inference, predictions are propagated in an auto-regressive manner, while the future thermal conditions are assumed known and injected at each successive prediction step.

\begin{figure}[htbp]
    \centering
    \includegraphics[width=\textwidth]{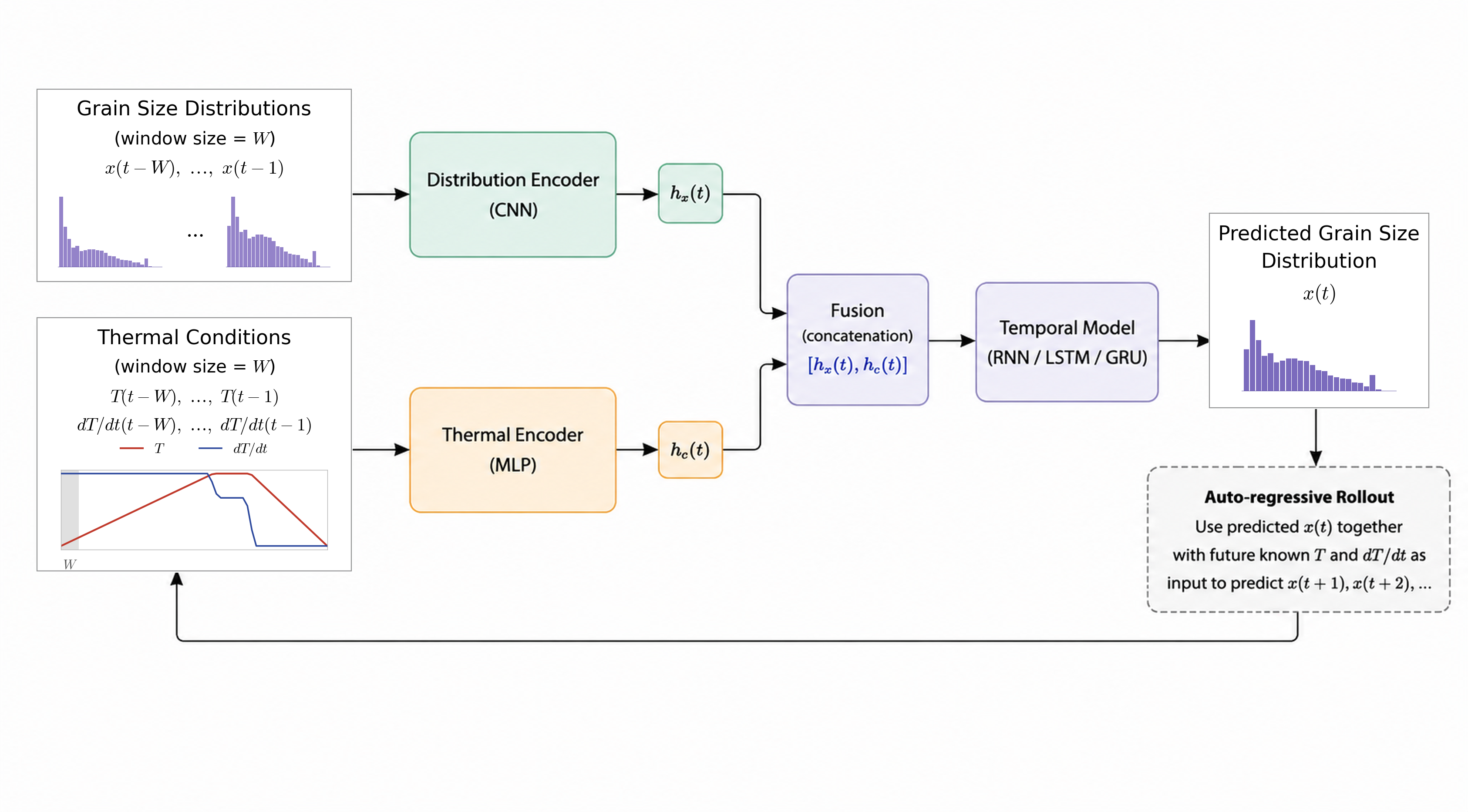}
    \caption{Schematic representation of the thermally conditioned surrogate model for grain size distribution forecasting.
    A temporal window of histograms is encoded using a CNN, while the thermal descriptors $T(t)$ and $\dot{T}(t)$ are processed through a dedicated MLP.
    The resulting feature vectors are concatenated and passed to a recurrent temporal model (RNN, GRU, or LSTM).
    The output projection predicts the next grain size distribution, which is reused during auto-regressive rollout with known future thermal conditions.}
    \label{fig:model_architecture}
\end{figure}

The microstructural pathway relies on a one-dimensional convolutional neural network (1D-CNN) \cite{chen2018deep, lecun2015deep} applied to the grain size histograms. In this representation, neighboring bins are inherently correlated, since peak displacement, distribution broadening, or local redistribution of grain populations generally affect adjacent size classes. One-dimensional convolutions therefore provide an efficient mechanism to capture short-range morphological dependencies along the grain size dimension. In the present implementation, two convolutional layers with 16 filters and a kernel size of 3 are employed, each followed by a LeakyReLU activation function with slope parameter $\alpha = 0.1$. The resulting convolutional features are flattened and projected through a fully connected layer into a compact 32-dimensional latent embedding. A dropout rate of 0.2 is applied to this embedding to improve robustness and prevent overfitting, yielding a representation that characterizes the microstructural state at each time step.

In parallel, the thermal conditions are encoded using a dedicated Multi-Layer Perceptron (MLP) \cite{geron2022hands}.
The instantaneous temperature $T(t)$ and thermal rate $\dot{T}(t)$ are standardized based on statistics computed from the training set and fed to two fully connected hidden layers with 16 units each.
LeakyReLU activation functions with slope parameter $\alpha = 0.1$ are used, and a dropout rate of 0.2 is applied.
A final fully connected layer then projects these features onto a 64-dimensional embedding that represents the local thermo-kinetic regime at each time step.

The latent representations extracted from the CNN and the MLP are concatenated to form a joint feature vector that combines morphological information with the thermal context.
This fused representation is then supplied to the recurrent component of the model.

To identify the most suitable temporal predictor for non-isothermal grain growth forecasting, three recurrent architectures were considered: a Simple Recurrent Neural Network (RNN), a Gated Recurrent Unit (GRU), and a Long Short-Term Memory (LSTM) network. Comprehensive tutorials on these deep models are available in~\cite{wikle2023statistical, fan2020selective, geron2022hands}.
These architectures were evaluated under identical conditions with the objective of comparing their ability to propagate microstructural evolution under auto-regressive forecasting and selecting the most appropriate formulation for subsequent analyses.
For all three variants, the same hyperparameter configuration was retained to ensure a fair comparison. Each recurrent architecture is implemented using three stacked recurrent layers with a hidden dimension of 128 units.
All models are trained for at most 100 epochs using the Adam optimizer with an initial learning rate of $1 \times 10^{-3}$ and a batch size of 32. The learning rate is halved when the validation loss stops decreasing for 4 epochs, and training is stopped after 10 epochs without improvement.
These common settings allow differences in predictive behavior to be attributed primarily to the recurrent formulation rather than to variations in training configuration.

Finally, the output of the recurrent block is projected through a dense layer onto the 30-bin normalized grain-size distribution.
A ReLU activation function is applied to enforce non-negativity of the predicted frequencies, ensuring physically admissible outputs during auto-regressive prediction.

Model training is performed in a supervised manner to predict successive grain size distributions, using the Mean Absolute Error (MAE) computed on the square roots of the relative frequencies as the loss function \cite{qi2020mean}. To reduce the mismatch between training, where the inputs are reference distributions, and auto-regressive inference, where the inputs are the model's own predictions, a pushforward training strategy is used: in addition to the next-step prediction, the model is unrolled over three successive steps using its own predictions as inputs, and the loss is evaluated at each of these steps.
All models are optimized within the TensorFlow/Keras framework \cite{abadi2016tensorflow} using the Adam optimizer \cite{kingma2017adammethodstochasticoptimization}, with identical training settings applied to all recurrent architectures in order to enable a fair comparison.
Hyperparameters were selected based on preliminary tests to ensure stable training behavior.

Prediction performance is evaluated by comparing the predicted and reference normalized grain-size distributions obtained after auto-regressive rollout. The error metrics are computed directly from the corresponding 30-bin frequency vectors. Performance is evaluated using the Mean Relative Error (MRE) and the coefficient of determination $R^2$ \cite{hodson2022root, chicco2021coefficient}. The MRE is defined as
\begin{equation}
\mathrm{MRE} = \frac{100}{N} \sum_{i\,:\,y_i>0} \frac{|y_i-\hat{y}_i|}{y_i},
\end{equation}
where $y_i$ and $\hat{y}_i$ are the reference and predicted relative frequencies and $N$ is the number of predicted values.

A schematic overview of the resulting architecture is presented in Fig.~\ref{fig:model_architecture}.

\subsection{Conditioning Strategies}
To investigate the role of thermal information, several conditioning strategies are evaluated using the same dataset and forecasting protocol:

\begin{itemize}
    \item C0 -- No $T$: no thermal descriptors are supplied, prediction relies only on the sequence of grain size distributions.

    \item C1 -- $T$ only: conditioning based solely on the instantaneous temperature $T(t)$.

    \item C2 -- $\dot{T}$ only: conditioning based solely on the thermal rate $\dot{T}(t)$.

    \item C3 -- $T + \dot{T}$: simultaneous conditioning by temperature and thermal rate, corresponding to the complete proposed framework.
\end{itemize}

This comparison makes it possible to distinguish the respective influence of absolute temperature, transient thermal effects, and their combined contribution to grain growth prediction under non-isothermal conditions.

\subsection{Generalization cases}
\label{subsec:generalization_cases}

In addition to the standard train--validation--test evaluation, four thermal histories not included in the learning dataset were considered to examine the response of the surrogate model under non-isothermal conditions. These cases represent thermal histories of increasing complexity, ranging from a simple reference cycle to multi-stage thermal paths involving successive heating and cooling stages.

\begin{figure}[htbp]
    \centering
    \begin{subfigure}[b]{0.85\textwidth}
        \centering
        \includegraphics[width=\textwidth]{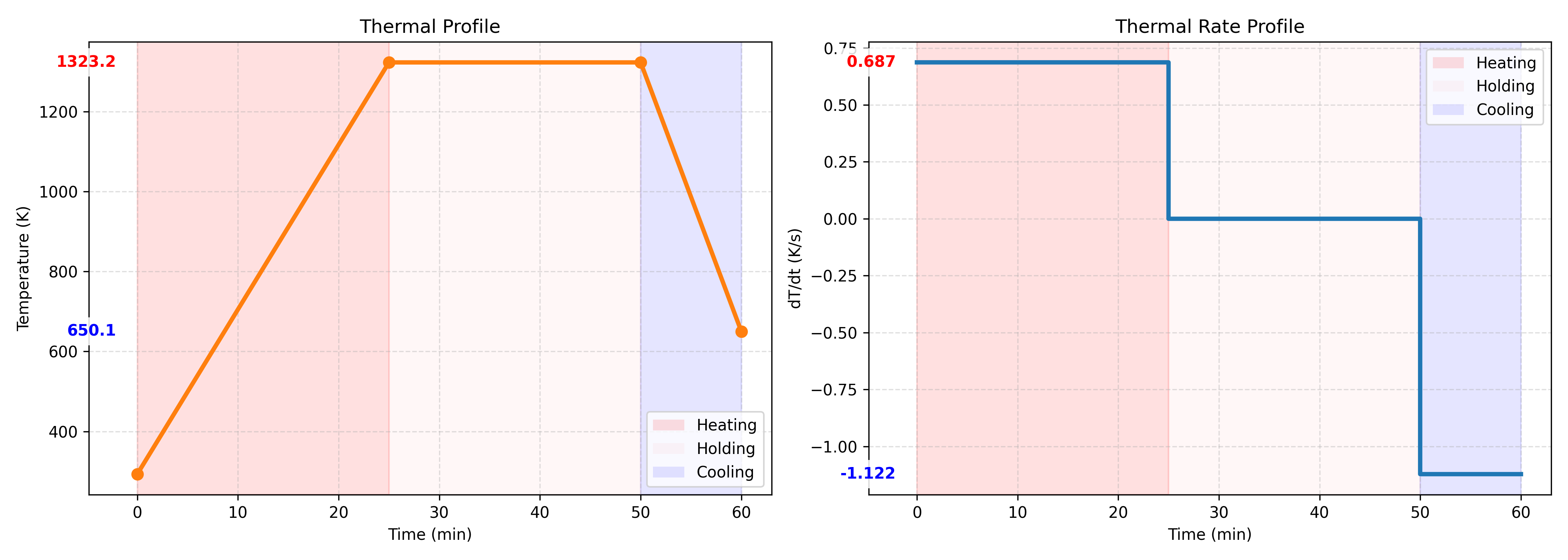}
        \caption{Case~1: baseline one-hour thermal cycle.}
        \label{fig:case1_generalization}
    \end{subfigure}

    \vspace{0.2cm}

    \begin{subfigure}[b]{0.85\textwidth}
        \centering
        \includegraphics[width=\textwidth]{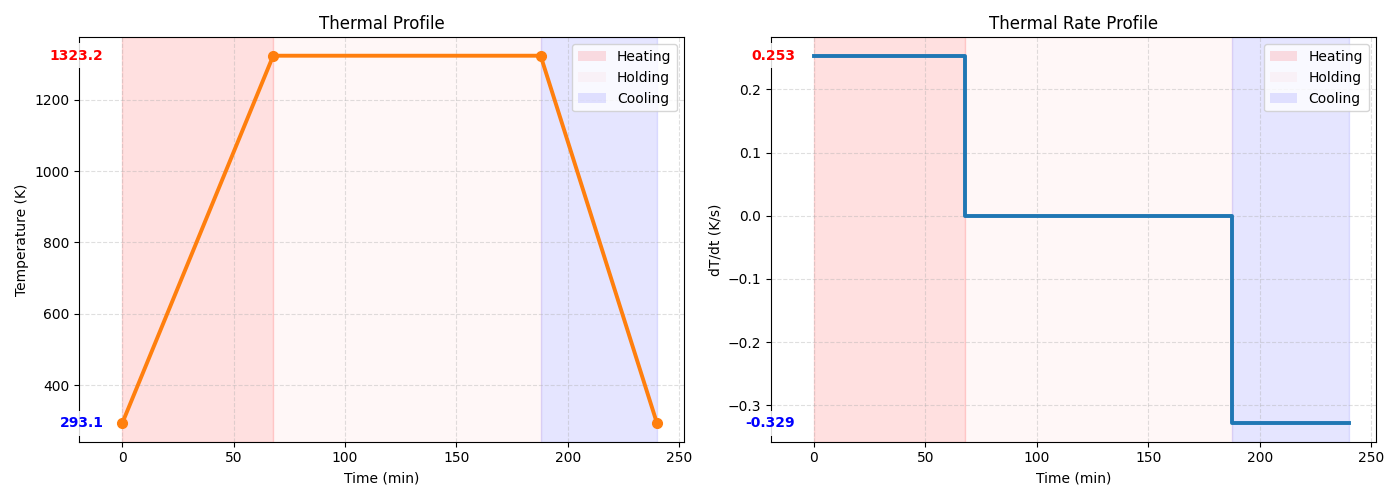}
        \caption{Case~2: extended four-hour thermal cycle with a prolonged isothermal holding stage.}
        \label{fig:case2_generalization}
    \end{subfigure}

    \vspace{0.2cm}

    \begin{subfigure}[b]{0.85\textwidth}
        \centering
        \includegraphics[width=\textwidth]{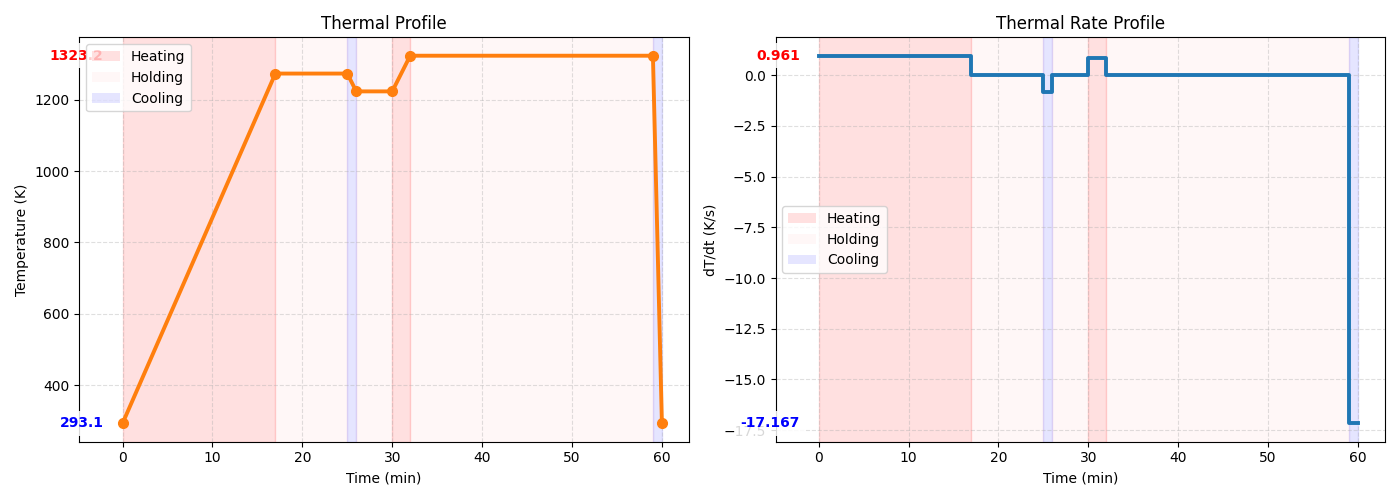}
        \caption{Case~3: multi-stage realistic thermal cycle.}
        \label{fig:case3_generalization}
    \end{subfigure}

    \vspace{0.2cm}

    \begin{subfigure}[b]{0.85\textwidth}
        \centering
        \includegraphics[width=\textwidth]{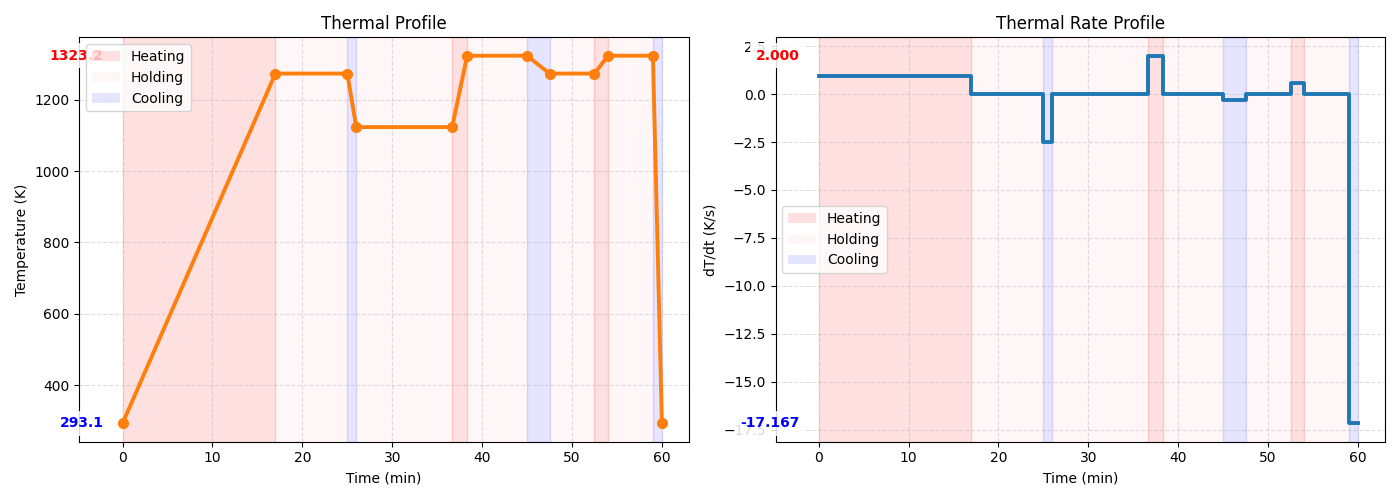}
        \caption{Case~4: complex multi-stage non-monotonic thermal cycle.}
        \label{fig:case4_generalization}
    \end{subfigure}

    \caption{Thermal histories of the four generalization cases excluded from training. For each case, the temperature evolution (left) and the corresponding thermal rate (right) are reported throughout the cycle.}
    \label{fig:generalization_cycles}
\end{figure}

\paragraph{Case 1: Baseline one-hour thermal cycle}

The first case represents a baseline heating--holding--cooling cycle with a total duration of one hour. The temperature increases from 293.15~K to 1323.15~K during the heating stage, followed by an isothermal holding period at 1323.15~K, and a final cooling stage down to 650.15~K. This case provides a reference non-isothermal scenario within a relatively simple thermo-kinetic regime. Its thermal history is shown in Fig.~\ref{fig:case1_generalization}.

\paragraph{Case 2: Extended four-hour thermal cycle}

The second case extends the thermal treatment to a four-hour cycle. The temperature increases from 293.15~K to 1323.15~K, followed by a prolonged isothermal holding stage, and a final cooling back to 293.15~K. Compared to the baseline case, this thermal cycle is characterized by a longer duration and an extended holding stage at high temperature. Its thermal history is shown in Fig.~\ref{fig:case2_generalization}.

\paragraph{Case 3: Multi-stage realistic thermal cycle}

The third case corresponds to a one-hour multi-stage thermal path. The temperature increases from 293.15~K to 1273.15~K and is held constant, then decreases to 1223.15~K during a short cooling stage and is held again, before being increased to 1323.15~K for a final holding stage. The cycle concludes with cooling to 293.15~K. Its thermal history is shown in Fig.~\ref{fig:case3_generalization}.

\paragraph{Case 4: Complex multi-stage non-monotonic thermal cycle}

The fourth case corresponds to a multi-stage non-monotonic thermal path involving successive heating, holding, cooling, and reheating stages. The temperature first increases from 293.15~K to 1273.15~K and is maintained during an initial holding stage. It is then reduced to 1123.15~K before being increased to 1323.15~K and held constant. A subsequent cooling stage lowers the temperature to 1273.15~K, followed by a final reheating stage back to 1323.15~K and a second holding period. The cycle concludes with cooling to 293.15~K. This non-monotonic profile was introduced to represent thermal treatments involving multiple successive heating and cooling stages over a broad temperature range. Its thermal history is shown in Fig.~\ref{fig:case4_generalization}. A variant of this cycle, in which the intermediate holding stage is performed at 1190.15~K instead of 1123.15~K, is also considered.

\section{Results}
\label{sec:results}
\subsection{Global prediction performance on unseen test sequences}

The overall prediction performance of the different conditioning strategies is summarized in Table~\ref{tab:global_perf}.
The reported results correspond to averages computed over all simulations of the one-hour dataset, each simulation being predicted in the $k$-fold cross-validation by a model that did not use it for training or validation, and evaluated over the full prediction horizon (from $t = 5$ to $t = 59$~min).

\begin{table}[htbp]
\centering
\caption{Global prediction performance averaged over all 86 one-hour simulations ($k$-fold cross-validation)}
\label{tab:global_perf}
\begin{tabular}{lcc}
\toprule
Configuration & MRE (\%) $\downarrow$ & $R^2 \uparrow$ \\
\midrule
C0 -- No thermal input & 45.26 & 0.8055 \\
C1 -- Temperature only ($T$) & 3.10 & 0.9969 \\
C2 -- Thermal rate only ($\dot{T}$) & 4.53 & 0.9905 \\
\textbf{C3 -- Temperature + thermal rate ($T + \dot{T}$)} & \textbf{2.99} & \textbf{0.9970} \\
\bottomrule
\end{tabular}
\end{table}

When no thermal information is provided to the model (C0), considerably larger prediction errors are obtained for all non-isothermal test sequences. The unconditioned configuration reaches an MRE of 45.26\% and an $R^2$ of 0.8055, clearly underperforming the thermally conditioned configurations. For 66 of the 86 simulations, the error exceeds 30\% at some point of the rollout.

Introducing either temperature (C1) or thermal rate (C2) substantially improves prediction accuracy compared with C0. Temperature conditioning leads to an MRE of 3.10\% and an $R^2$ of 0.9969, while thermal-rate conditioning gives an MRE of 4.53\% and an $R^2$ of 0.9905.

The best performance is obtained when temperature and thermal rate are provided simultaneously (C3), with an MRE of 2.99\% and an $R^2$ of 0.9970. Over the 86 simulations, the MRE of individual simulations ranges from 0.78\% to 8.17\%. Among the four conditioning strategies, C3 therefore provides the closest agreement with the reference predictions.

\subsection{Influence of the recurrent formulation within the hybrid framework}

The effect of the recurrent formulation is examined in Table~\ref{tab:recurrent_comp}.
In all cases, the recurrent block (Simple RNN, GRU, or LSTM) is embedded within the same hybrid architecture, combining a CNN for grain size distribution encoding and an MLP for thermal descriptor processing.

\begin{table}[htbp]
\centering
\caption{Comparison of recurrent formulations within the hybrid architecture (CNN + MLP + recurrent).}
\label{tab:recurrent_comp}
\begin{tabular}{lcc}
\toprule
Recurrent block & MRE (\%) $\downarrow$ & $R^2 \uparrow$ \\
\midrule
Simple RNN & 10.33 & 0.9812 \\
GRU & 4.52 & 0.9935 \\
\textbf{LSTM} & \textbf{2.99} & \textbf{0.9970} \\
\bottomrule
\end{tabular}
\end{table}

The Simple RNN shows notably higher prediction errors than the two gated architectures, with an MRE of 10.33\% and an $R^2$ of 0.9812. For 14 of the 86 simulations, its error exceeds 30\% at some point of the rollout. In contrast, both the GRU and LSTM provide substantially lower errors and maintain a high agreement with the reference evolutions. Among the three recurrent formulations, the LSTM achieves the best overall performance, with an MRE of 2.99\% and an $R^2$ of 0.9970, compared with 4.52\% and 0.9935 for the GRU, respectively.

\begin{figure}[htbp]
    \centering
    \begin{subfigure}[t]{0.48\textwidth}
        \centering
        \includegraphics[width=\linewidth]{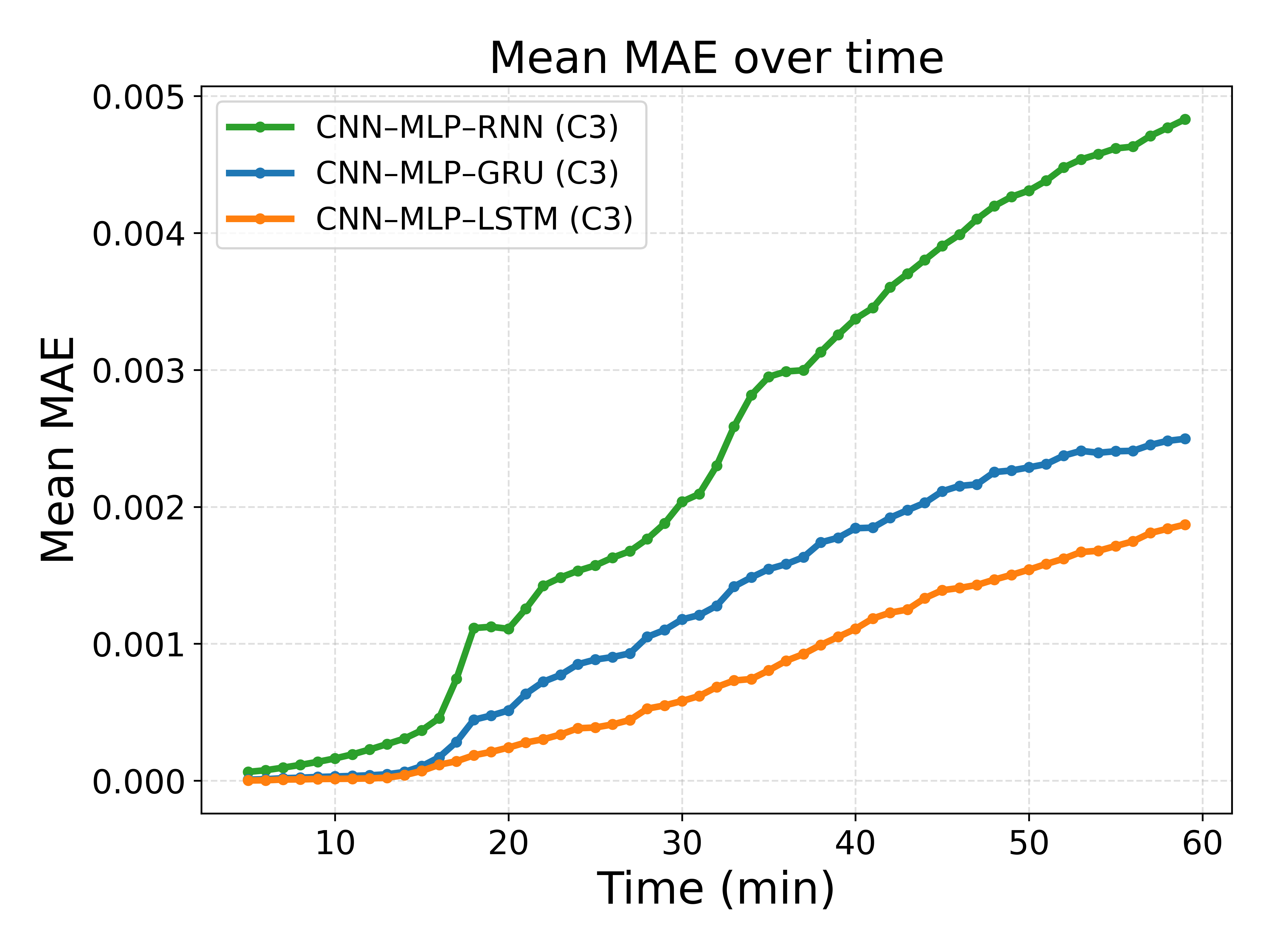}
        \caption{}
    \end{subfigure}
    \hfill
    \begin{subfigure}[t]{0.48\textwidth}
        \centering
        \includegraphics[width=\linewidth]{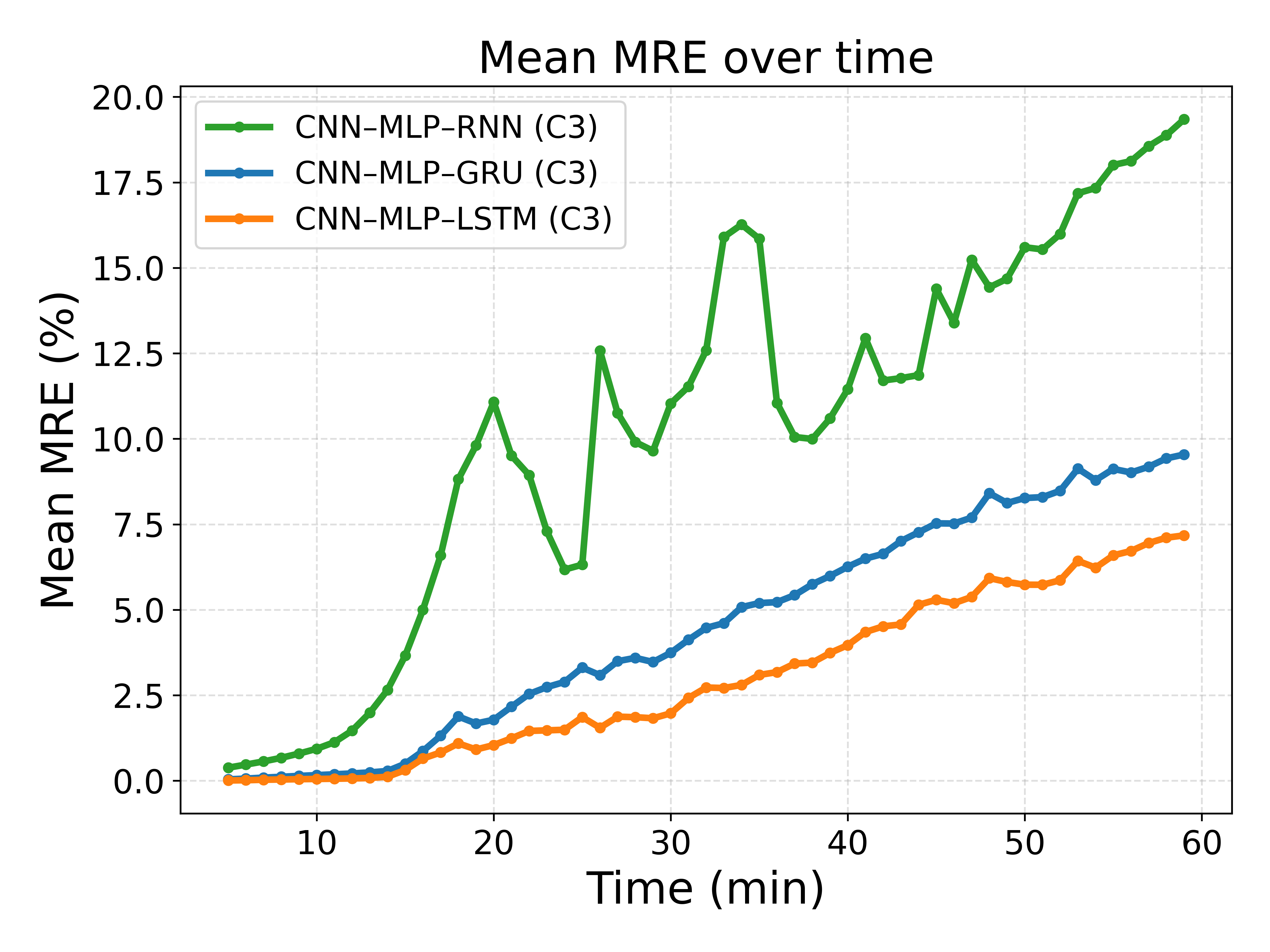}
        \caption{}
    \end{subfigure}

    \vspace{0.2cm}

    \begin{subfigure}[t]{0.48\textwidth}
        \centering
        \includegraphics[width=\linewidth]{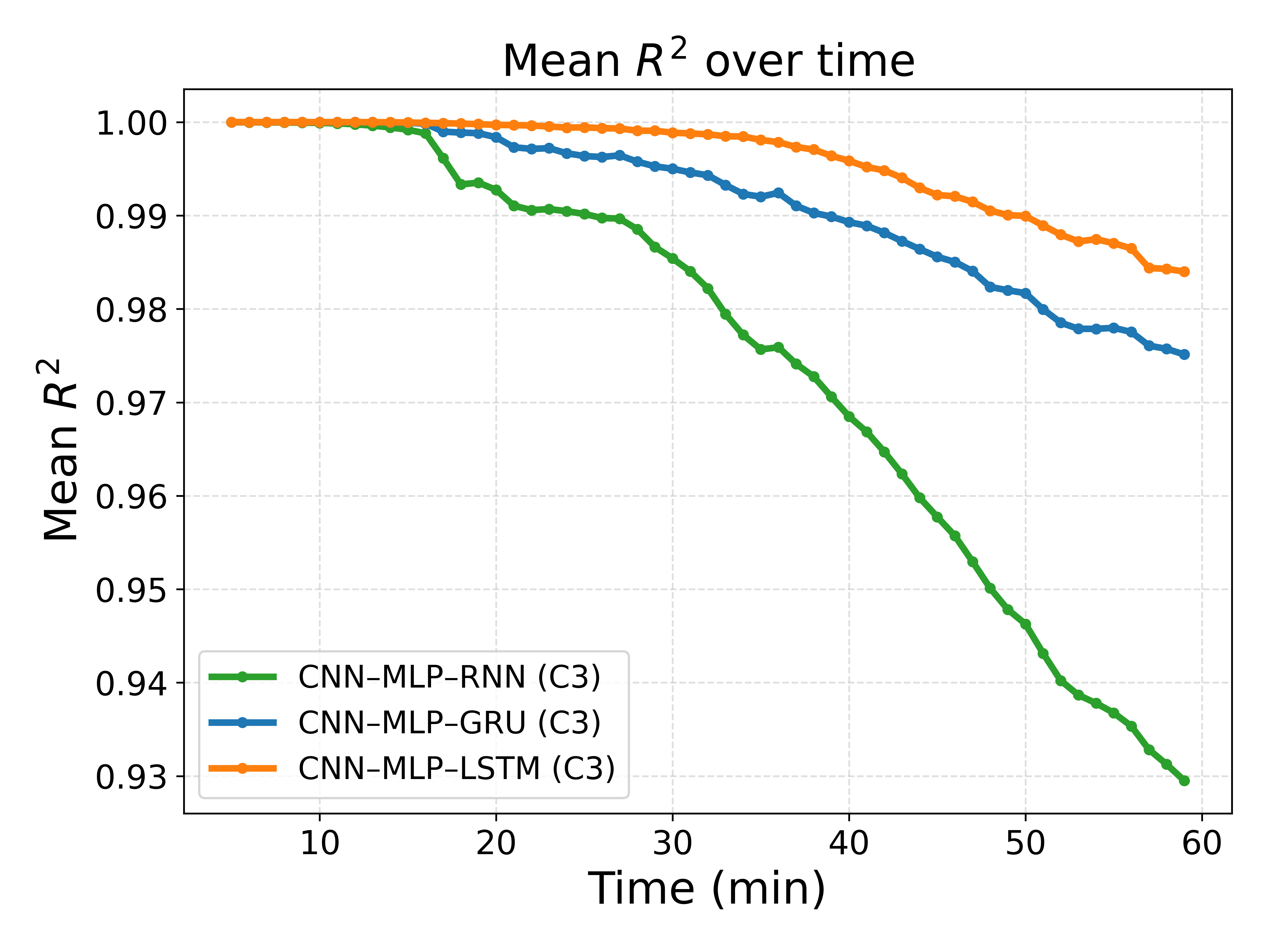}
        \caption{}
    \end{subfigure}
    \caption{Comparison of the recurrent formulations within the hybrid architecture (CNN + MLP + recurrent block) over the 86 one-hour simulations: mean (a) MAE, (b) MRE and (c) $R^2$ over time.}
    \label{fig:recurrent_time}
\end{figure}

The temporal evolution of the errors of the three recurrent formulations is shown in Fig.~\ref{fig:recurrent_time}. The error of the Simple RNN increases from the beginning of the heating stage and shows strong oscillations, whereas the GRU and the LSTM remain close during the first part of the cycle. The gap between the GRU and the LSTM then increases progressively: at the end of the cycle, the mean MRE reaches about 19\% for the Simple RNN and 9.5\% for the GRU, compared with 7.2\% for the LSTM, and the mean $R^2$ is about 0.93, 0.975 and 0.984, respectively. The advantage of the LSTM therefore lies mainly in its ability to limit the accumulation of errors during the auto-regressive rollout.

\origsubsection{Contribution of the hybrid architecture components}

To isolate the contribution of the architectural components beyond the recurrent block itself, Table~\ref{tab:arch_comp} compares two LSTM-based configurations under identical thermal conditioning ($T + \dot{T}$).

In the \emph{LSTM only} configuration, grain size distributions and thermal descriptors are injected directly into the LSTM without prior processing. In the hybrid configuration, the grain size distributions are first encoded by the CNN and the thermal descriptors by the MLP, and the resulting features are combined before the LSTM.

\begin{table}[htbp]
\centering
\caption{Impact of architectural components under identical thermal conditioning (one-hour dataset, $k$-fold cross-validation).}
\label{tab:arch_comp}
\begin{tabular}{lccccc}
\toprule
Configuration & CNN & MLP & MRE (\%) $\downarrow$ & Max.\ MRE (\%) $\downarrow$ & $R^2 \uparrow$ \\
\midrule
LSTM only & -- & -- & 3.80 & 10.75 & 0.9942 \\
\textbf{CNN + MLP + LSTM} & \checkmark & \checkmark & \textbf{2.99} & \textbf{8.17} & \textbf{0.9970} \\
\bottomrule
\end{tabular}
\end{table}

The comparison shows that encoding the grain size distributions with the CNN and the thermal descriptors with the MLP before the LSTM improves the prediction performance. The MRE decreases from 3.80\% to 2.99\%, the largest error among the 86 simulations decreases from 10.75\% to 8.17\%, and the $R^2$ increases from 0.9942 to 0.9970. In addition, two simulations exceed an MRE of 30\% at some point of the rollout with the LSTM-only configuration, compared with none for the hybrid architecture.

\subsection{Phase-wise error analysis}

To better understand how prediction errors are distributed across different thermo-kinetic regimes, the Mean Relative Error is evaluated separately for the heating phase, the isothermal plateau, and the cooling phase.
The results, averaged over all test sequences, are reported in Table~\ref{tab:phase_mre}.

\begin{table}[htbp]
\centering
\caption{Phase-wise Mean Relative Error (averages over the 86 one-hour simulations).}
\label{tab:phase_mre}
\begin{tabular}{lccc}
\toprule
Configuration & Heating $\downarrow$ & Plateau $\downarrow$ & Cooling $\downarrow$ \\
\midrule
C0 -- No thermal input & 41.26 & 39.78 & 56.50 \\
C1 -- $T$ only & 0.48 & 3.76 & 5.61 \\
C2 -- $\dot{T}$ only & 1.44 & 5.34 & 7.48 \\
\textbf{C3 -- $T + \dot{T}$} & \textbf{0.44} & \textbf{3.59} & \textbf{5.49} \\
\bottomrule
\end{tabular}
\end{table}

Clear differences are observed between the three thermal stages. For the thermally conditioned configurations, the errors remain relatively low during heating and increase during the isothermal plateau and cooling. This increase is most pronounced for C2, with an MRE of 7.48\% during cooling, compared with 5.61\% for C1. In contrast, C3 maintains the lowest error throughout the three stages, with MRE values of 0.44\%, 3.59\%, and 5.49\% during heating, holding, and cooling, respectively.

\subsection{Temporal evolution and physical consistency of the selected model}

Following the global and phase-wise performance analyses, the temporal behavior of the selected model is examined over the full prediction horizon. The joint conditioning strategy (C3) combined with the LSTM recurrent architecture is considered in more detail, as it provided the most accurate predictions in the previous analyses.

\begin{figure}[htbp]
    \centering

    \begin{subfigure}[t]{0.48\textwidth}
        \centering
        \includegraphics[width=\linewidth]{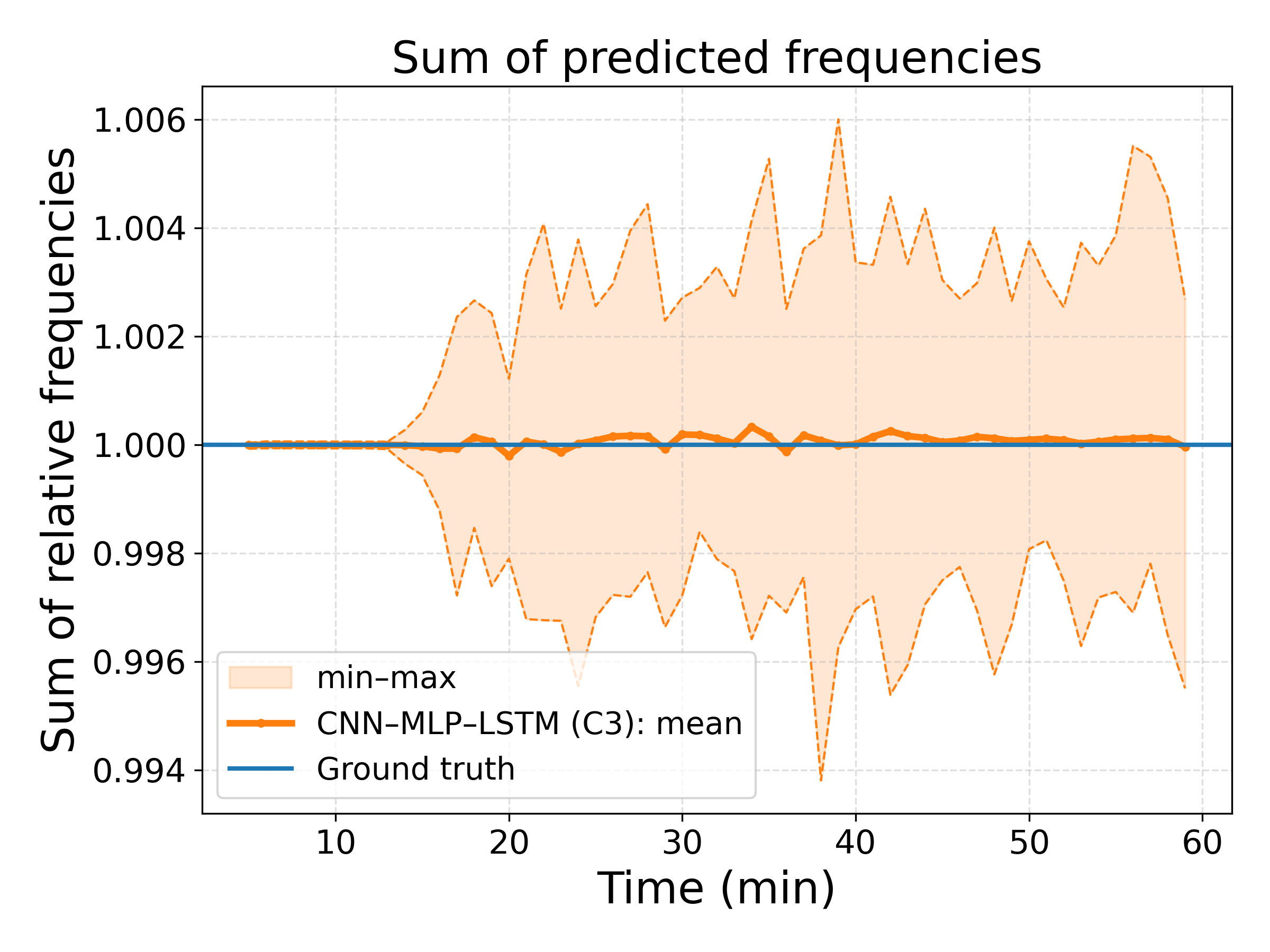}
        \caption{}
        \label{fig:temporal_grain_count}
    \end{subfigure}
    \hfill
    \begin{subfigure}[t]{0.48\textwidth}
        \centering
        \includegraphics[width=\linewidth]{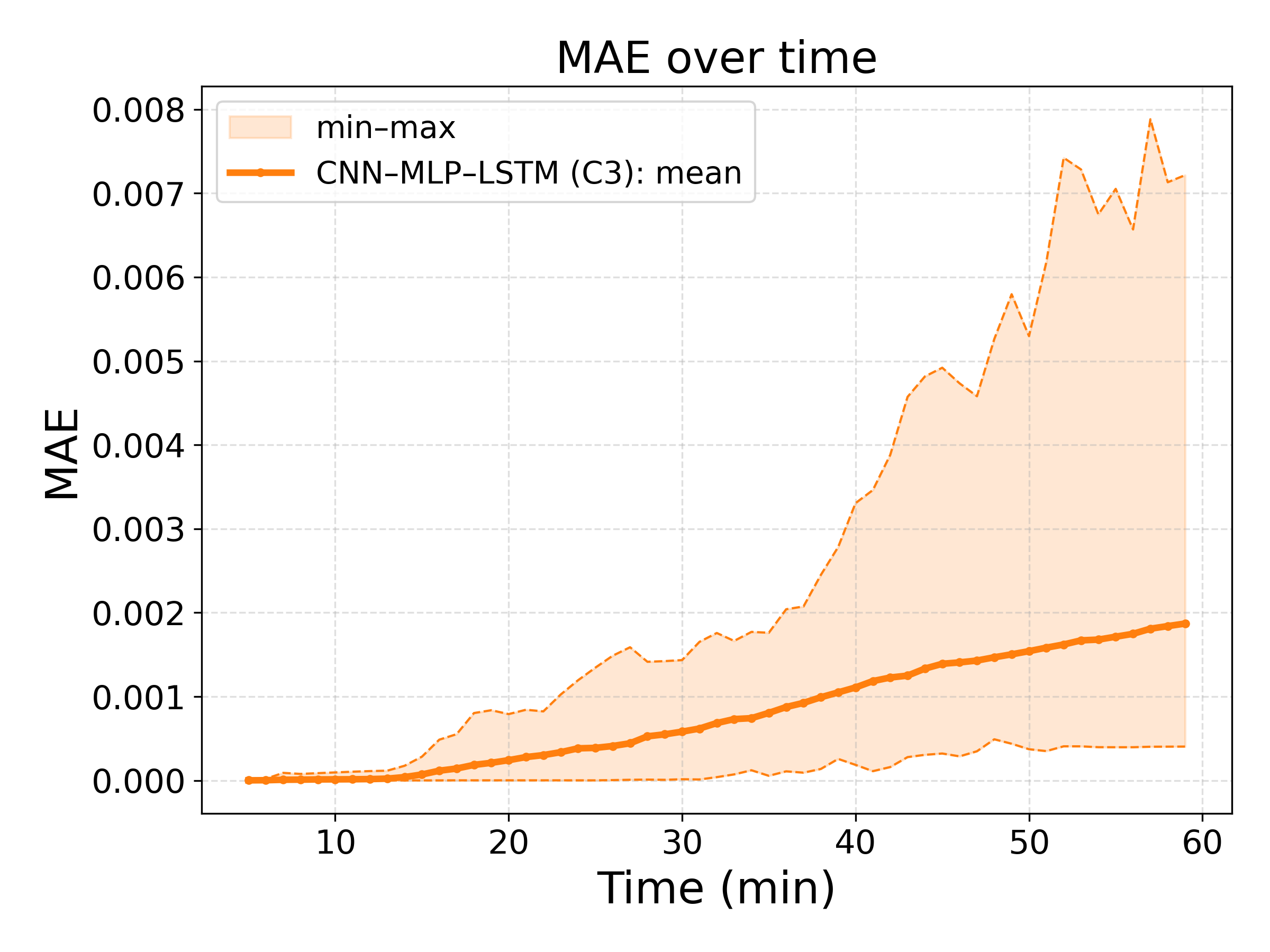}
        \caption{}
        \label{fig:temporal_mae}
    \end{subfigure}

    \vspace{0.2cm}

    \begin{subfigure}[t]{0.48\textwidth}
        \centering
        \includegraphics[width=\linewidth]{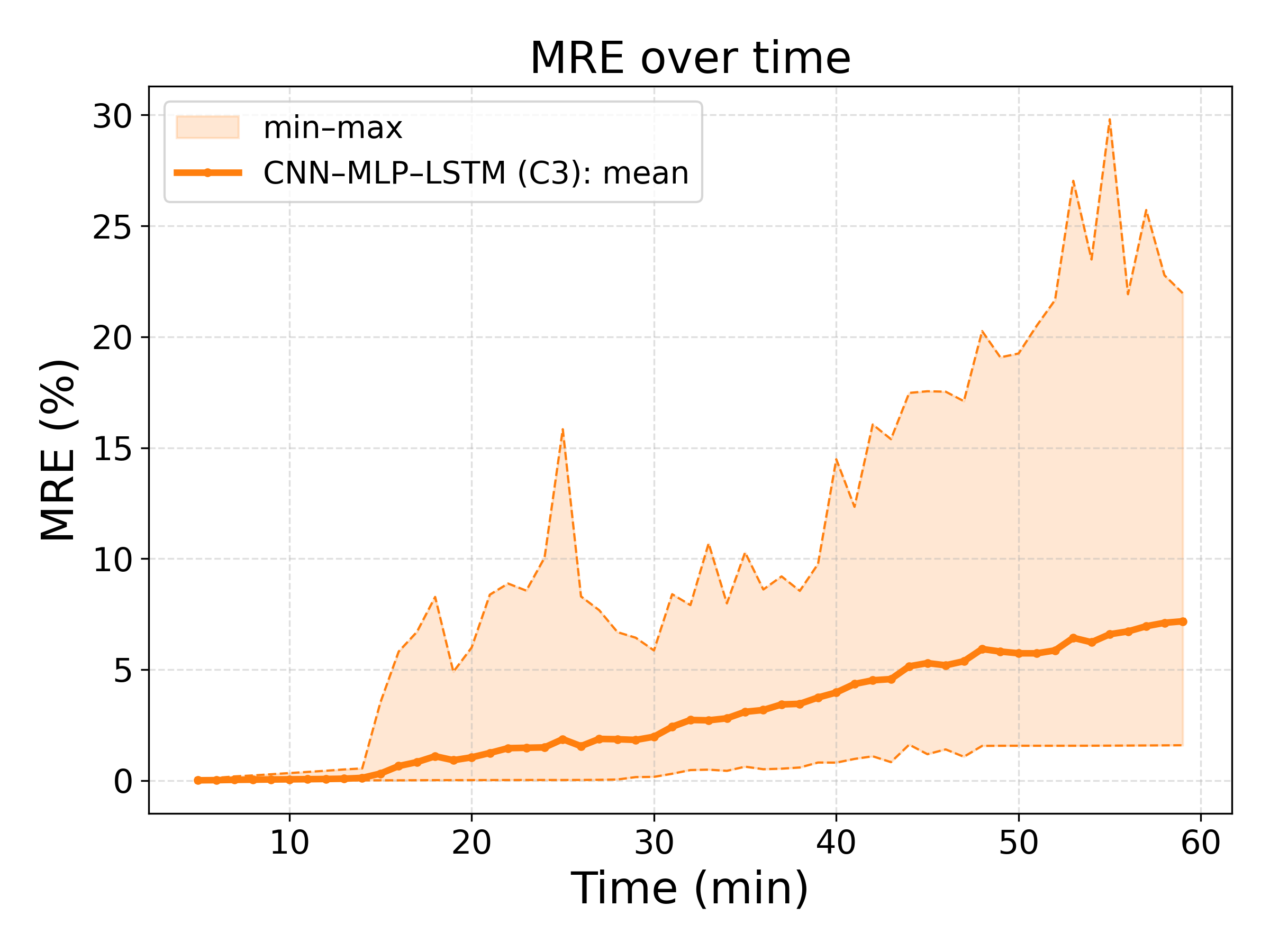}
        \caption{}
        \label{fig:temporal_mre}
    \end{subfigure}
    \hfill
    \begin{subfigure}[t]{0.48\textwidth}
        \centering
        \includegraphics[width=\linewidth]{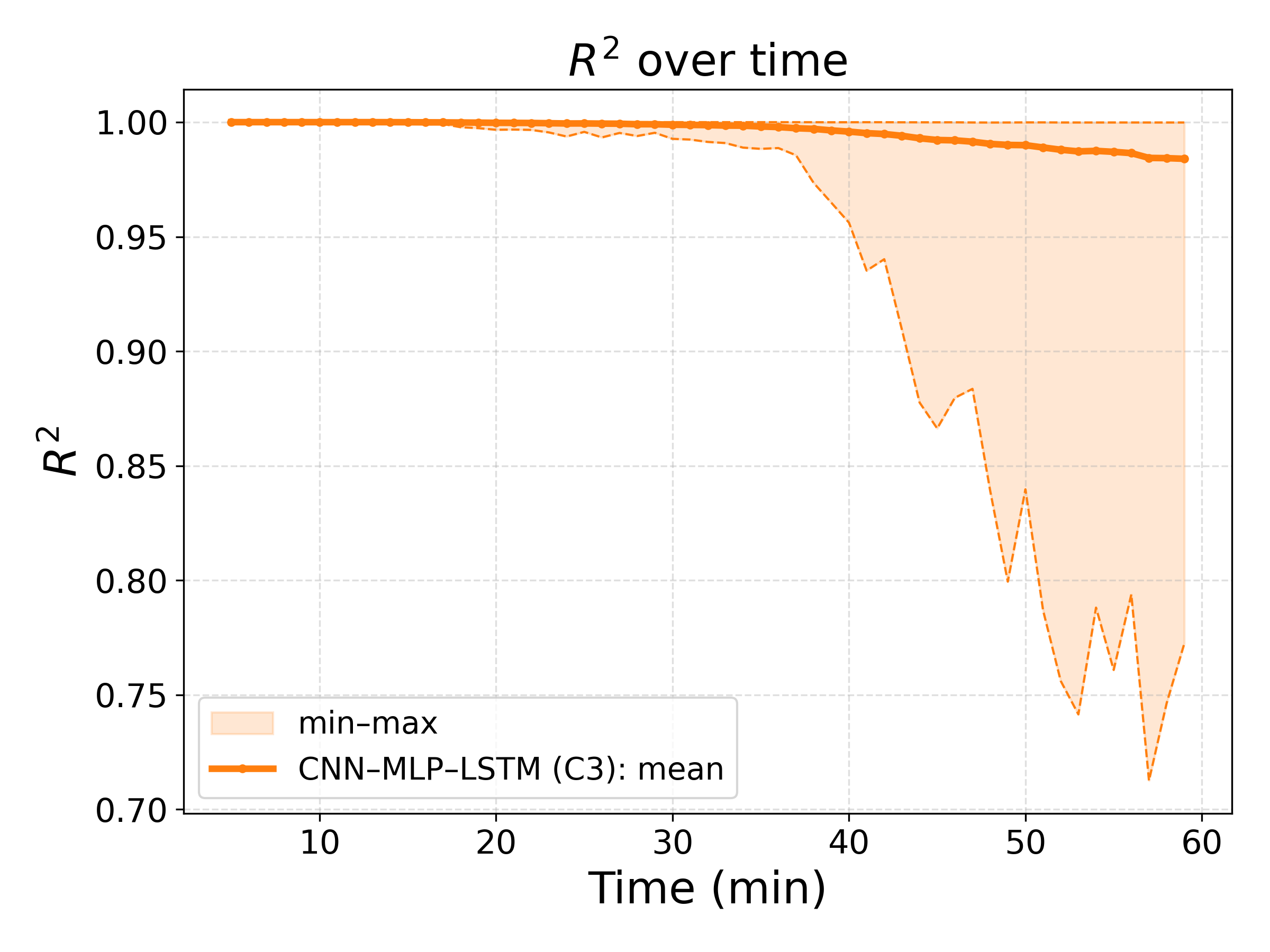}
        \caption{}
        \label{fig:temporal_r2}
    \end{subfigure}
\caption{Temporal evolution of forecasting accuracy and physical consistency for the selected hybrid CNN--MLP--LSTM model (C3) over the 86 one-hour simulations: mean (line) and minimum--maximum envelope (shaded area).
(a) Sum of predicted relative frequencies.
(b) Mean Absolute Error (MAE) over time.
(c) Mean Relative Error (MRE) over time.
(d) Coefficient of determination ($R^2$) over time.}

    \label{fig:temporal_stability_c3}
\end{figure}

Figure~\ref{fig:temporal_stability_c3}(a) reports the temporal evolution of the sum of predicted relative frequencies over the 86 simulations (mean, minimum and maximum). The reference distributions remain equal to unity by construction, while the predicted values remain close to one throughout the prediction horizon. The predicted distributions therefore remain close to the expected normalization during the auto-regressive rollout.

The temporal evolution of the Mean Absolute Error and Mean Relative Error is presented in Figs.~\ref{fig:temporal_stability_c3}(b) and~\ref{fig:temporal_stability_c3}(c), respectively. Both error measures increase progressively as the prediction horizon extends, the mean MRE reaching about 7\% at the end of the cycle. The maximum envelope shows occasional peaks at the end of the rollout, which remain below 30\% for all simulations. These peaks occur mainly for the smallest domains with the widest initial distributions (2~mm, $\sigma = 32~\mu$m), which contain about 500 grains at the end of the simulation: several bins then contain only a few grains, so that a difference of a single grain produces a large relative error in that bin.

The evolution of the coefficient of determination is shown in Fig.~\ref{fig:temporal_stability_c3}(d).
Although a gradual decrease in $R^2$ is observed as the prediction progresses, the values remain high over the full sequence, indicating that the predicted trajectories retain a strong correlation with the reference evolution, including at later stages.

\begin{figure}[htbp]
    \centering

    \begin{subfigure}[t]{0.48\textwidth}
        \centering
        \includegraphics[width=\linewidth]{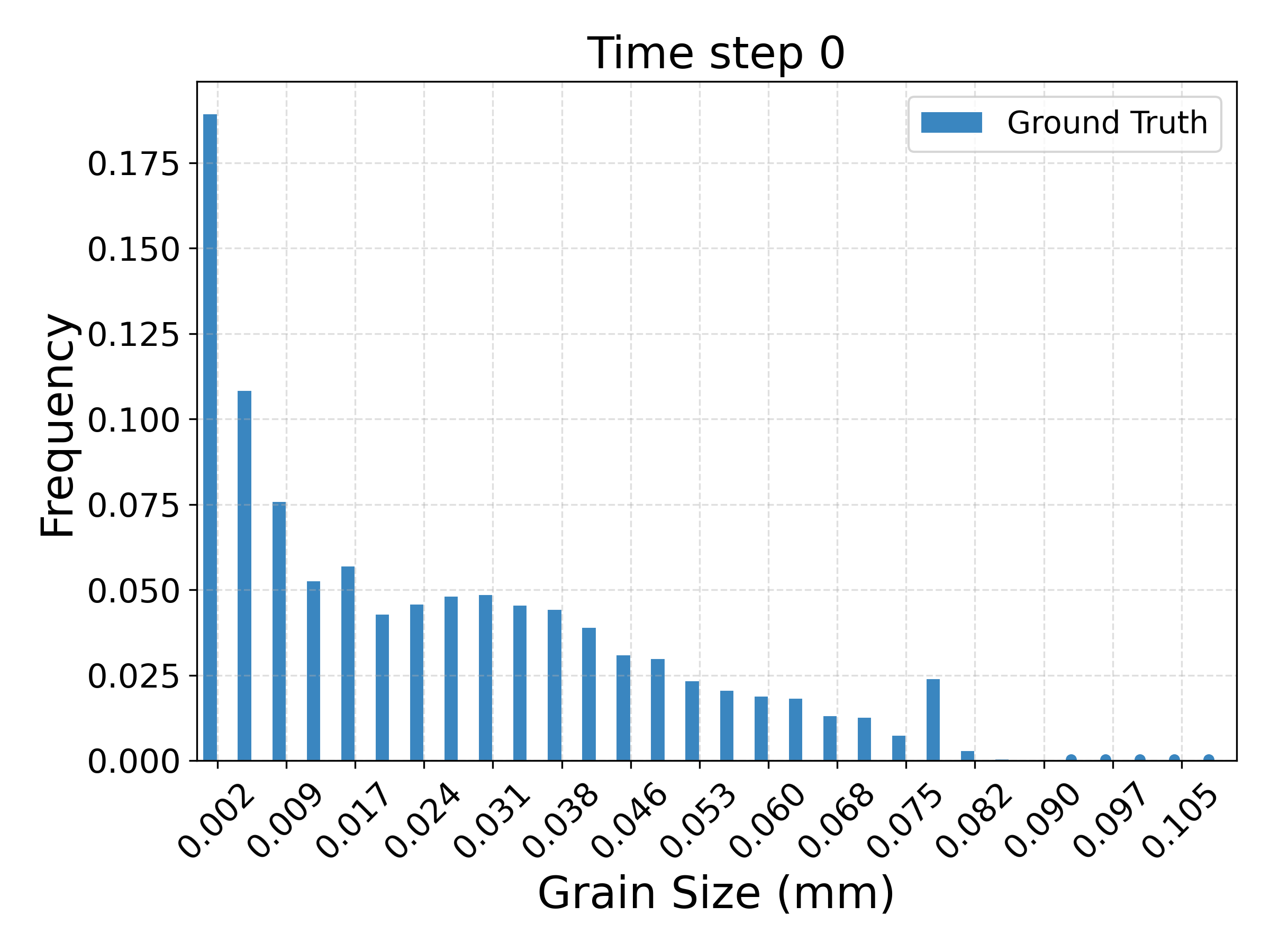}
        \caption{}
    \end{subfigure}
    \hfill
    \begin{subfigure}[t]{0.48\textwidth}
        \centering
        \includegraphics[width=\linewidth]{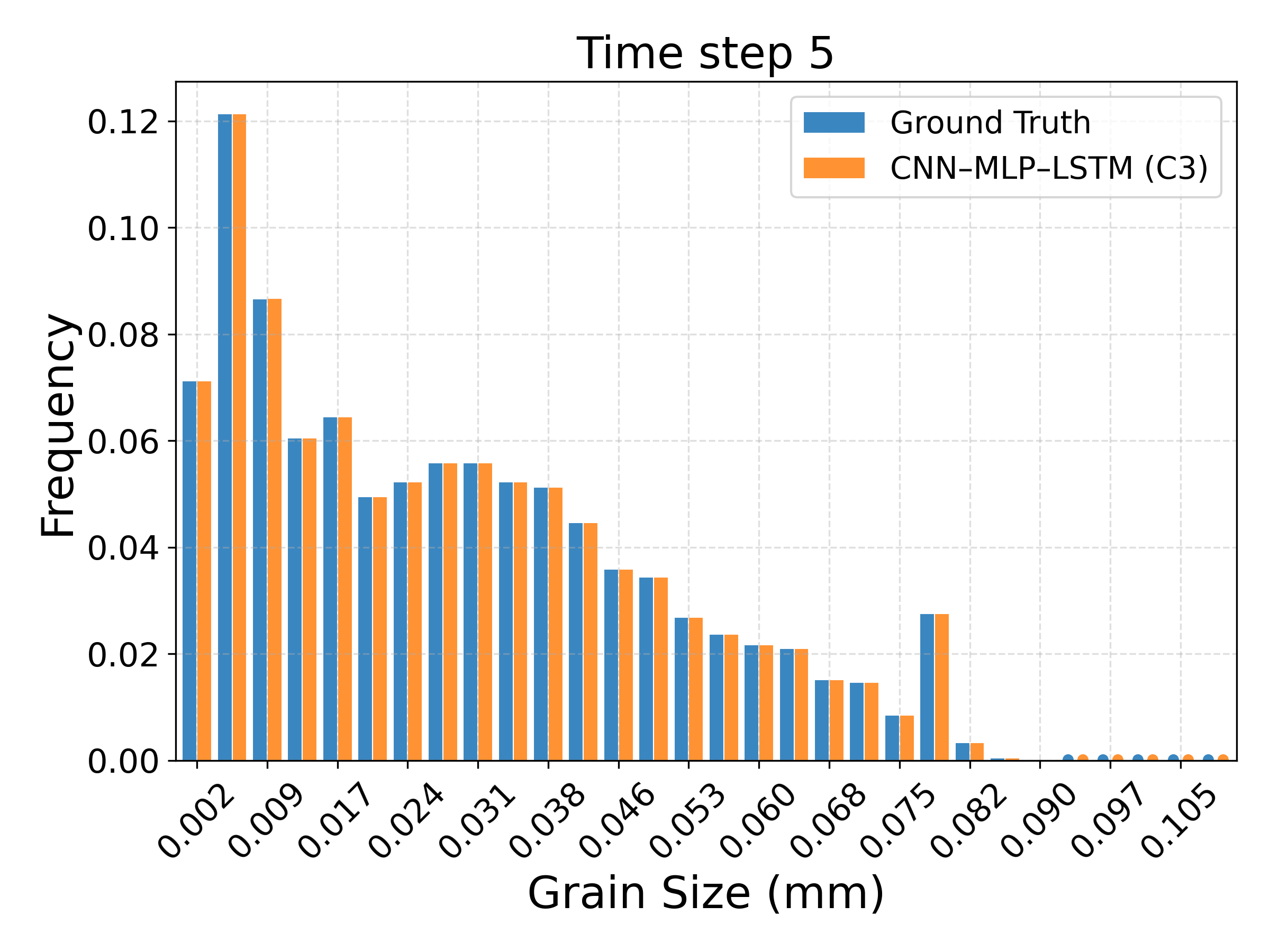}
        \caption{}
    \end{subfigure}

    \vspace{0.2cm}

    \begin{subfigure}[t]{0.48\textwidth}
        \centering
        \includegraphics[width=\linewidth]{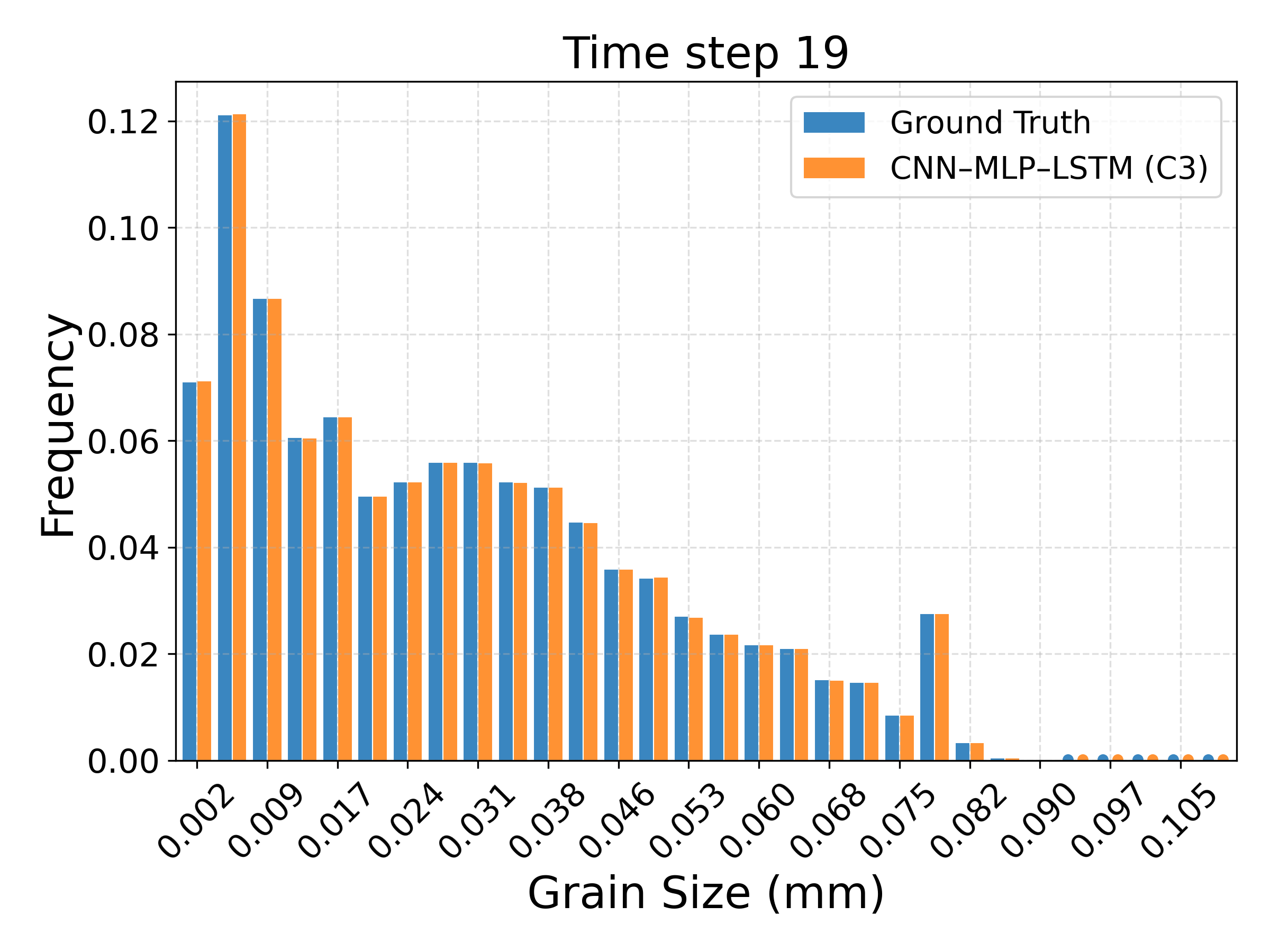}
        \caption{}
    \end{subfigure}
    \hfill
    \begin{subfigure}[t]{0.48\textwidth}
        \centering
        \includegraphics[width=\linewidth]{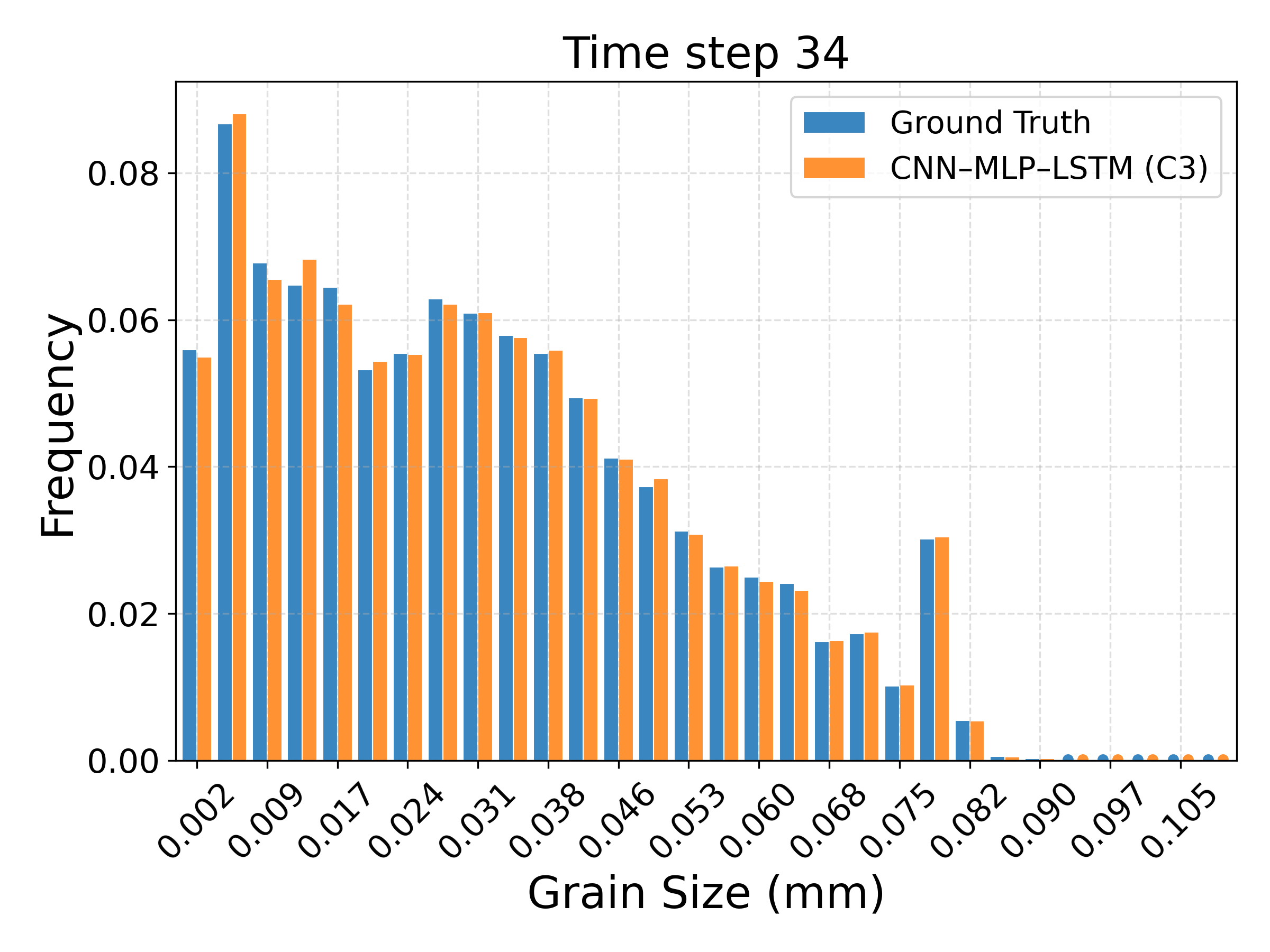}
        \caption{}
    \end{subfigure}

    \vspace{0.2cm}

    \begin{subfigure}[t]{0.55\textwidth}
        \centering
        \includegraphics[width=\linewidth]{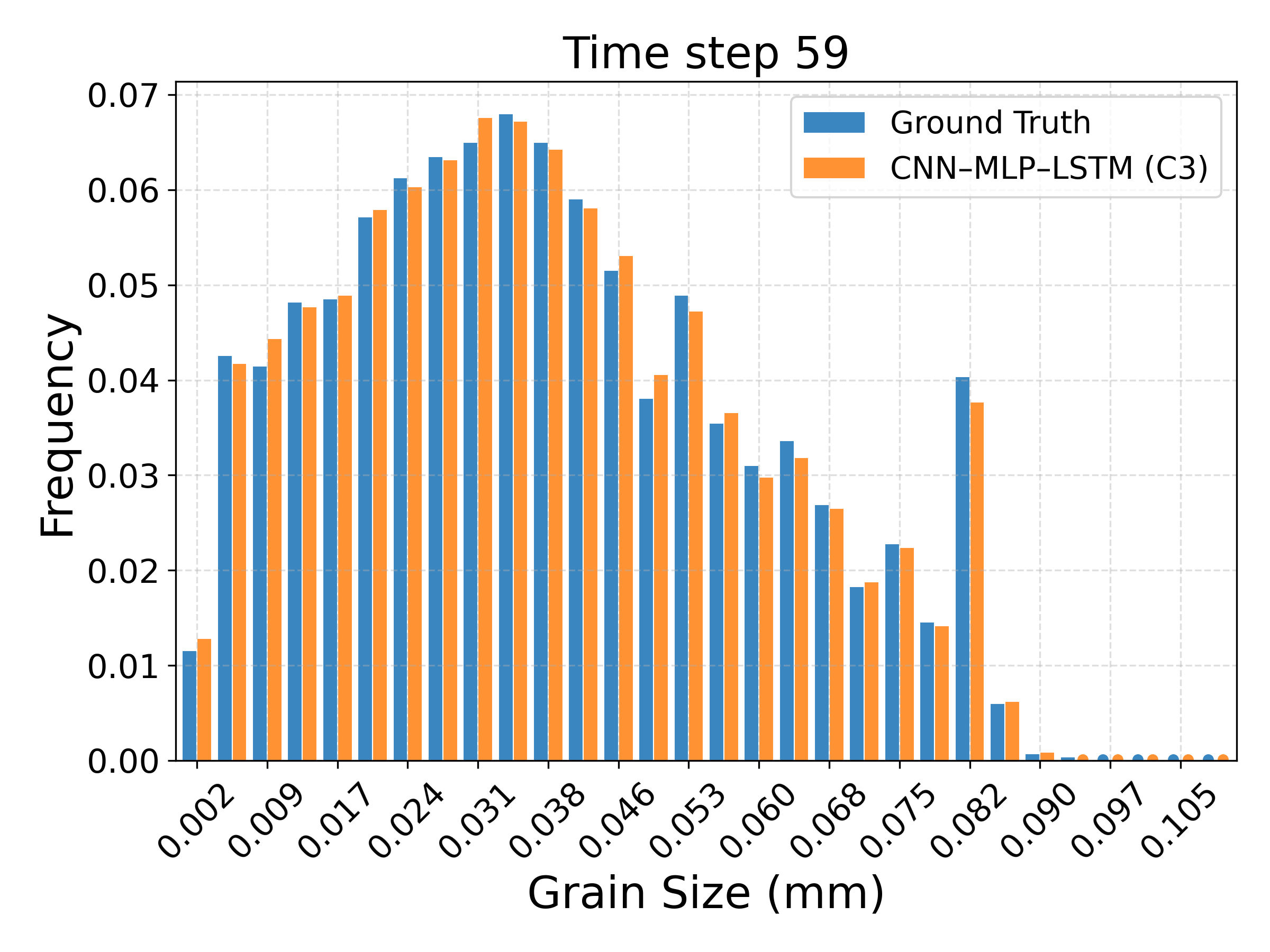}
        \caption{}
    \end{subfigure}

\caption{Comparison of predicted (CNN--MLP--LSTM) and TRM normalized grain-size distributions for a representative test sequence at selected prediction steps: (a) the initial state, (b) the first predicted time step, (c) an intermediate stage of the thermal cycle, (d) a later stage of grain growth, and (e) the final state of the sequence.}

    \label{fig:temporal_distribution_example}
\end{figure}

To further illustrate the temporal behavior of the selected model, Fig.~\ref{fig:temporal_distribution_example} shows the evolution of the normalized grain-size distribution for a representative sequence (the simulation with the median MRE, 2.71\%). The selected snapshots correspond to the initial state, an early prediction step, the end of heating, the end of the isothermal holding stage, and the final state after cooling. The progressive displacement of the distribution toward larger grain sizes is observed throughout the thermal cycle and remains consistent with the reference evolution. From the first predicted step to the final cooling stage, the main characteristics of the distribution are maintained, including the location of the dominant peak and the overall spread of grain sizes. Small differences become visible at later times, particularly for the largest grain classes. These differences remain limited compared with the overall evolution of the distribution.

\begin{figure}[htbp]
    \centering
    \includegraphics[width=0.9\textwidth]{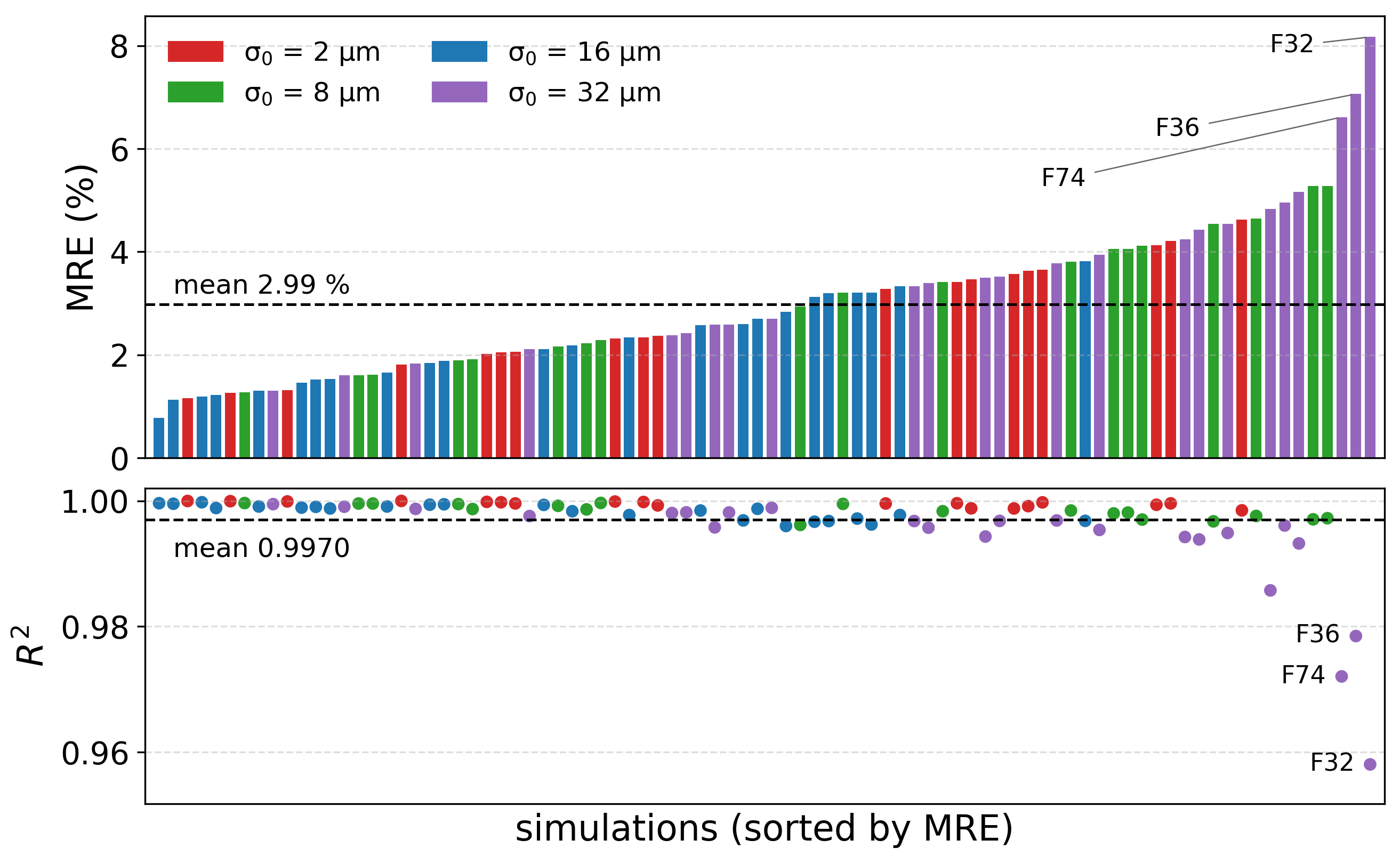}
    \caption{MRE (top) and $R^2$ (bottom) of each of the 86 one-hour simulations ($k$-fold cross-validation), sorted by increasing MRE and coloured by the standard deviation $\sigma_0$ of the initial distribution. The dashed lines give the mean values.}
    \label{fig:per_simulation}
\end{figure}

The errors of the individual simulations are shown in Fig.~\ref{fig:per_simulation}, where the 86 simulations are sorted by increasing MRE. The MRE remains below 5\% for 80 of the 86 simulations and below 8.2\% for all of them, and the $R^2$ is higher than 0.99 for 82 simulations. The three largest errors are obtained for F32, F36 and F74, which all correspond to the widest initial distributions ($\sigma_0 = 32~\mu$m) in the smallest domains (2 and 3~mm). These simulations contain few grains at the end of the cycle (about 500 to 1100, compared with a median of about 3200 over the dataset), so that several bins contain only a few grains and the relative error of these bins increases. The mean values reported in Table~\ref{tab:global_perf} are therefore representative of the whole dataset and are not driven by a few well-predicted simulations.

\origsubsection{Four-hour thermal cycles}

\begin{figure}[htbp]
    \centering
    \begin{subfigure}[t]{0.48\textwidth}
        \centering
        \includegraphics[width=\linewidth]{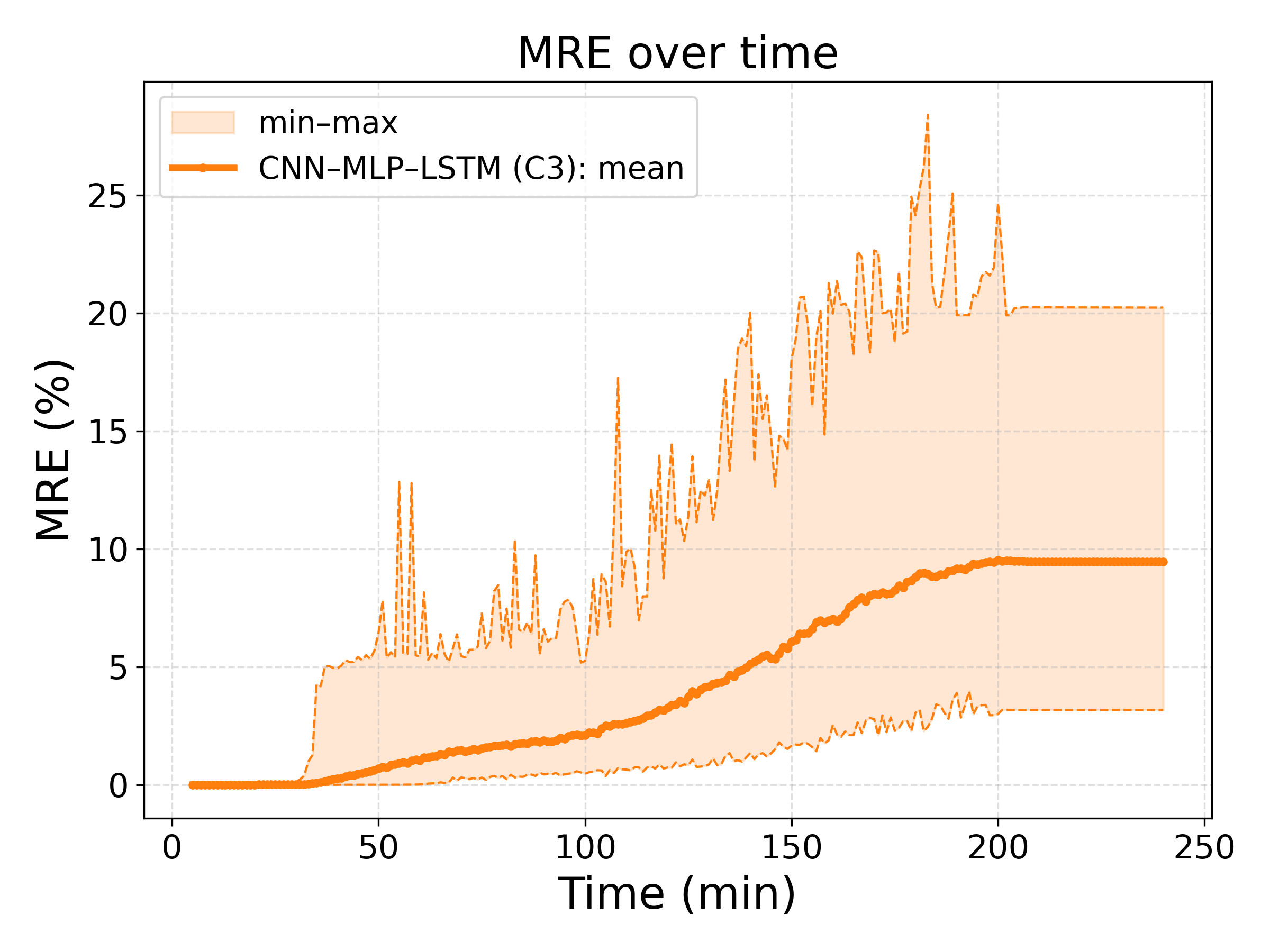}
        \caption{}
    \end{subfigure}
    \hfill
    \begin{subfigure}[t]{0.48\textwidth}
        \centering
        \includegraphics[width=\linewidth]{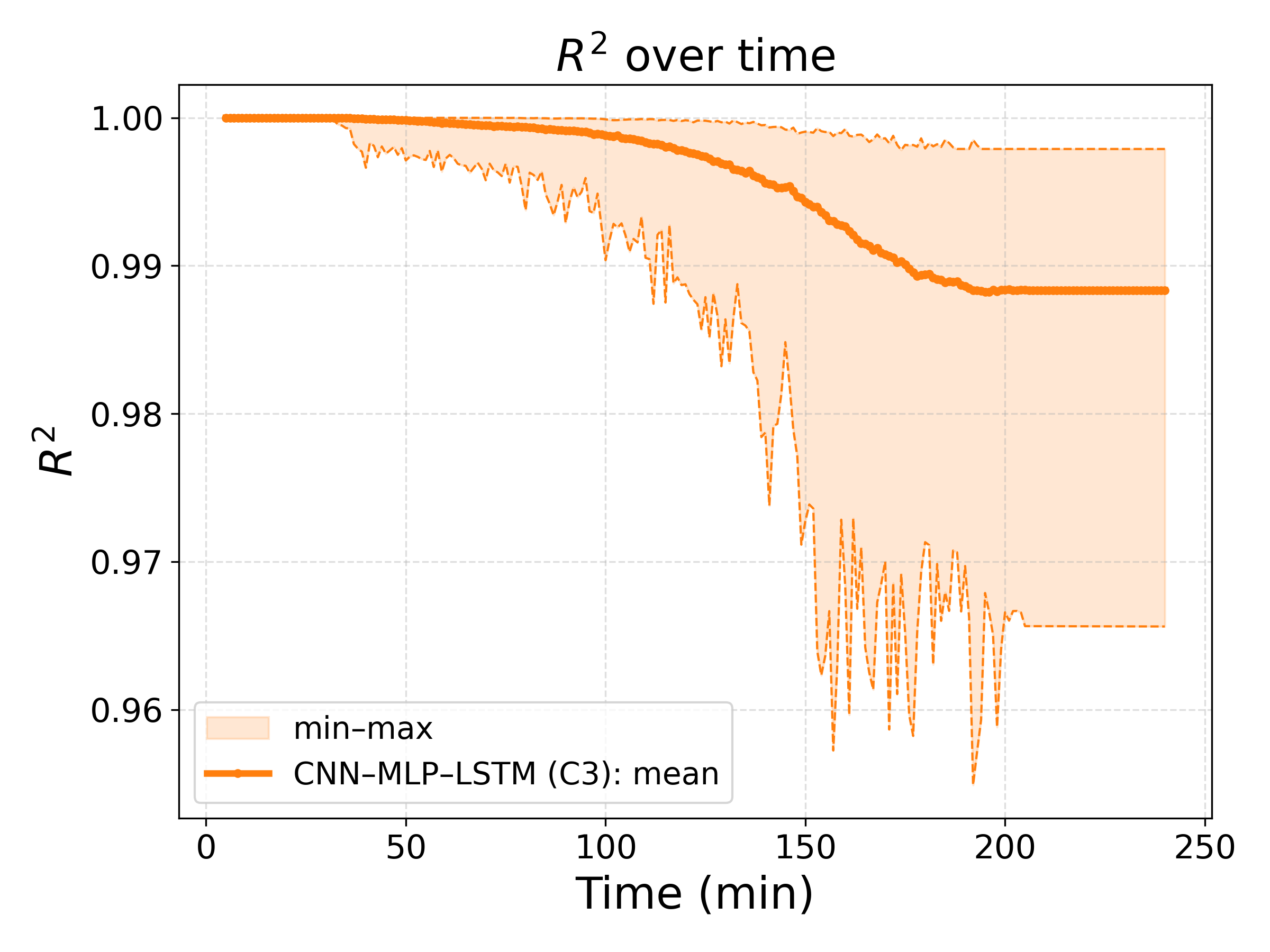}
        \caption{}
    \end{subfigure}
    \caption{Four-hour thermal cycles: (a) MRE and (b) $R^2$ over time for the 208 simulations ($k$-fold cross-validation): mean (line) and minimum--maximum envelope (shaded area).}
    \label{fig:temporal_4h}
\end{figure}

\begin{figure}[htbp]
    \centering
    \begin{subfigure}[t]{0.48\textwidth}
        \centering
        \includegraphics[width=\linewidth]{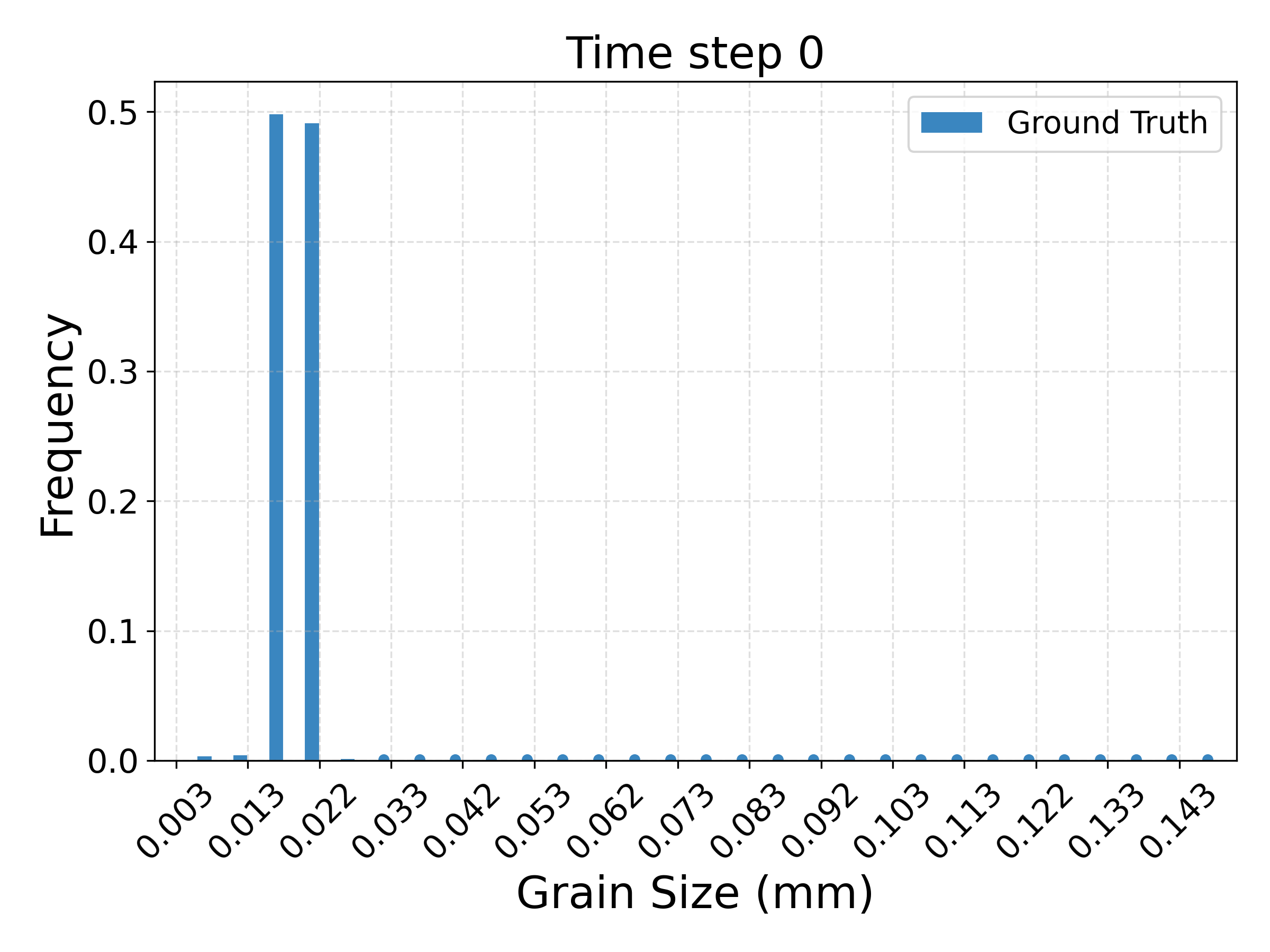}
        \caption{}
    \end{subfigure}
    \hfill
    \begin{subfigure}[t]{0.48\textwidth}
        \centering
        \includegraphics[width=\linewidth]{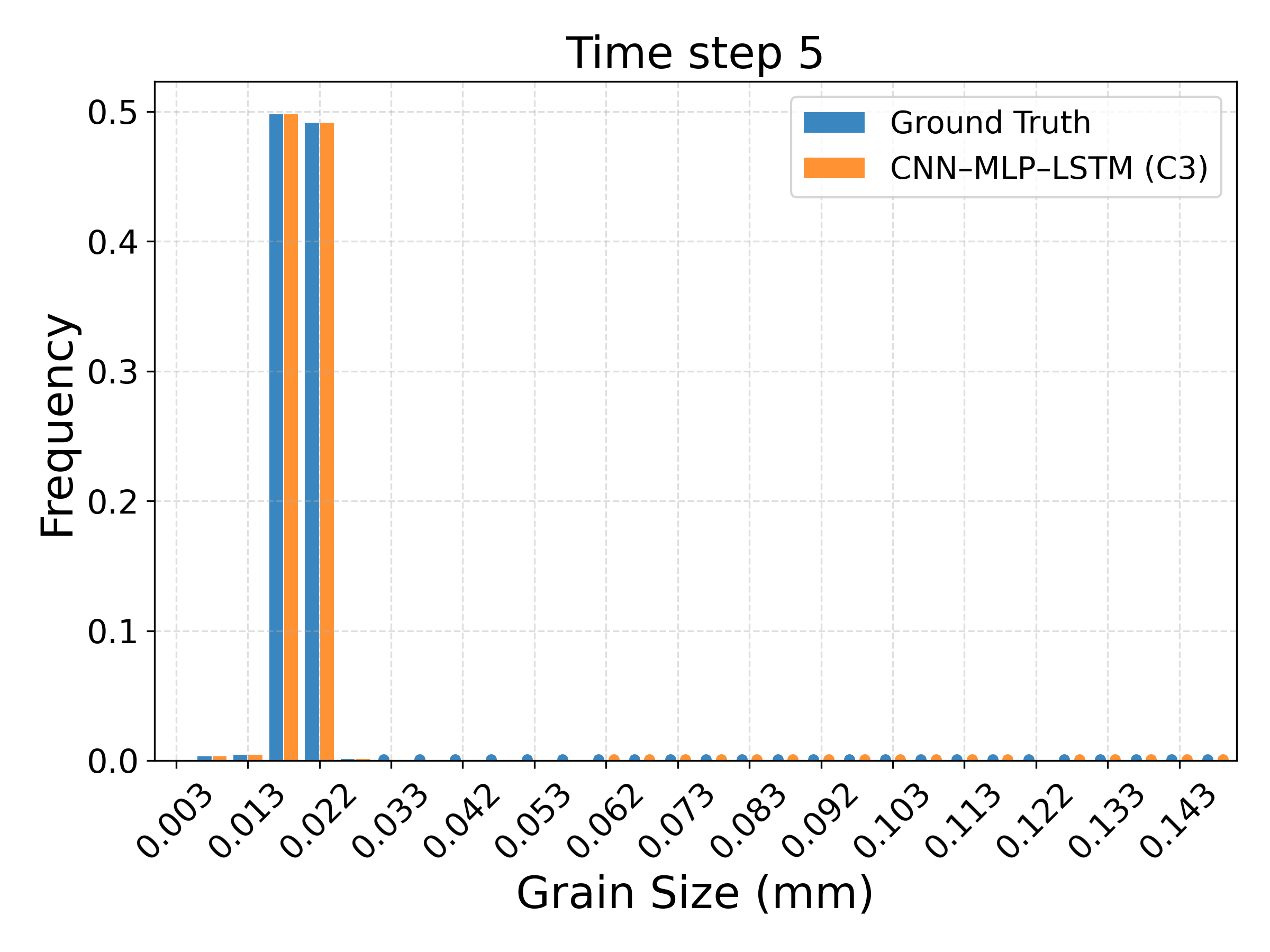}
        \caption{}
    \end{subfigure}

    \vspace{0.2cm}

    \begin{subfigure}[t]{0.48\textwidth}
        \centering
        \includegraphics[width=\linewidth]{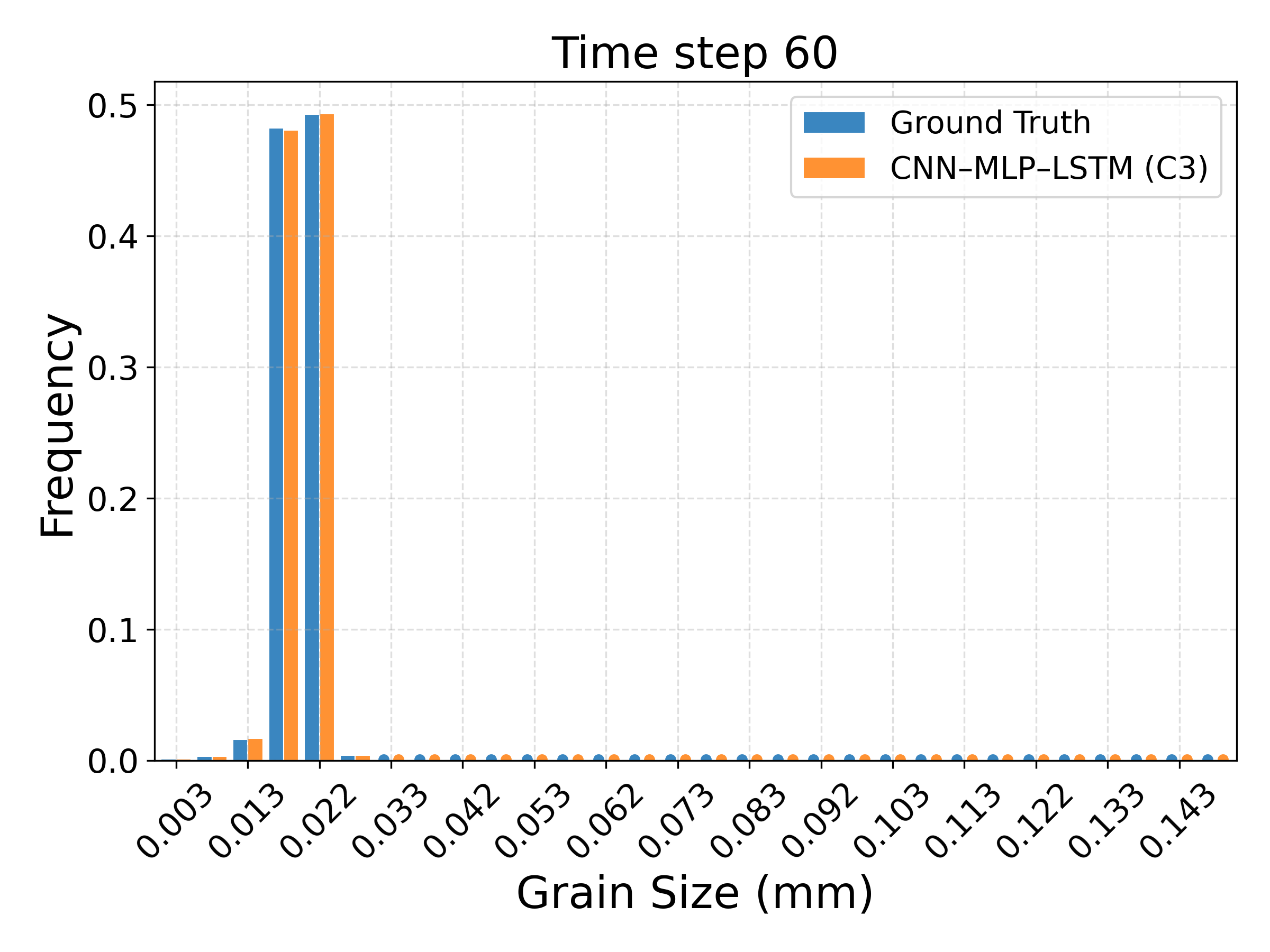}
        \caption{}
    \end{subfigure}
    \hfill
    \begin{subfigure}[t]{0.48\textwidth}
        \centering
        \includegraphics[width=\linewidth]{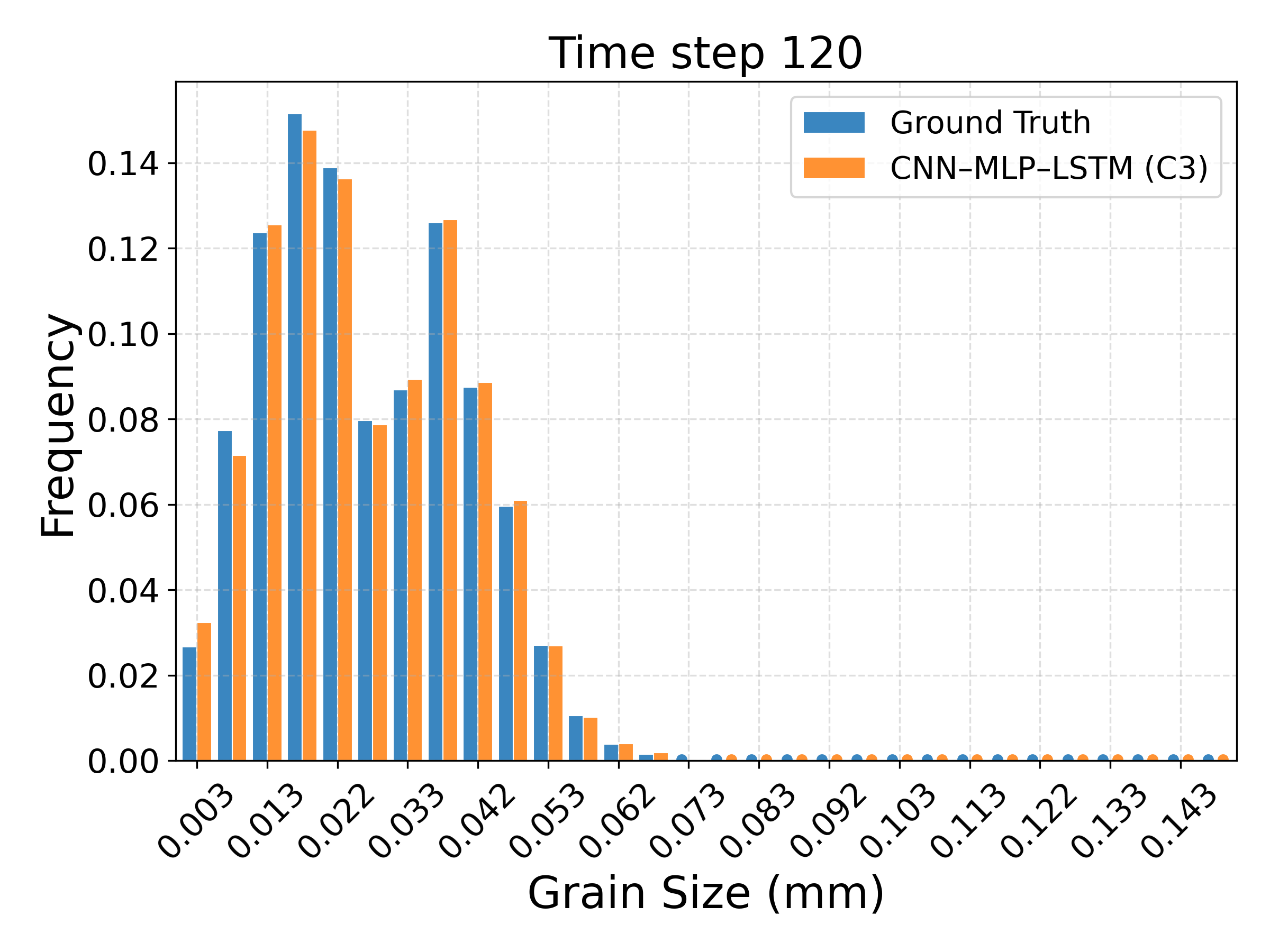}
        \caption{}
    \end{subfigure}

    \vspace{0.2cm}

    \begin{subfigure}[t]{0.48\textwidth}
        \centering
        \includegraphics[width=\linewidth]{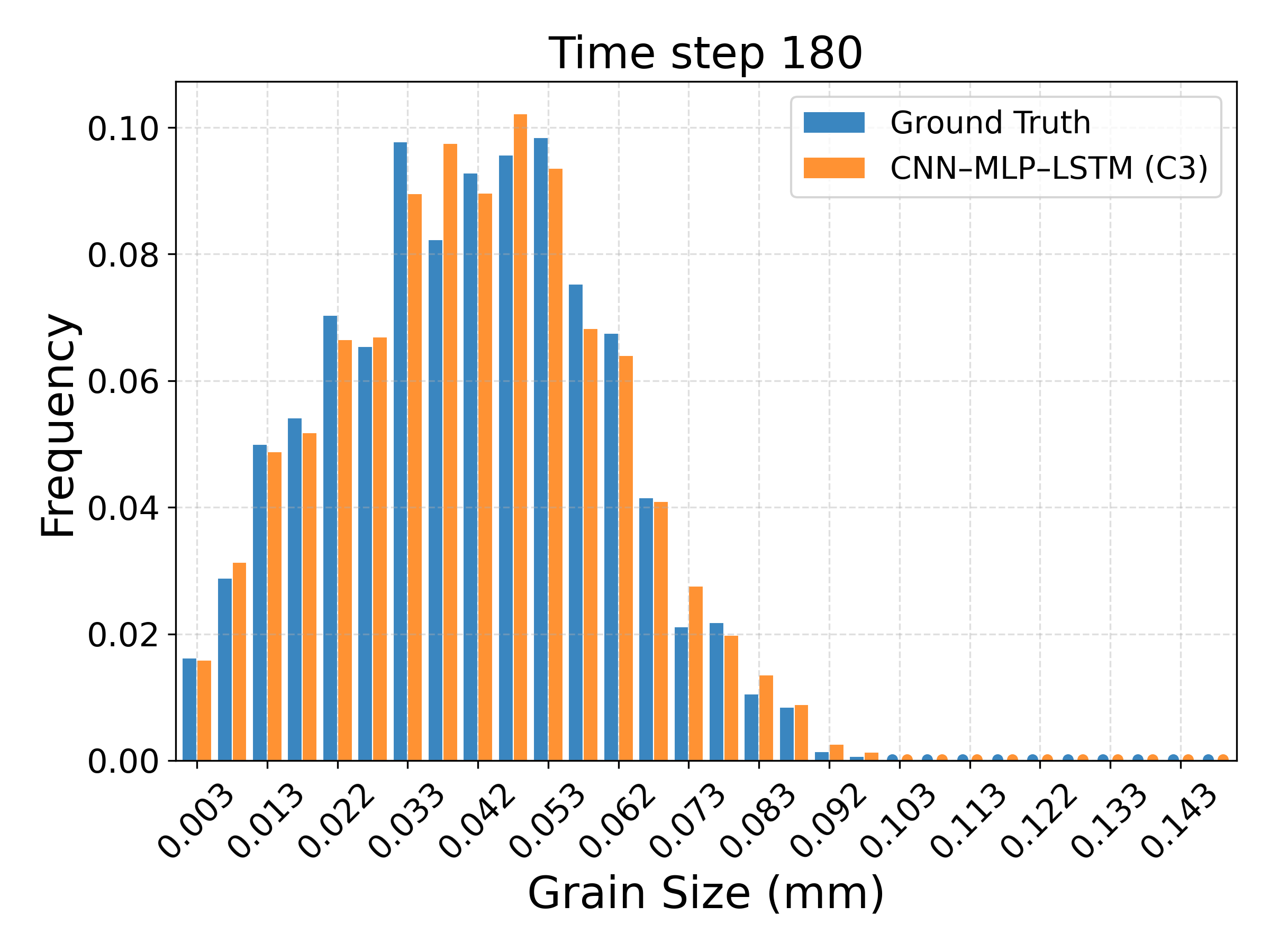}
        \caption{}
    \end{subfigure}
    \hfill
    \begin{subfigure}[t]{0.48\textwidth}
        \centering
        \includegraphics[width=\linewidth]{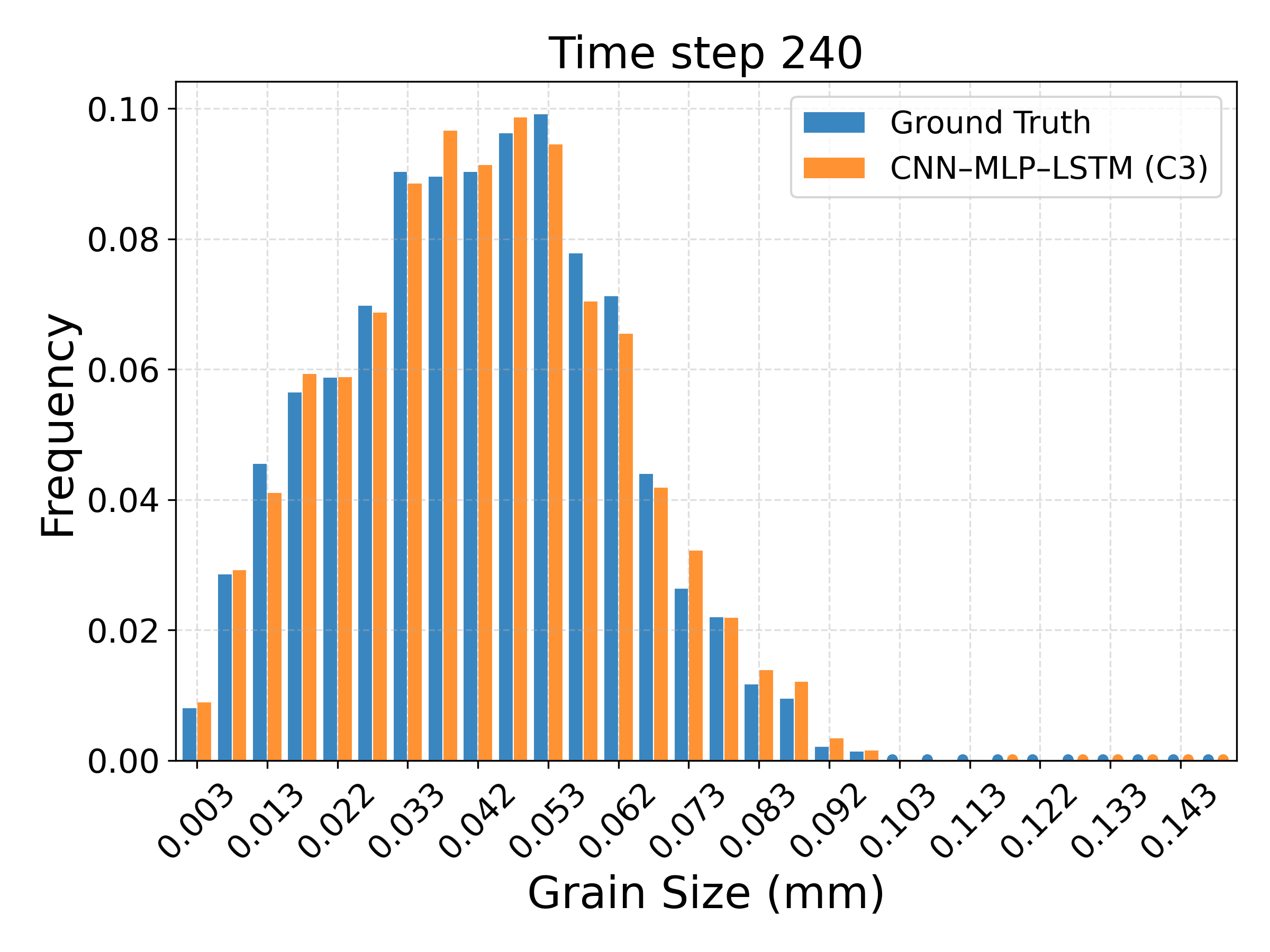}
        \caption{}
    \end{subfigure}

    \vspace{0.2cm}

    \caption{Four-hour thermal cycles: comparison of predicted (CNN--MLP--LSTM) and TRM normalized grain-size distributions for a representative simulation at (a) the initial state, (b) the first predicted time step, (c) $t = 60$~min, (d) $t = 120$~min, (e) $t = 180$~min, and (f) the final state ($t = 240$~min).}
    \label{fig:distribution_4h}
\end{figure}

The same model, trained and evaluated in the same way on the four-hour dataset (208 simulations, $k = 5$), gives an MRE of 4.51\% (from 2.07\% to 8.20\% depending on the simulation) and an $R^2$ of 0.9975. No simulation exceeds an MRE of 30\% at any time step of the 236 predicted steps. As for the one-hour cycles, the error is lowest during heating (0.27\%) and highest during cooling (9.44\%), with 4.00\% during the holding stages (Fig.~\ref{fig:temporal_4h}). The evolution of the grain-size distribution of a representative four-hour simulation (the simulation with the median MRE) is shown in Fig.~\ref{fig:distribution_4h}: the predicted distributions follow the reference ones over the 236 predicted steps.

\begin{figure}[htbp]
    \centering
    \includegraphics[width=0.9\textwidth]{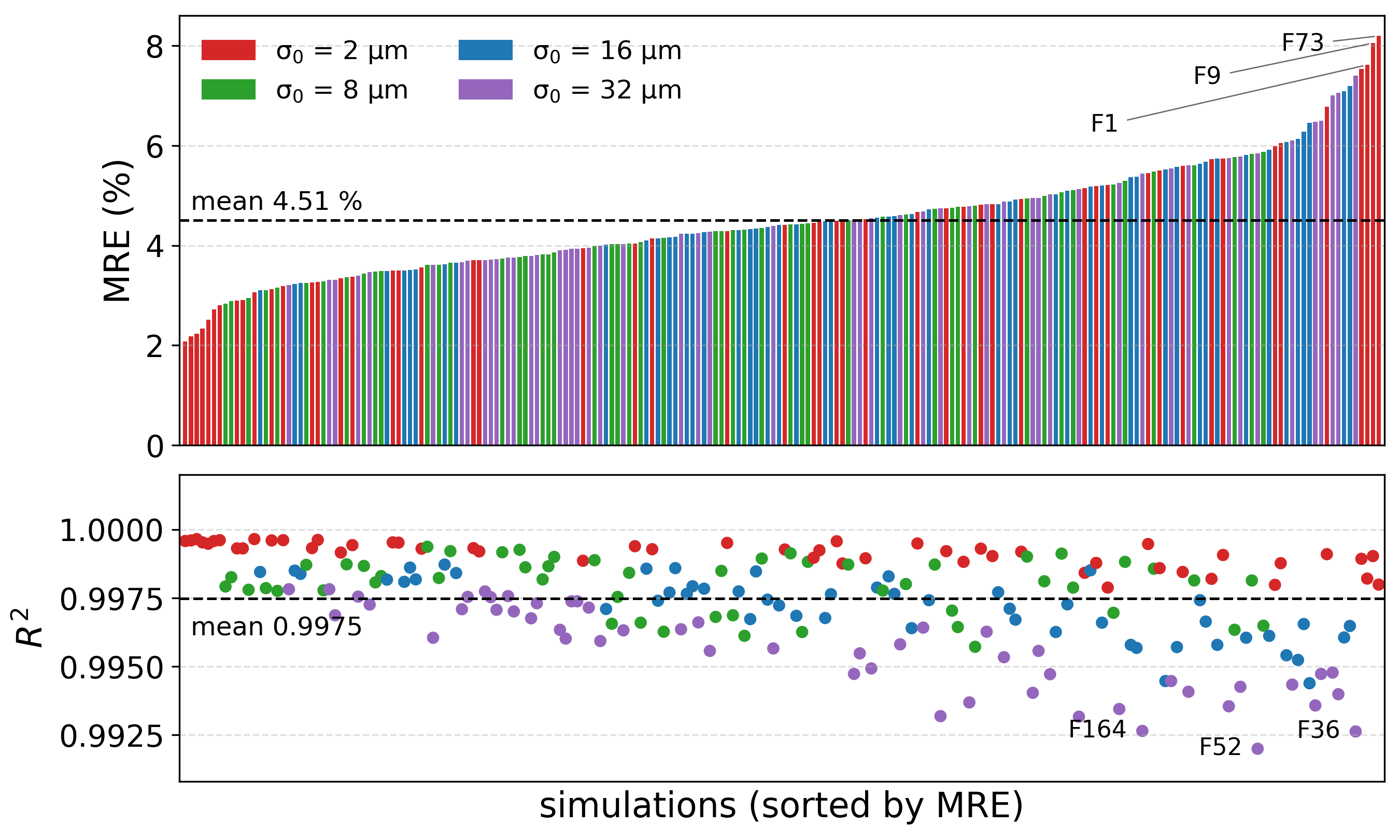}
    \caption{MRE (top) and $R^2$ (bottom) of each of the 208 four-hour simulations ($k$-fold cross-validation), sorted by increasing MRE and coloured by the standard deviation $\sigma_0$ of the initial distribution. The dashed lines give the mean values.}
    \label{fig:per_simulation_4h}
\end{figure}

The errors of the 208 four-hour simulations are shown individually in Fig.~\ref{fig:per_simulation_4h}. The MRE remains below 5\% for 150 simulations and below 8.2\% for all of them, and the $R^2$ is higher than 0.992 for every simulation. In contrast with the one-hour cycles, the errors depend only weakly on the initial distribution: the mean MRE is 4.41\%, 4.23\%, 4.76\% and 4.63\% for $\sigma_0 = 2$, 8, 16 and 32~$\mu$m, respectively. The largest MRE values are obtained for narrow initial distributions ($\sigma_0 = 2~\mu$m), whose $R^2$ nevertheless remains above 0.998, while the lowest $R^2$ values correspond to the widest initial distributions ($\sigma_0 = 32~\mu$m).

\origsubsection{Generalization to unseen thermal histories}
\label{subsec:generalization}

The generalization capability of the proposed CNN--MLP--LSTM framework is assessed using four thermal histories explicitly excluded from the training and validation datasets. For these tests, the model is trained once on all simulations of the corresponding dataset (one-hour dataset for Cases 1, 3 and 4, four-hour dataset for Case 2), and the first five minutes of each unseen simulation are given before the auto-regressive rollout. These cases involve temperature levels and temporal paths not previously encountered by the model and progressively introduce longer and more complex thermal cycles. For each unseen cycle, all available simulations (different initial distributions $\sigma_0$ on the same thermal path) are predicted, and the results are reported in Table~\ref{tab:generalization}. The distributions shown in the following figures correspond to the simulations marked with an asterisk.

{\setlength{\intextsep}{2pt}
\begin{table}[H]
\centering
\caption{Prediction performance on the ten simulations of the four unseen thermal cycles. The simulation shown in the corresponding figure is marked with an asterisk; $^\dagger$: intermediate holding stage at 1190.15~K instead of 1123.15~K.}
\label{tab:generalization}
{\small
\begin{tabular}{llcccc}
\toprule
Case & Model & $a$ (mm) & $\sigma_0$ ($\mu$m) & MRE (\%) $\downarrow$ & $R^2 \uparrow$ \\
\midrule
Case 1 -- one-hour cycle & 1 h & 3 & 16$^*$ & 1.91 & 0.9988 \\
\midrule
Case 2 -- four-hour cycle & 4 h & 5 & 8$^*$ & 10.40 & 0.9922 \\
 & & & 16 & 14.66 & 0.9825 \\
 & & & 32 & 12.70 & 0.9705 \\
\midrule
Case 3 -- multi-stage cycle & 1 h & 3 & 2 & 8.84 & 0.9919 \\
 & & & 8$^*$ & 6.82 & 0.9916 \\
 & & & 16 & 7.36 & 0.9815 \\
 & & & 32 & 11.17 & 0.9404 \\
\midrule
Case 4 -- non-monotonic cycle & 1 h & 3 & 8$^*$ & 6.09 & 0.9930 \\
Case 4 -- variant$^\dagger$ & 1 h & 3 & 8 & 5.73 & 0.9922 \\
\bottomrule
\end{tabular}
}
\end{table}}

\subsubsection{Case 1: One-hour unseen thermal cycle}

\begin{figure}[htbp]
    \centering

    \begin{subfigure}[t]{0.48\textwidth}
        \centering
        \includegraphics[width=\linewidth]{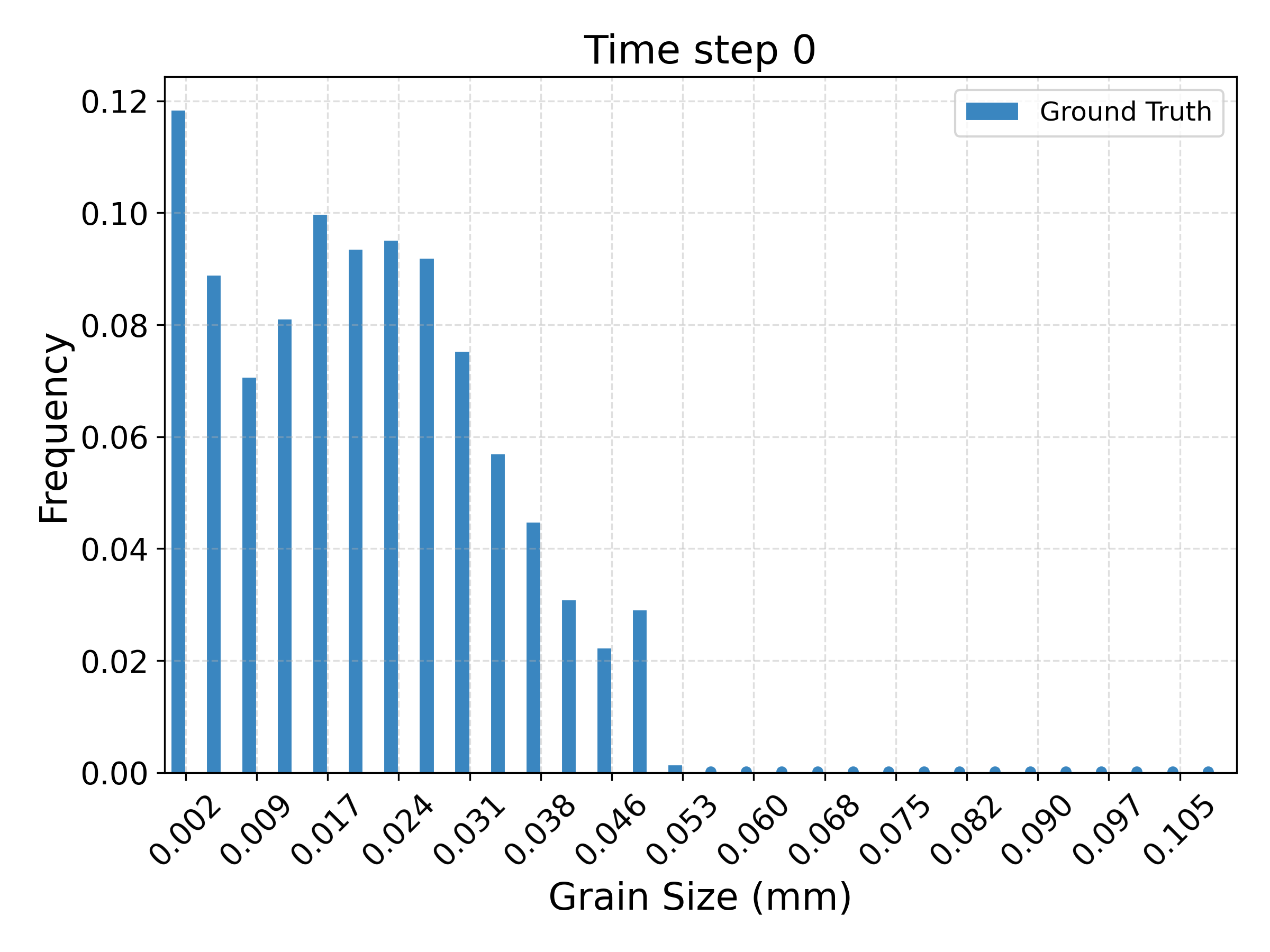}
        \caption{}
    \end{subfigure}
    \hfill
    \begin{subfigure}[t]{0.48\textwidth}
        \centering
        \includegraphics[width=\linewidth]{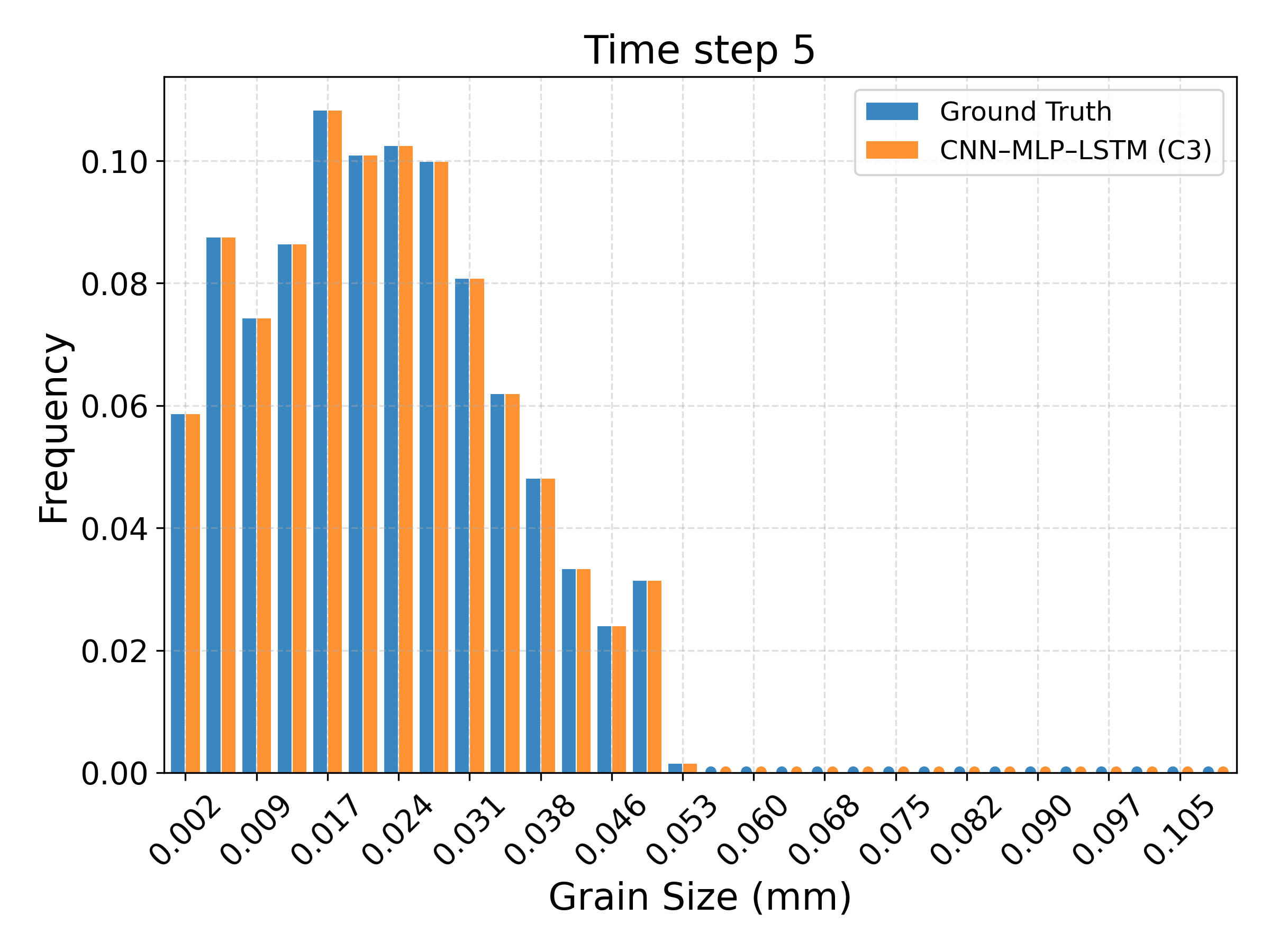}
        \caption{}
    \end{subfigure}

    \vspace{0.2cm}

    \begin{subfigure}[t]{0.48\textwidth}
        \centering
        \includegraphics[width=\linewidth]{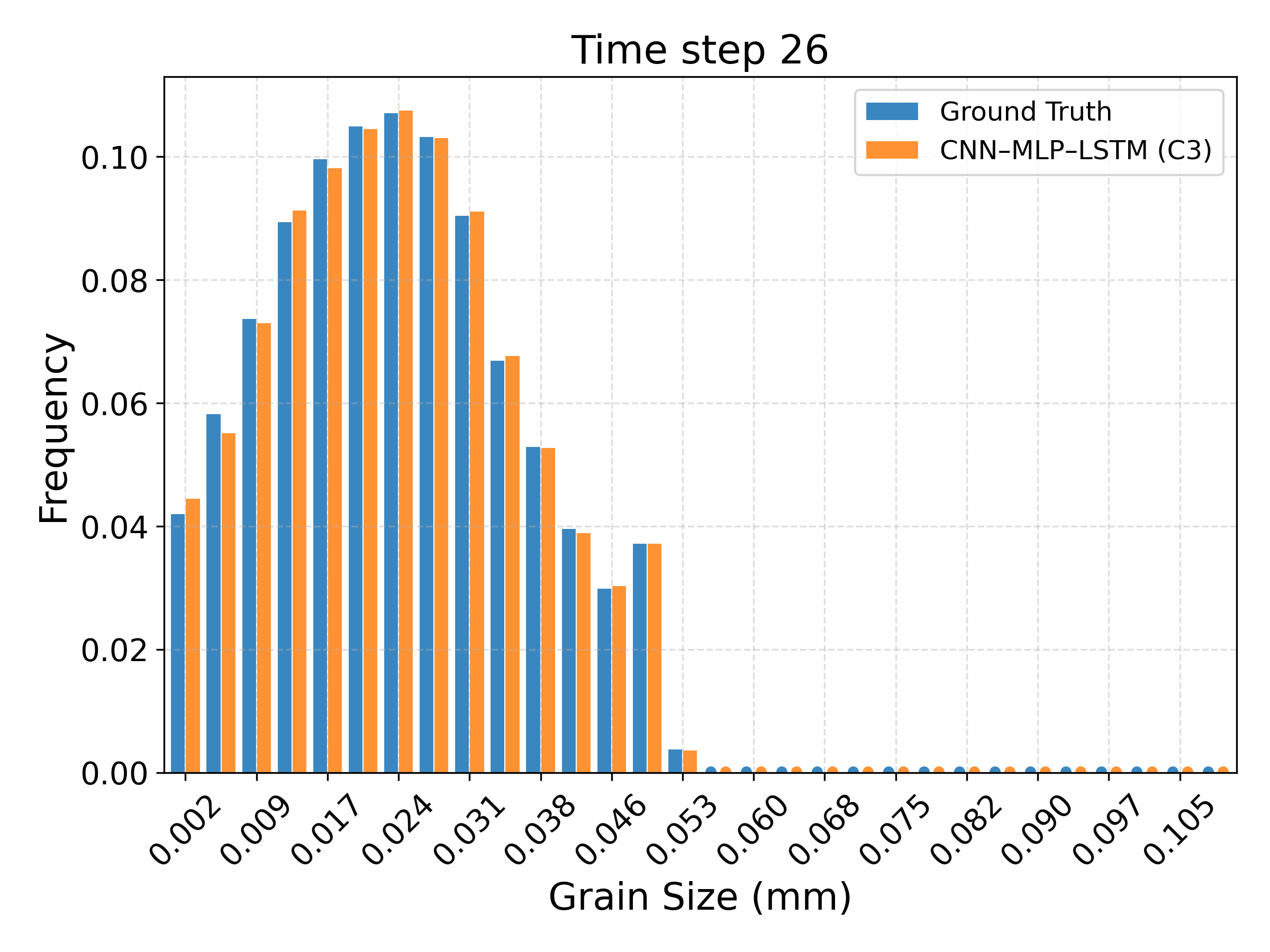}
        \caption{}
    \end{subfigure}
    \hfill
    \begin{subfigure}[t]{0.48\textwidth}
        \centering
        \includegraphics[width=\linewidth]{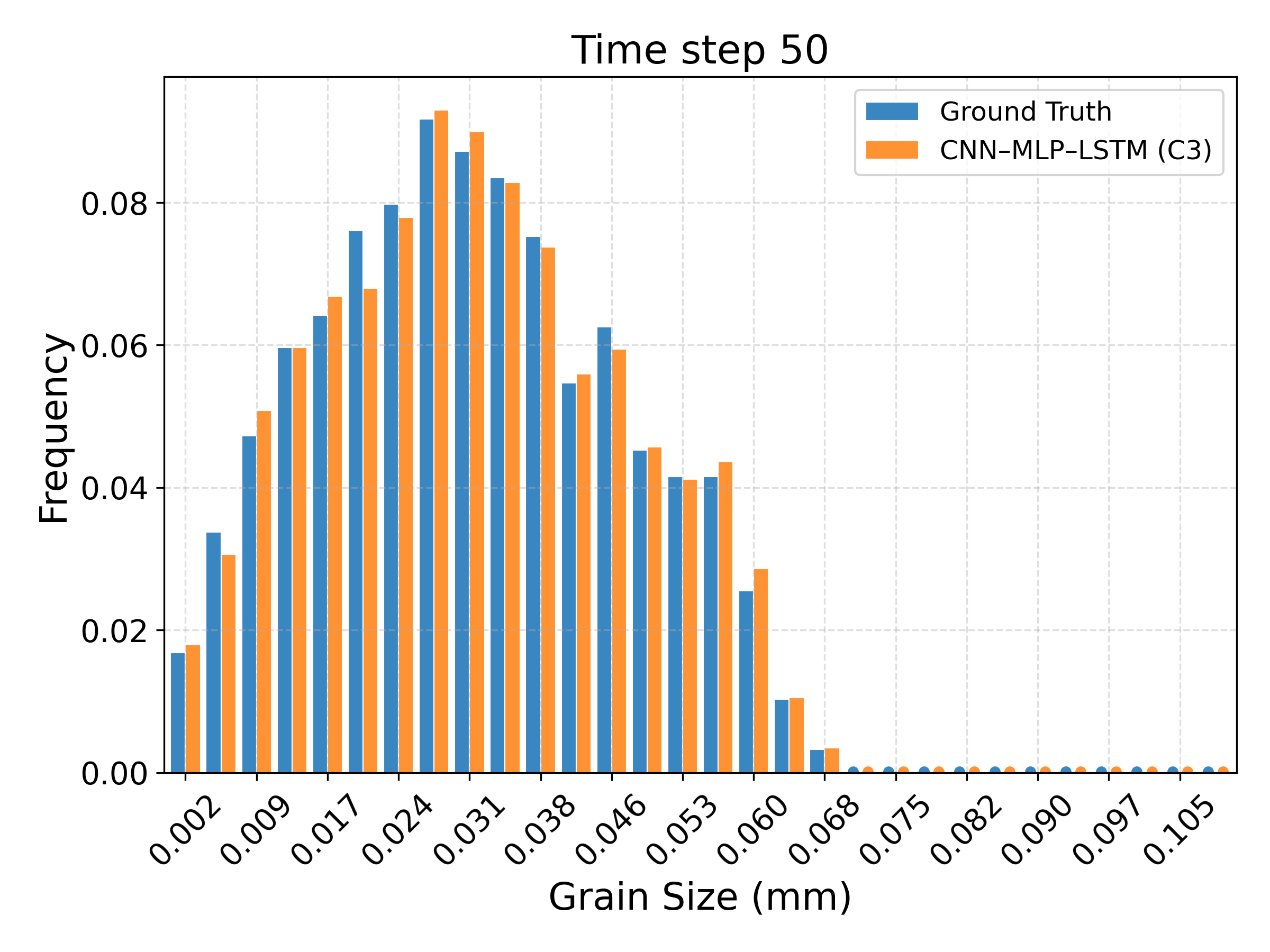}
        \caption{}
    \end{subfigure}

    \vspace{0.2cm}

    \begin{subfigure}[t]{0.55\textwidth}
        \centering
        \includegraphics[width=\linewidth]{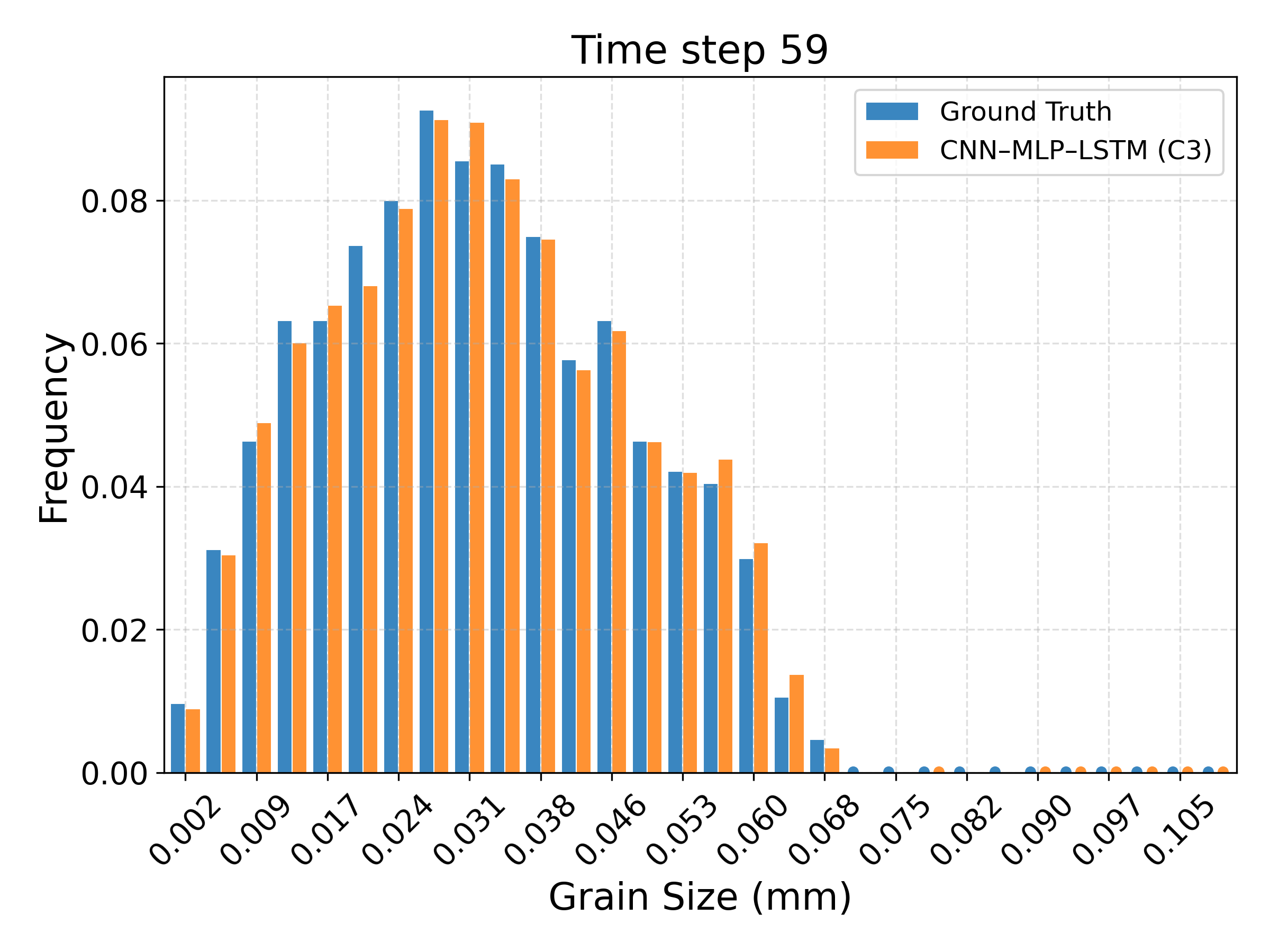}
        \caption{}
    \end{subfigure}
\caption{Case~1: Comparison of reference (TRM) and predicted (CNN--MLP--LSTM) normalized grain-size distributions for a representative one-hour unseen thermal cycle at (a) the initial state, (b) the first predicted time step, (c) the end of heating, (d) the end of the isothermal holding stage, and (e) the final state after cooling.}

    \label{fig:gen_case1}
\end{figure}

For the first unseen thermal cycle, the predicted distributions remain in close agreement with the reference distributions at the selected prediction steps (Fig.~\ref{fig:gen_case1}). The displacement of the dominant peak toward larger grain sizes is reproduced throughout the cycle, with only small differences appearing in the tails of the distribution at the final stage. The model reaches an MRE of 1.91\% and an $R^2$ of 0.9988 (Table~\ref{tab:generalization}).

\subsubsection{Case 2: Four-hour unseen thermal cycle}

\begin{figure}[htbp]
    \centering

    \begin{subfigure}[t]{0.48\textwidth}
        \centering
        \includegraphics[width=\linewidth]{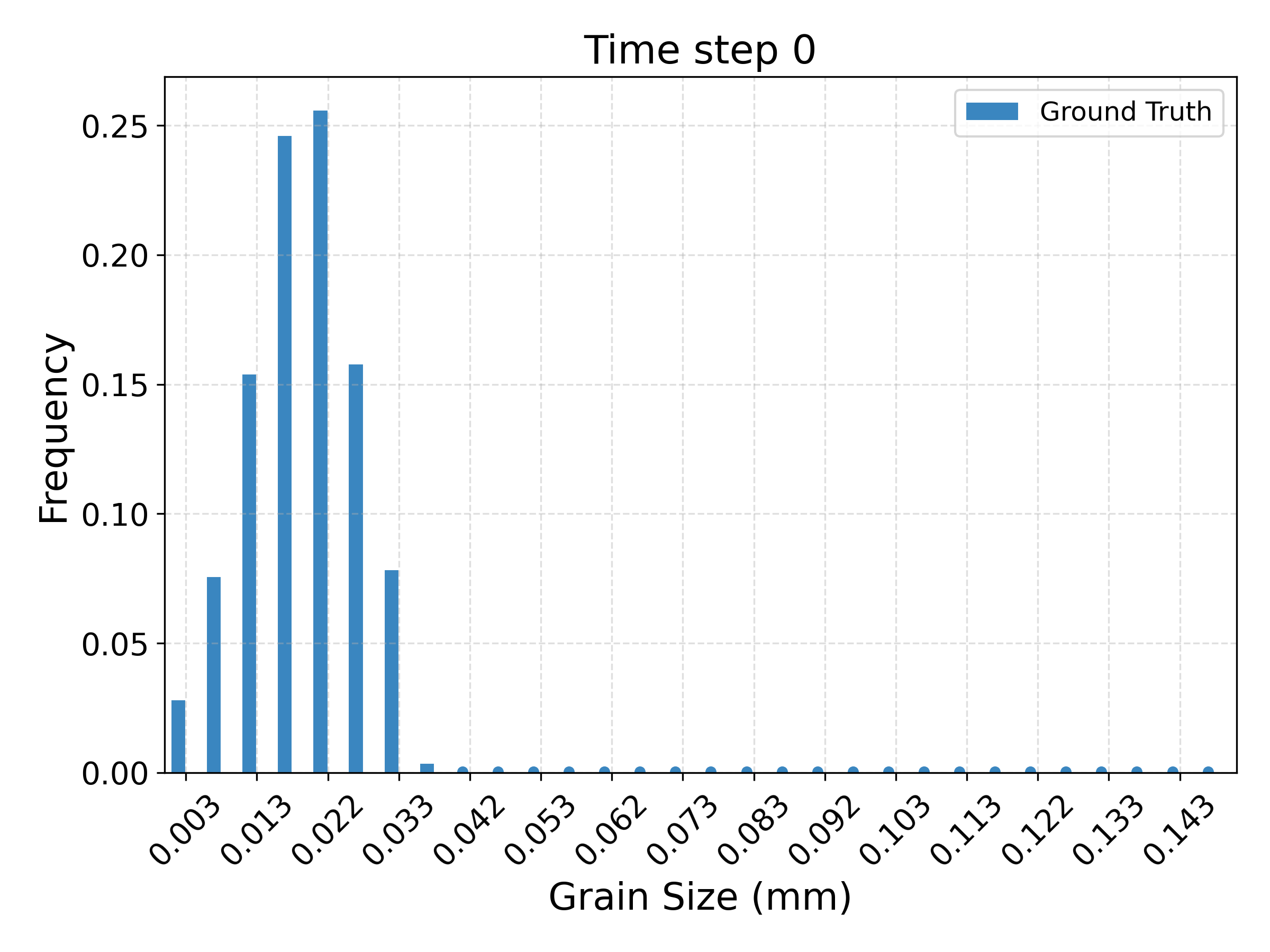}
        \caption{}
    \end{subfigure}
    \hfill
    \begin{subfigure}[t]{0.48\textwidth}
        \centering
        \includegraphics[width=\linewidth]{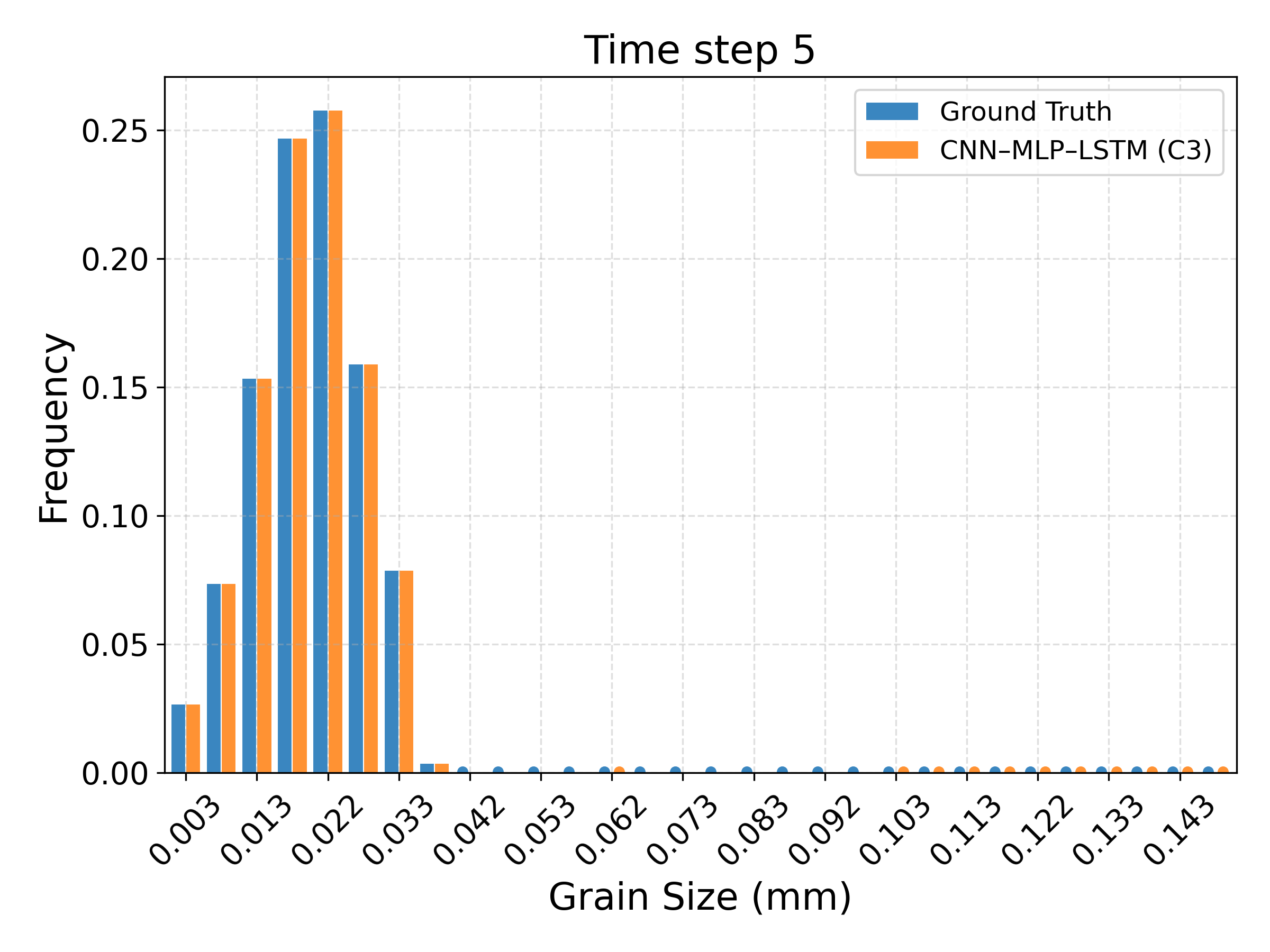}
        \caption{}
    \end{subfigure}

    \vspace{0.2cm}

    \begin{subfigure}[t]{0.48\textwidth}
        \centering
        \includegraphics[width=\linewidth]{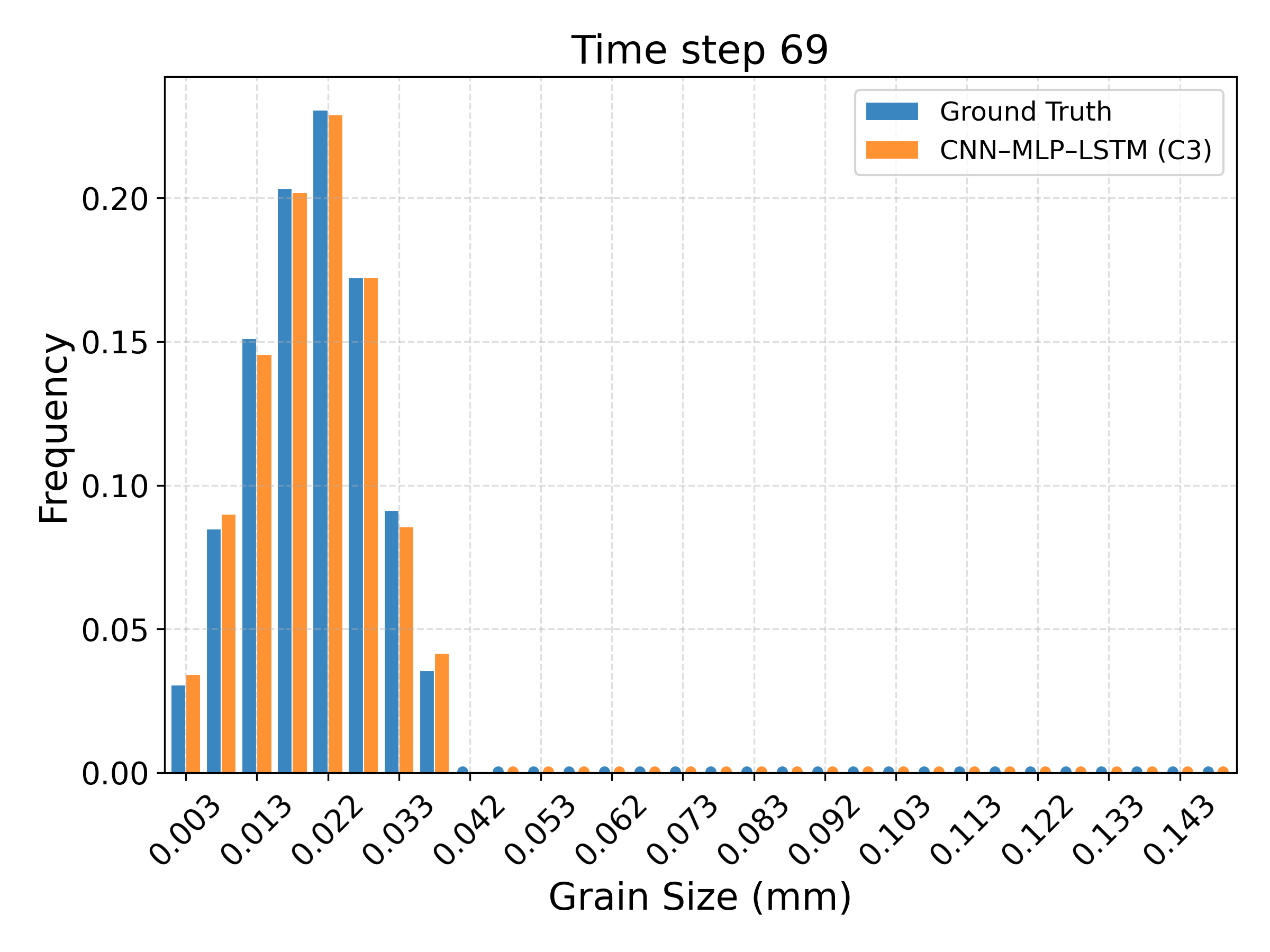}
        \caption{}
    \end{subfigure}
    \hfill
    \begin{subfigure}[t]{0.48\textwidth}
        \centering
        \includegraphics[width=\linewidth]{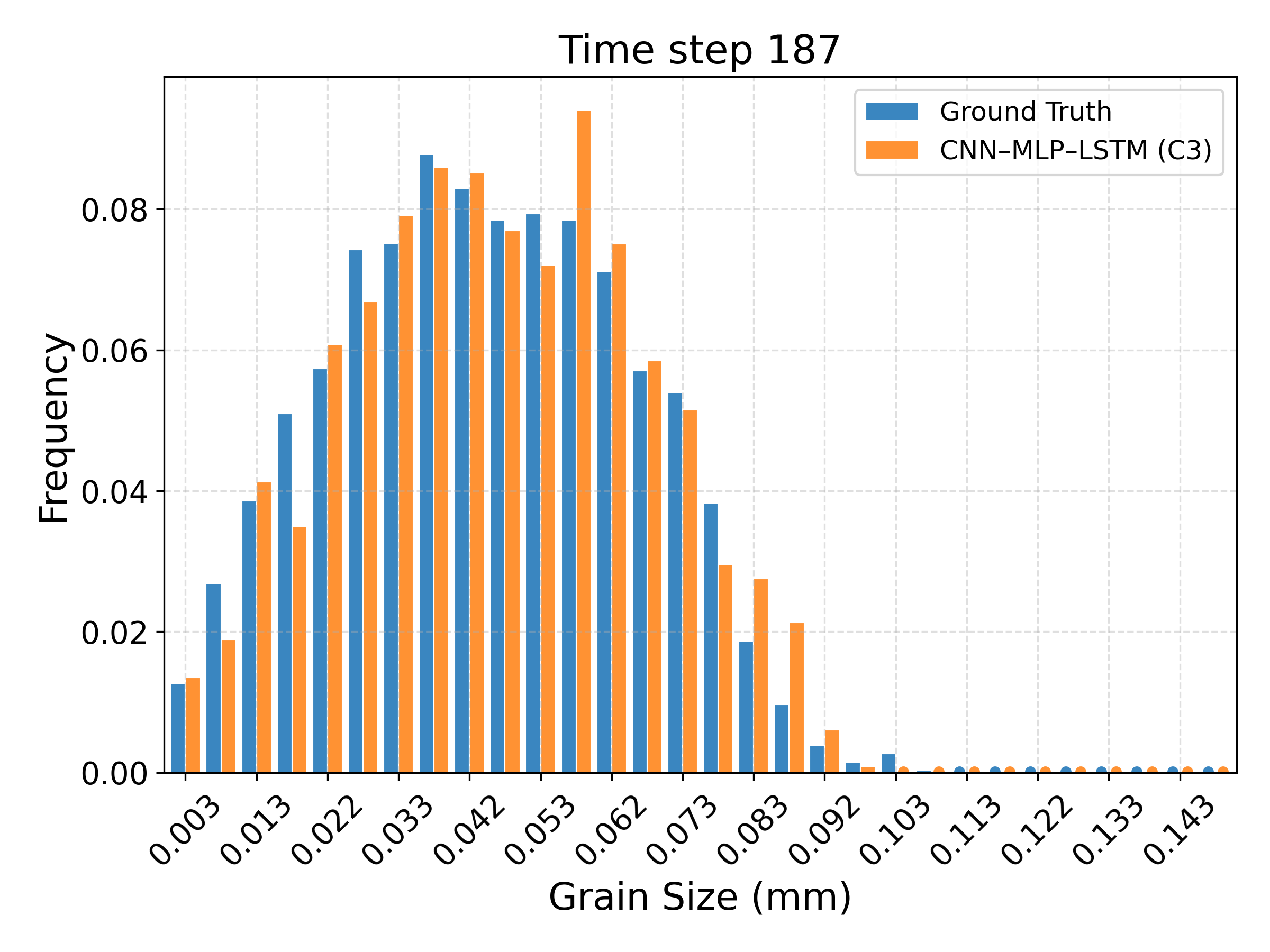}
        \caption{}
    \end{subfigure}

    \vspace{0.2cm}

    \begin{subfigure}[t]{0.55\textwidth}
        \centering
        \includegraphics[width=\linewidth]{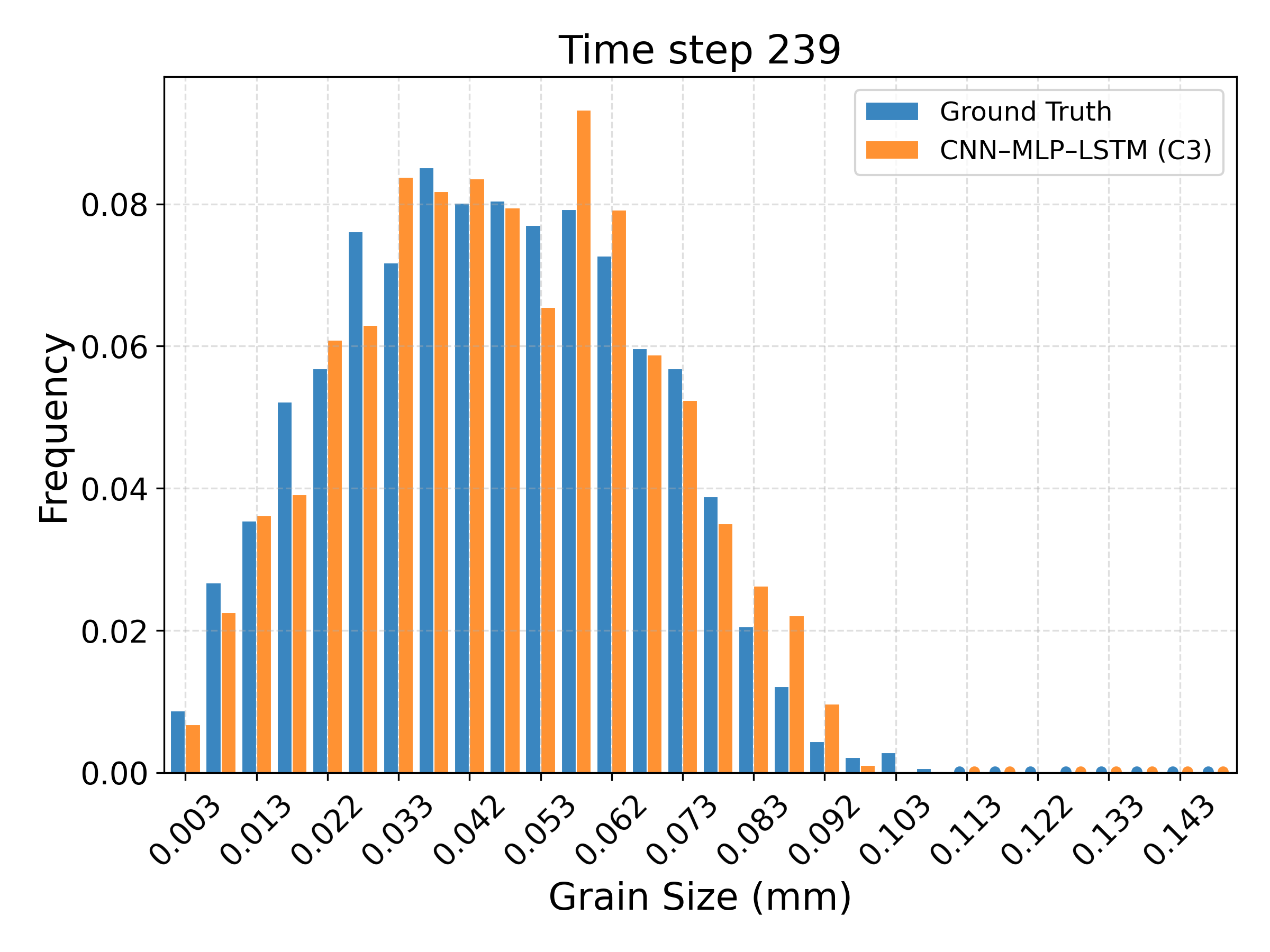}
        \caption{}
    \end{subfigure}

\caption{Case~2: Comparison of reference (TRM) and predicted (CNN--MLP--LSTM) normalized grain-size distributions for a representative four-hour unseen thermal cycle at (a) the initial state, (b) the first predicted time step, (c) the end of heating, (d) the end of the extended isothermal holding stage, and (e) the final state after cooling.}

    \label{fig:gen_case2}
\end{figure}

The second case extends the prediction to a four-hour thermal cycle. Despite the longer prediction horizon, the predicted distributions continue to follow the reference evolution during heating, holding, and cooling (Fig.~\ref{fig:gen_case2}). Differences become more visible at the later stages, particularly for the largest grain-size classes. An MRE of 10.40\% and an $R^2$ of 0.9922 are obtained with the model trained on the four-hour dataset. This cycle also involves a domain size (5~mm) that is not present in the four-hour training data. For the two other initial distributions of this cycle ($\sigma_0 = 16$ and 32~$\mu$m), the MRE reaches 14.66\% and 12.70\%, with $R^2$ values of 0.9825 and 0.9705 (Table~\ref{tab:generalization}).

\subsubsection{Case 3: Multi-stage realistic thermal cycle}

\begin{figure}[htbp]
\centering

\begin{subfigure}[t]{0.48\textwidth}
    \centering
    \includegraphics[width=\linewidth]{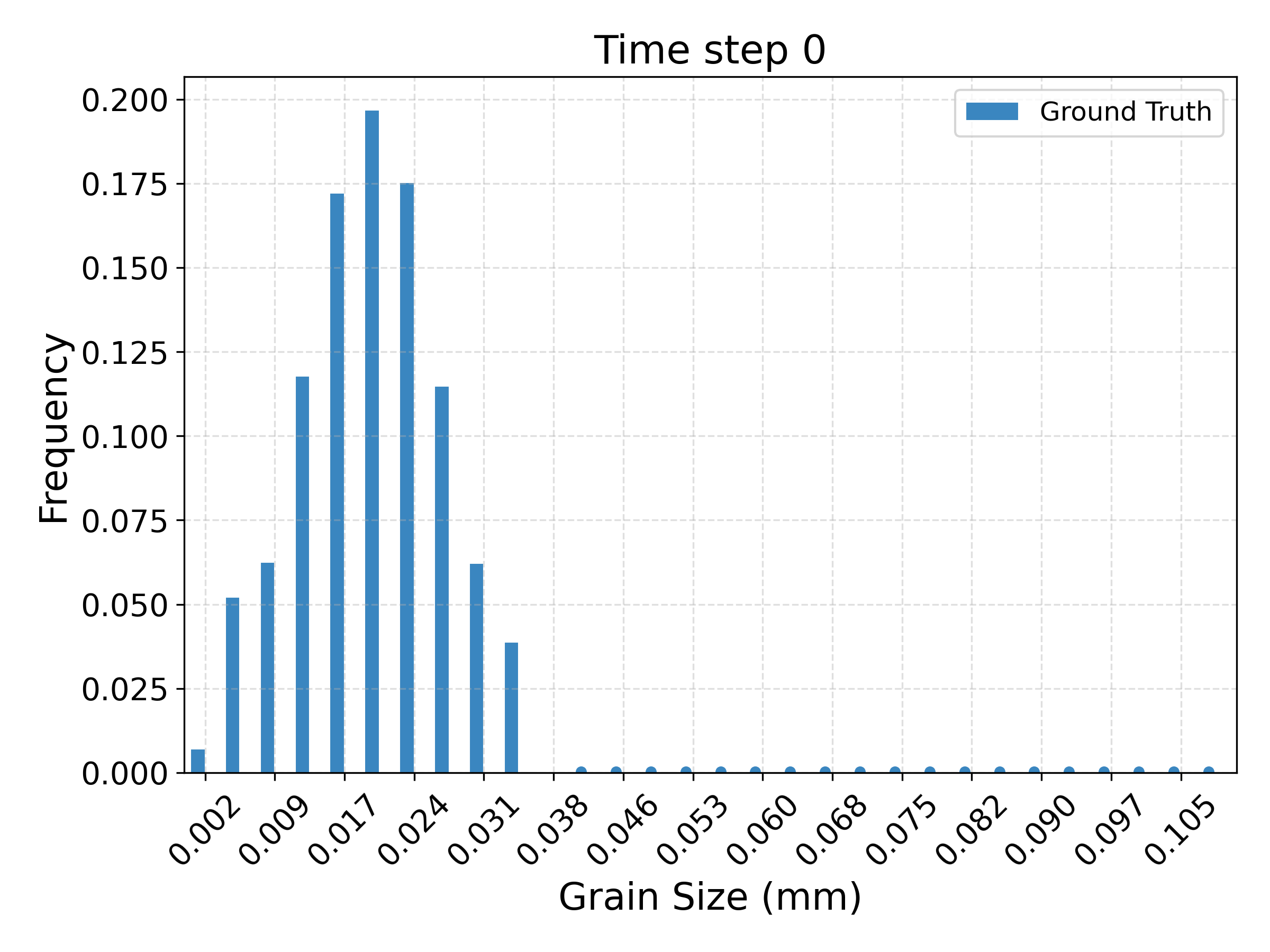}
    \caption{}
\end{subfigure}
\hfill
\begin{subfigure}[t]{0.48\textwidth}
    \centering
    \includegraphics[width=\linewidth]{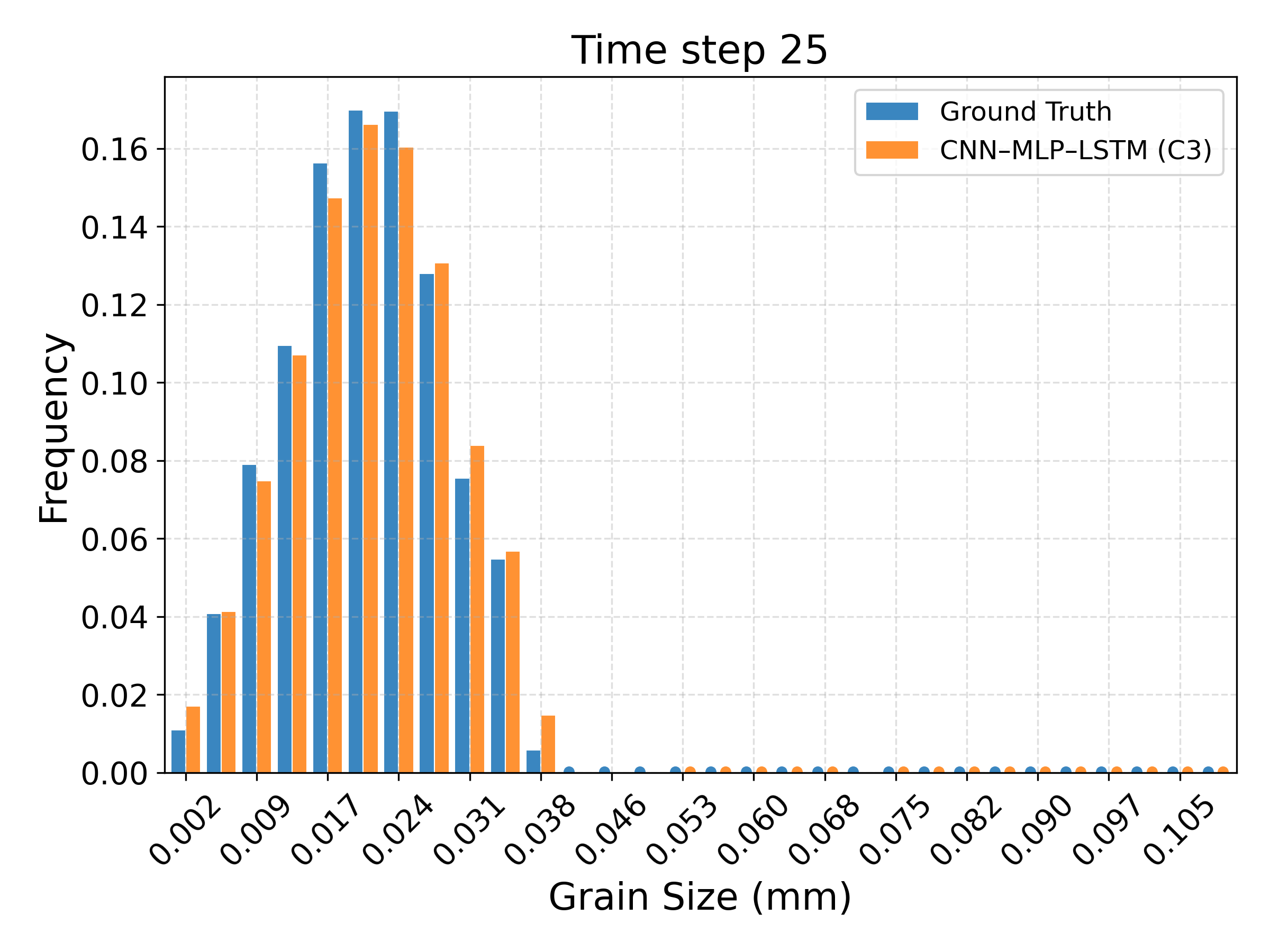}
    \caption{}
\end{subfigure}

\vspace{0.3cm}

\begin{subfigure}[t]{0.48\textwidth}
    \centering
    \includegraphics[width=\linewidth]{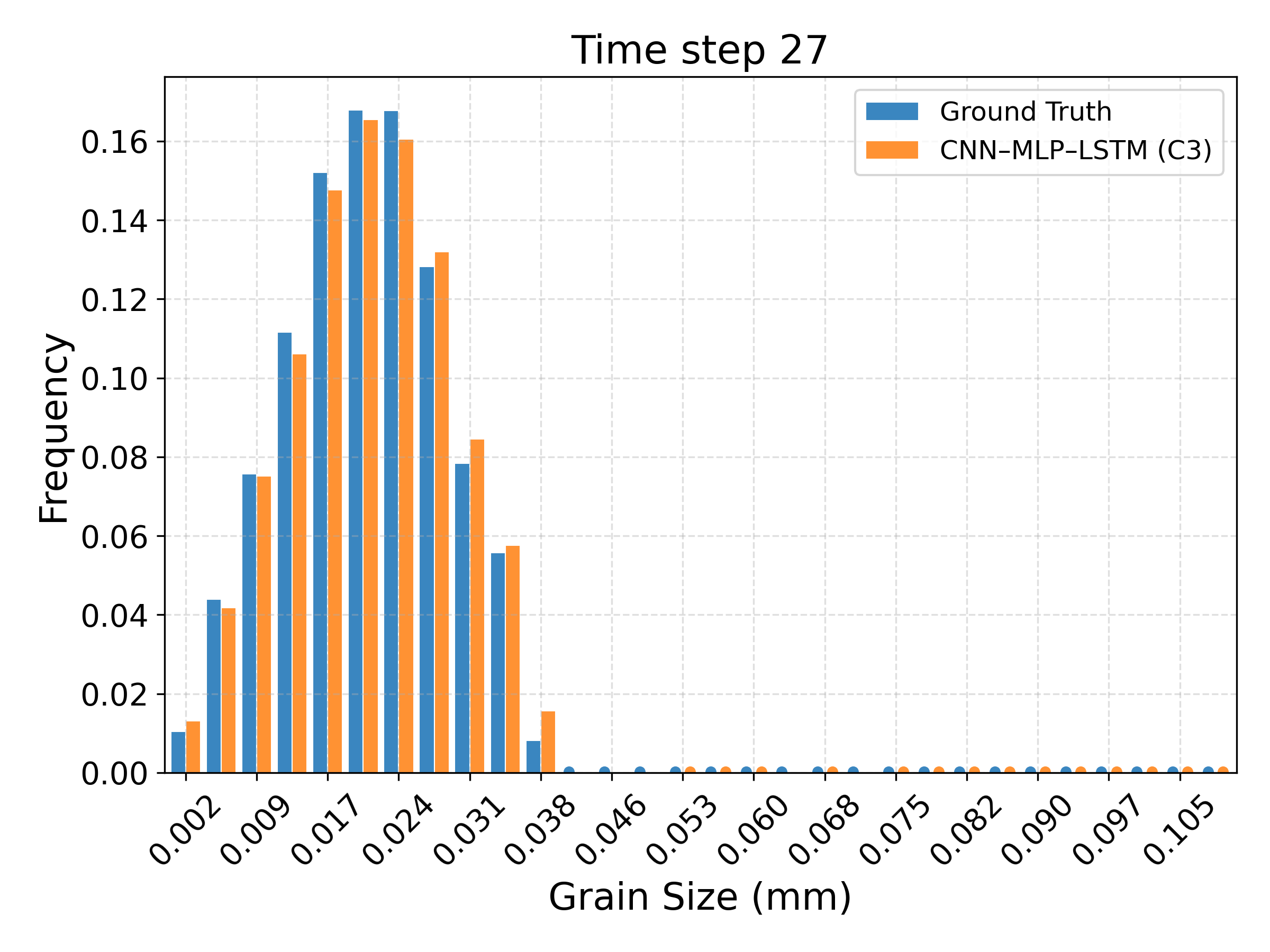}
    \caption{}
\end{subfigure}
\hfill
\begin{subfigure}[t]{0.48\textwidth}
    \centering
    \includegraphics[width=\linewidth]{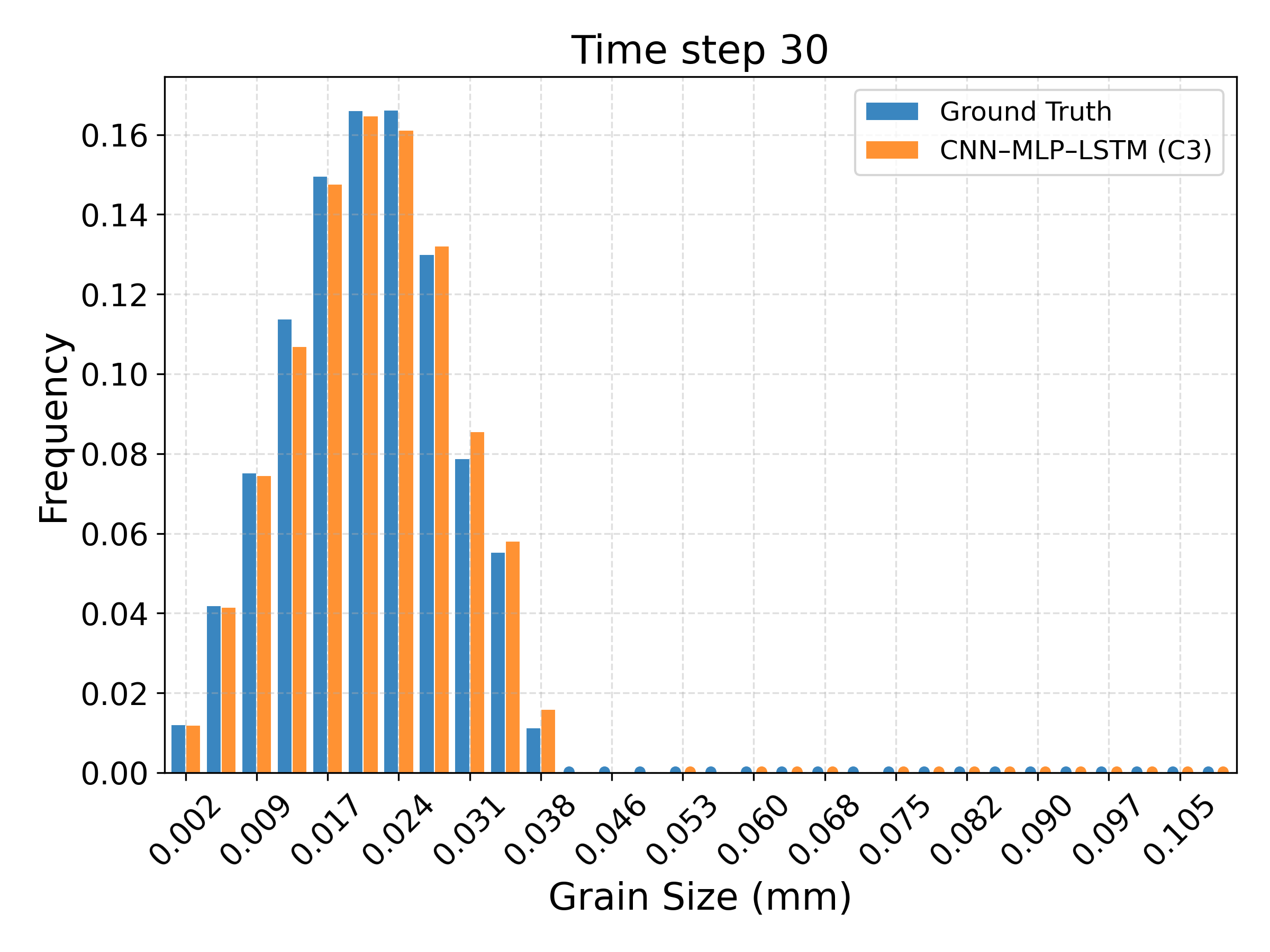}
    \caption{}
\end{subfigure}

\vspace{0.3cm}

\begin{subfigure}[t]{0.48\textwidth}
    \centering
    \includegraphics[width=\linewidth]{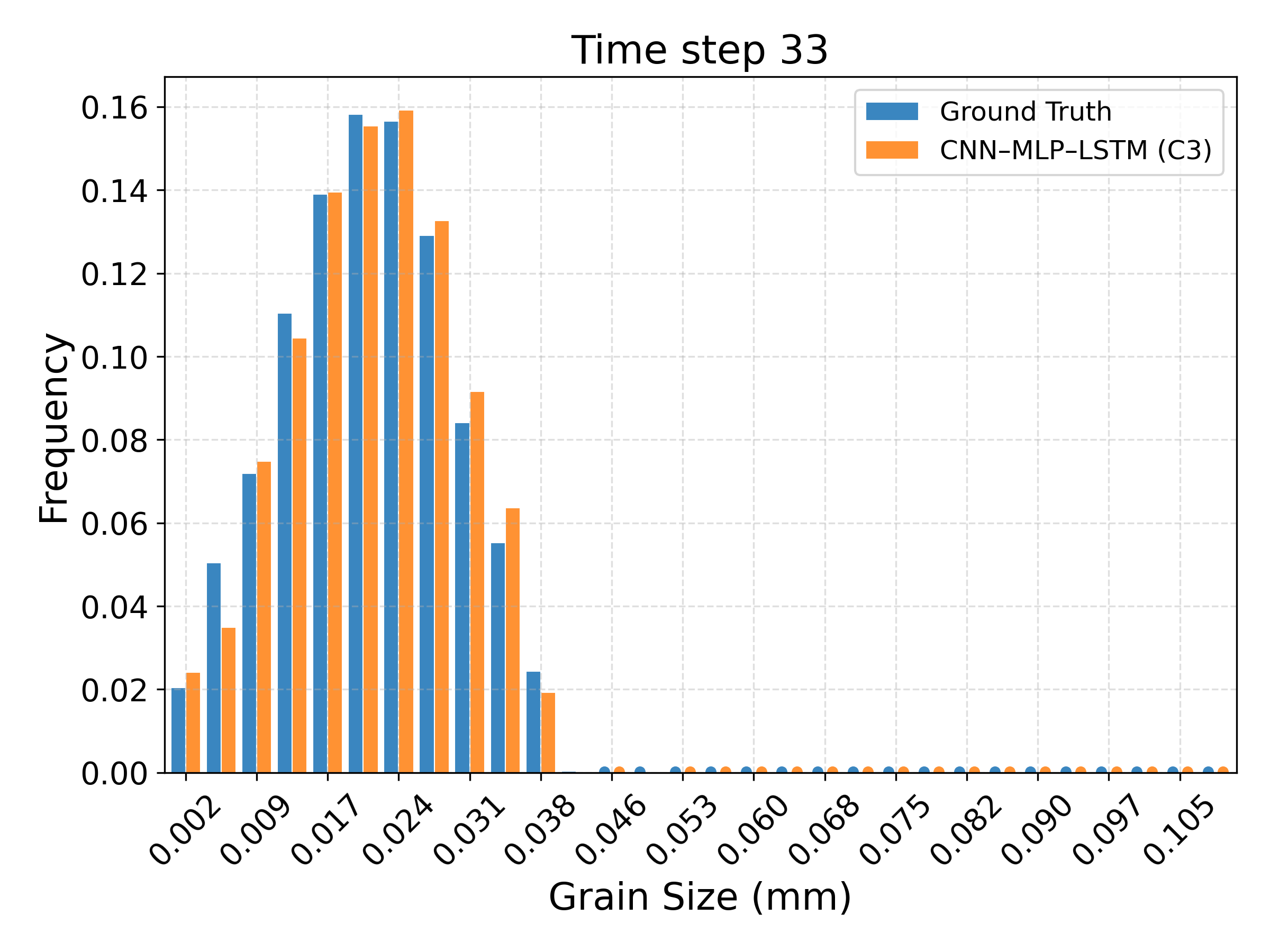}
    \caption{}
\end{subfigure}
\hfill
\begin{subfigure}[t]{0.48\textwidth}
    \centering
    \includegraphics[width=\linewidth]{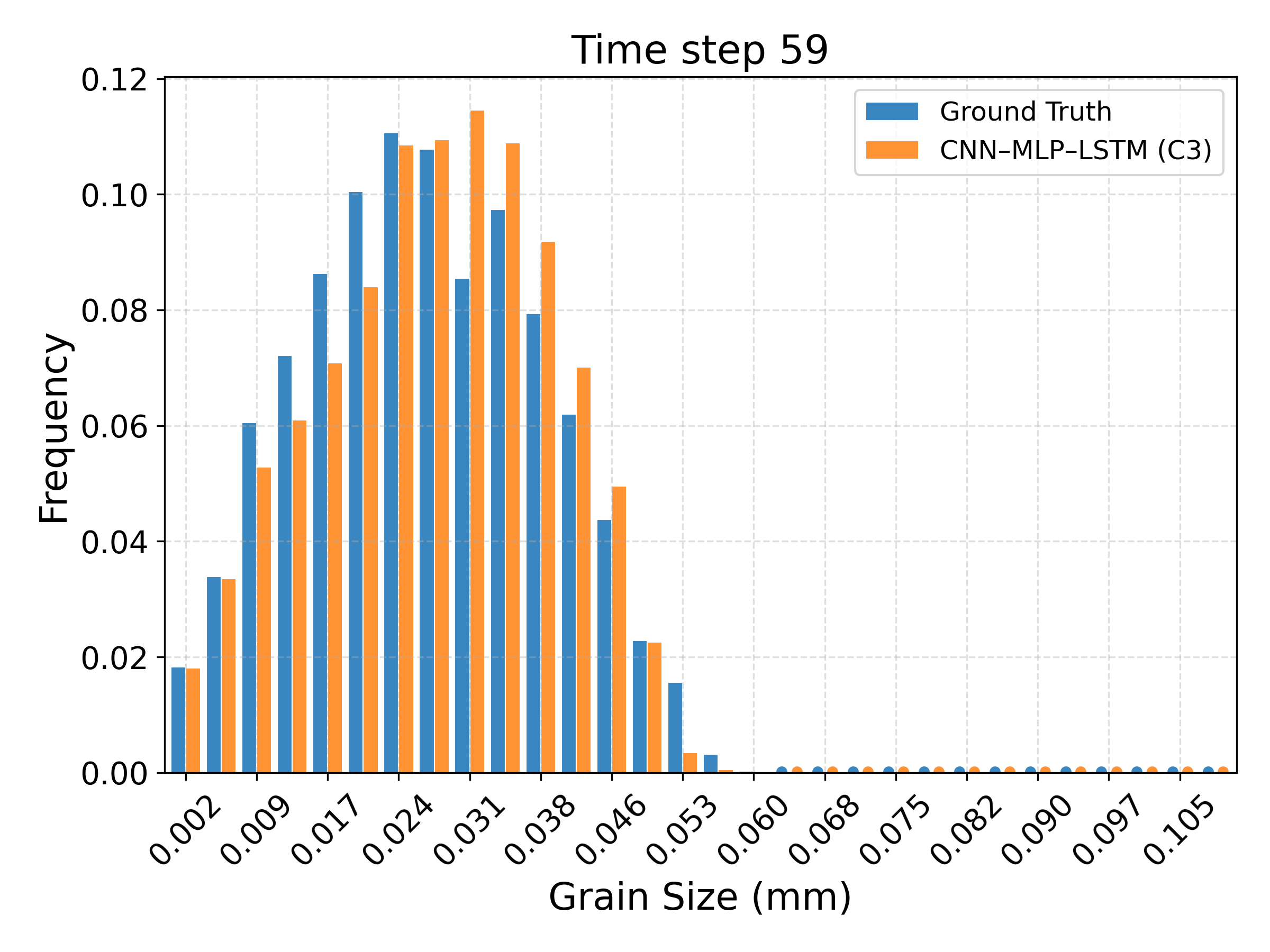}
    \caption{}
\end{subfigure}

\caption{Case~3: Comparison of reference (TRM) and predicted (CNN--MLP--LSTM) normalized grain-size distributions for a representative multi-stage unseen thermal cycle at (a) the initial state, (b) after the first isothermal plateau, (c) after partial cooling, (d) following reheating and holding, (e) at the end of the second heating stage, and (f) after the final plateau.}

\label{fig:gen_case3}
\end{figure}

\begin{figure}[htbp]
\centering

\begin{subfigure}[t]{0.40\textwidth}
\centering
\includegraphics[width=\linewidth]{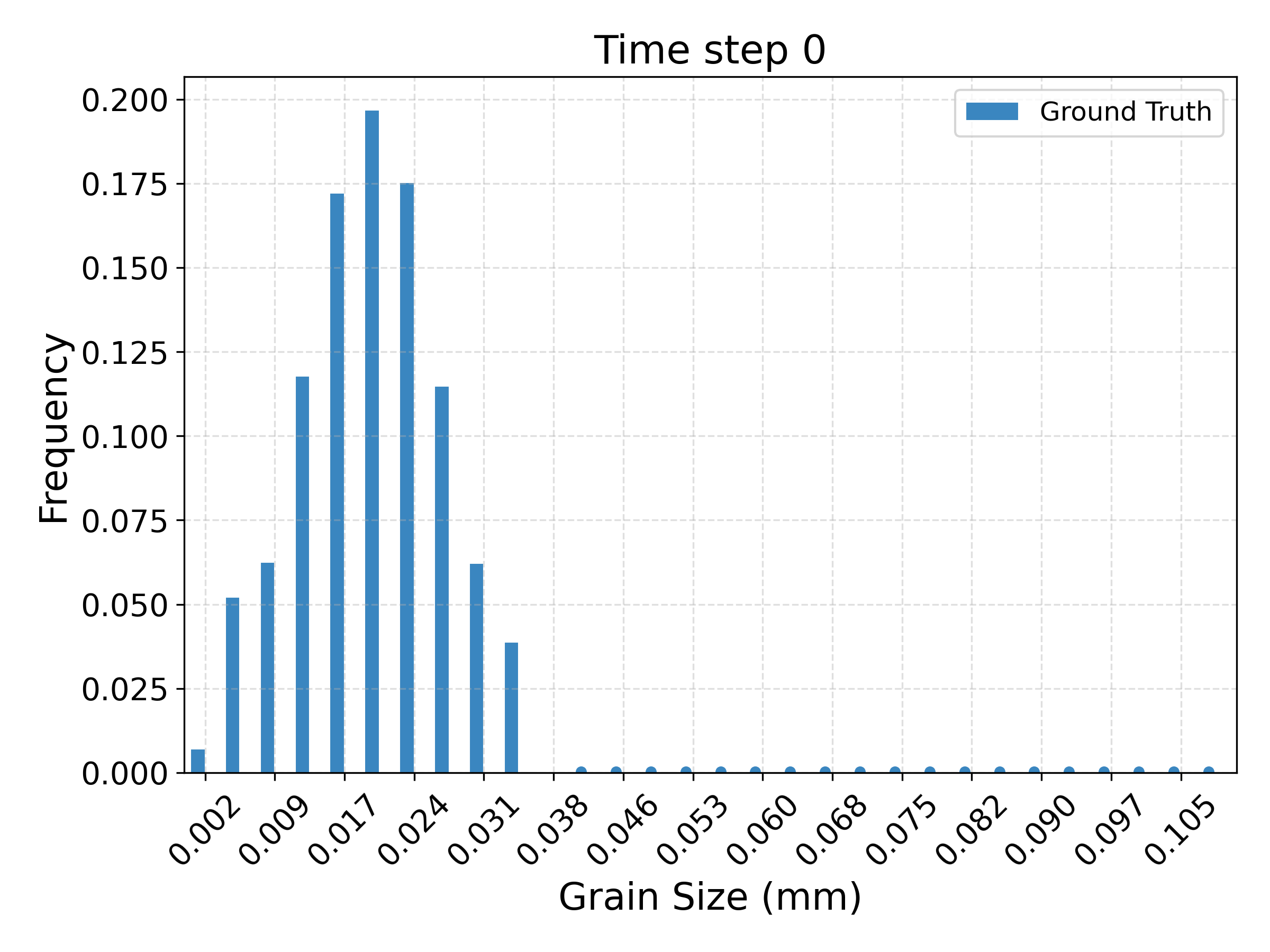}
\caption{}
\end{subfigure}
\hspace{0.01\textwidth}
\begin{subfigure}[t]{0.40\textwidth}
\centering
\includegraphics[width=\linewidth]{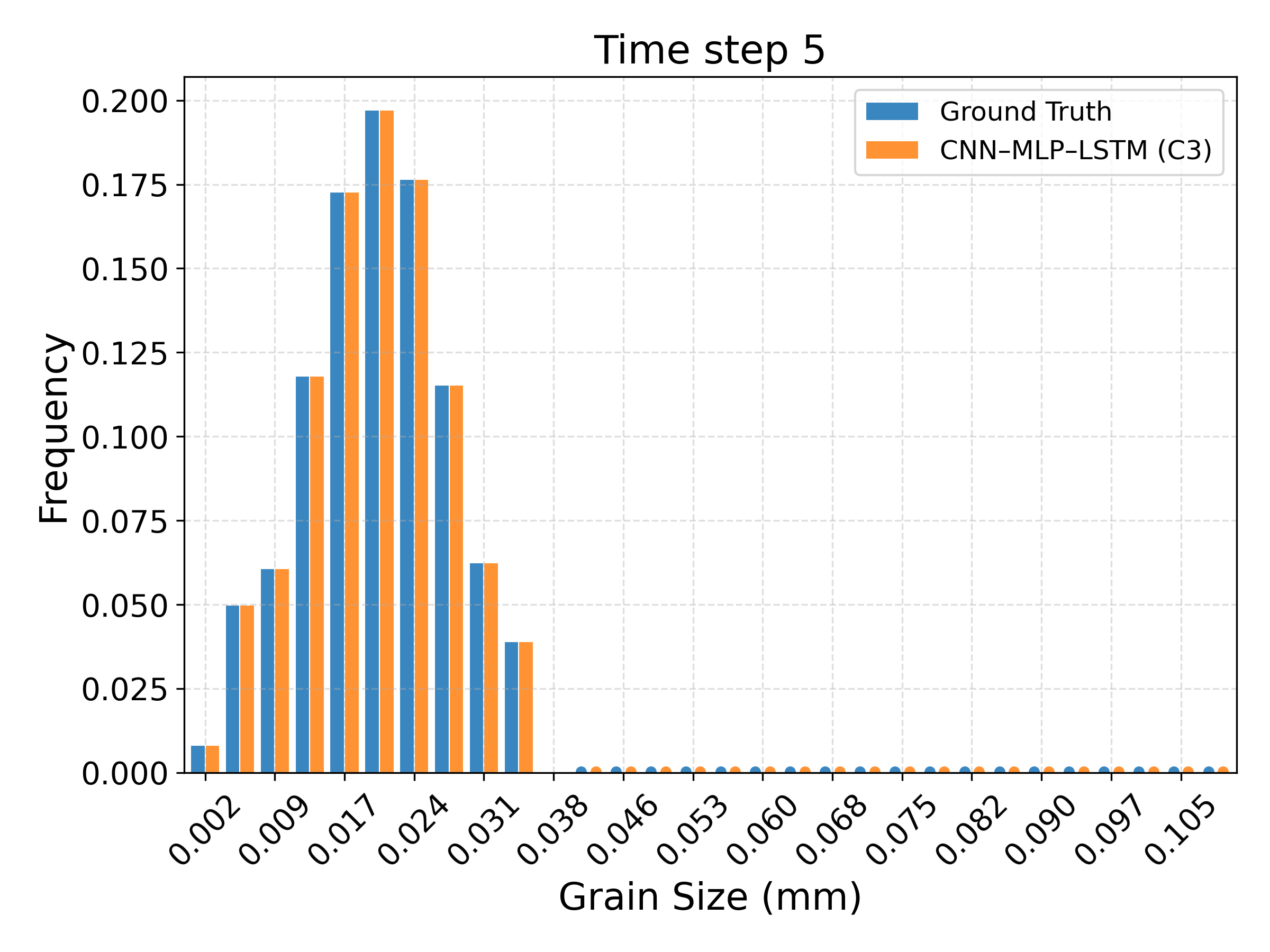}
\caption{}
\end{subfigure}

\vspace{0.2cm}

\begin{subfigure}[t]{0.40\textwidth}
\centering
\includegraphics[width=\linewidth]{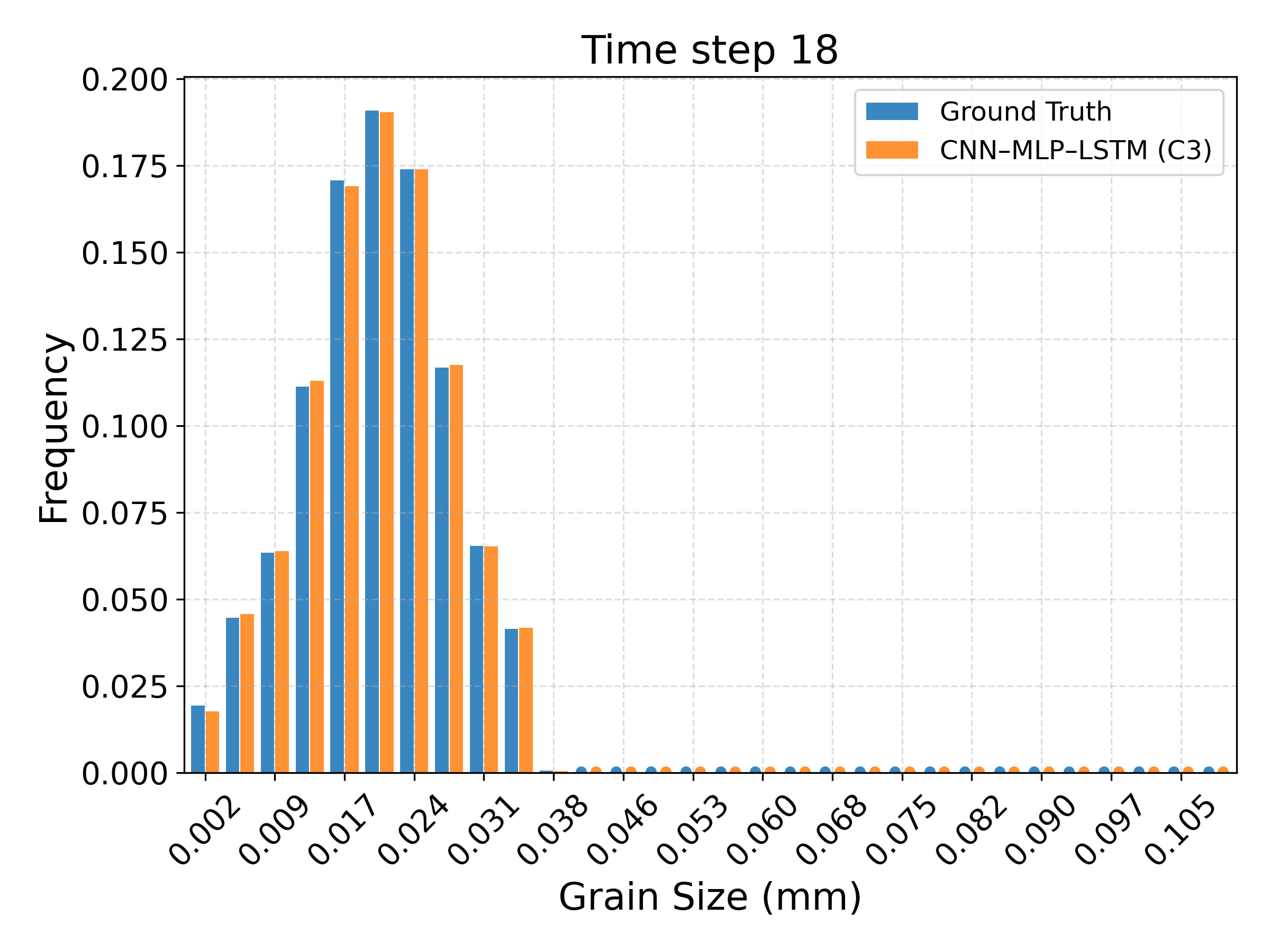}
\caption{}
\end{subfigure}
\hspace{0.01\textwidth}
\begin{subfigure}[t]{0.40\textwidth}
\centering
\includegraphics[width=\linewidth]{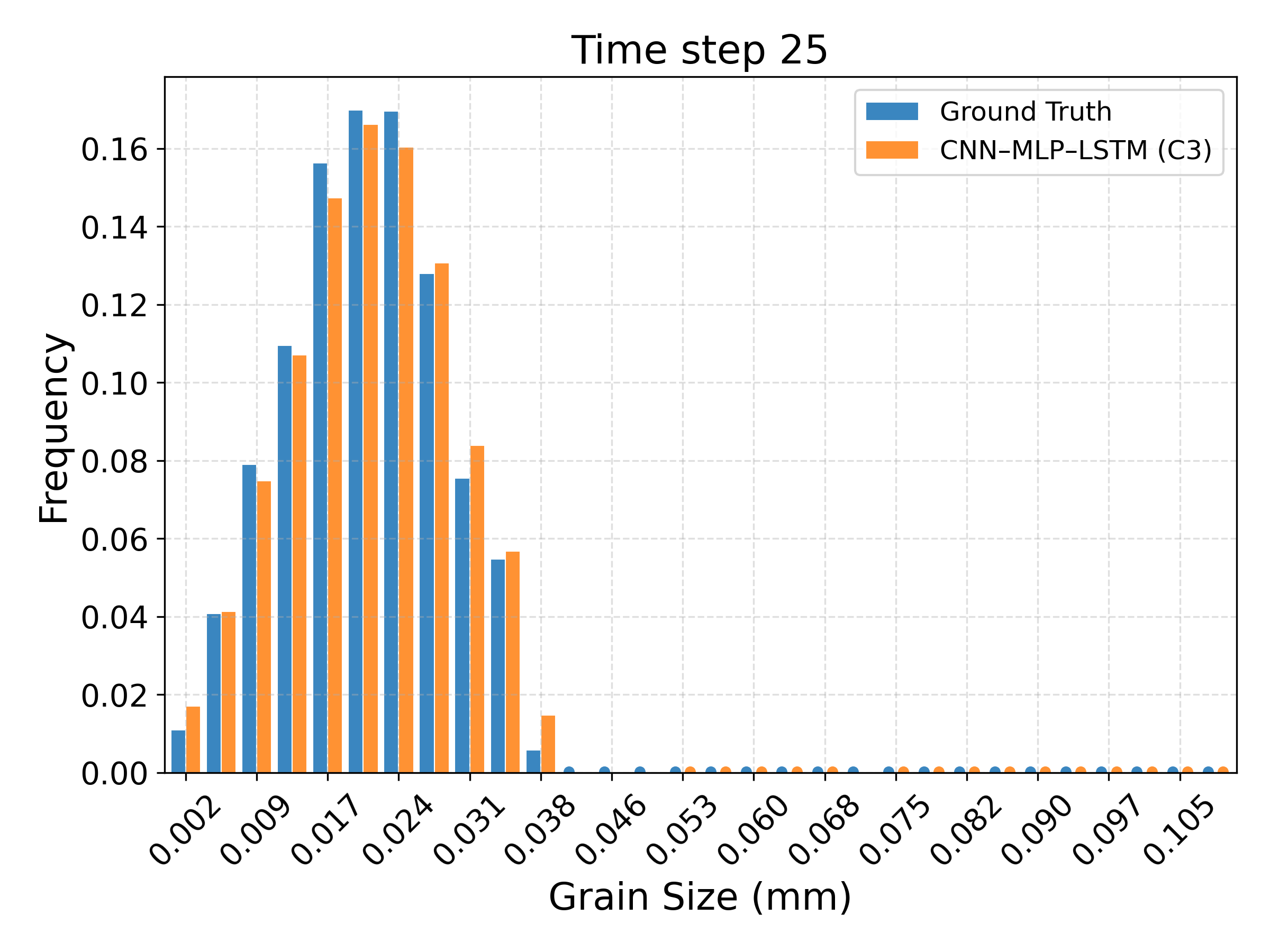}
\caption{}
\end{subfigure}

\vspace{0.2cm}

\begin{subfigure}[t]{0.40\textwidth}
\centering
\includegraphics[width=\linewidth]{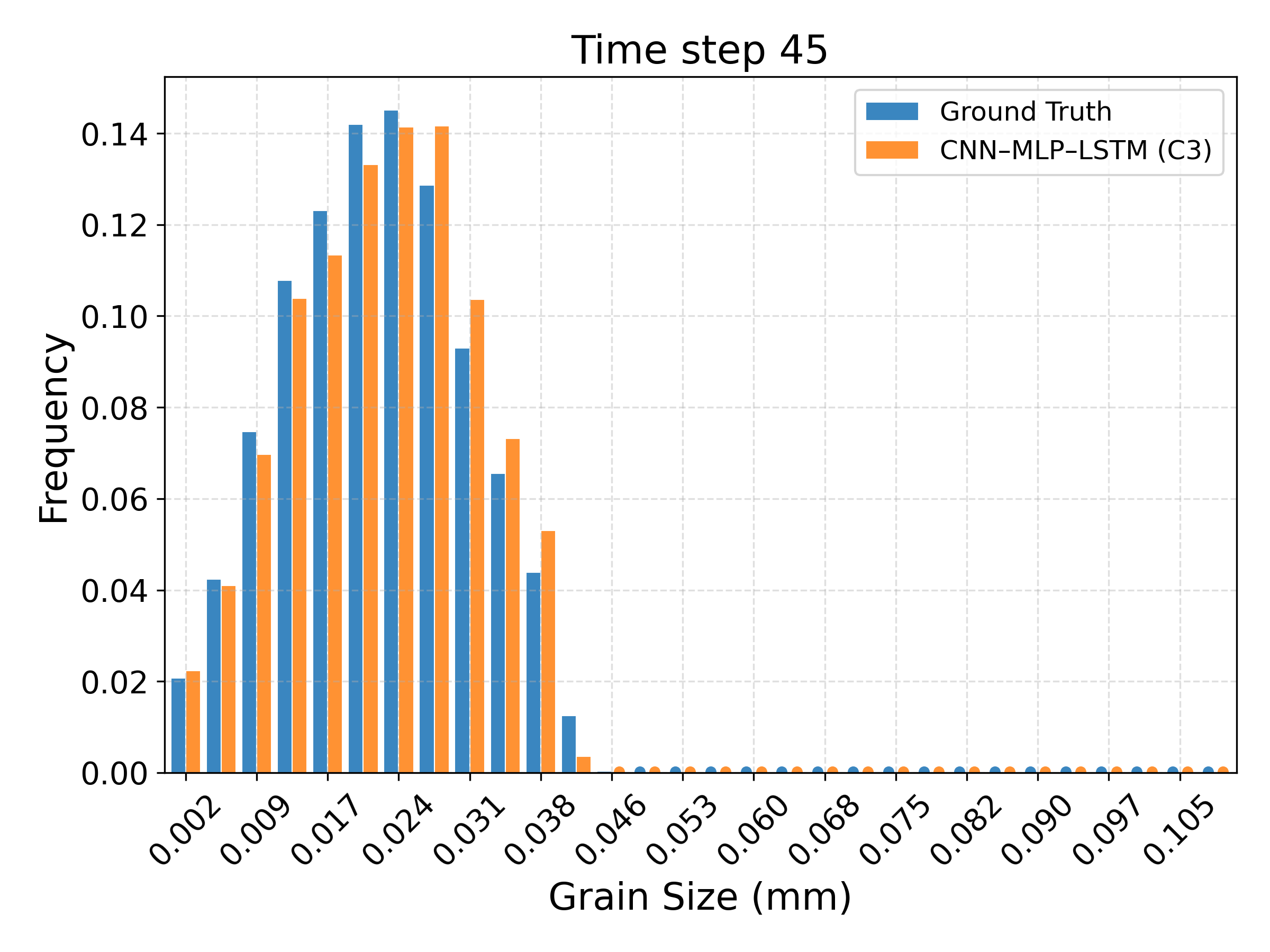}
\caption{}
\end{subfigure}
\hspace{0.01\textwidth}
\begin{subfigure}[t]{0.40\textwidth}
\centering
\includegraphics[width=\linewidth]{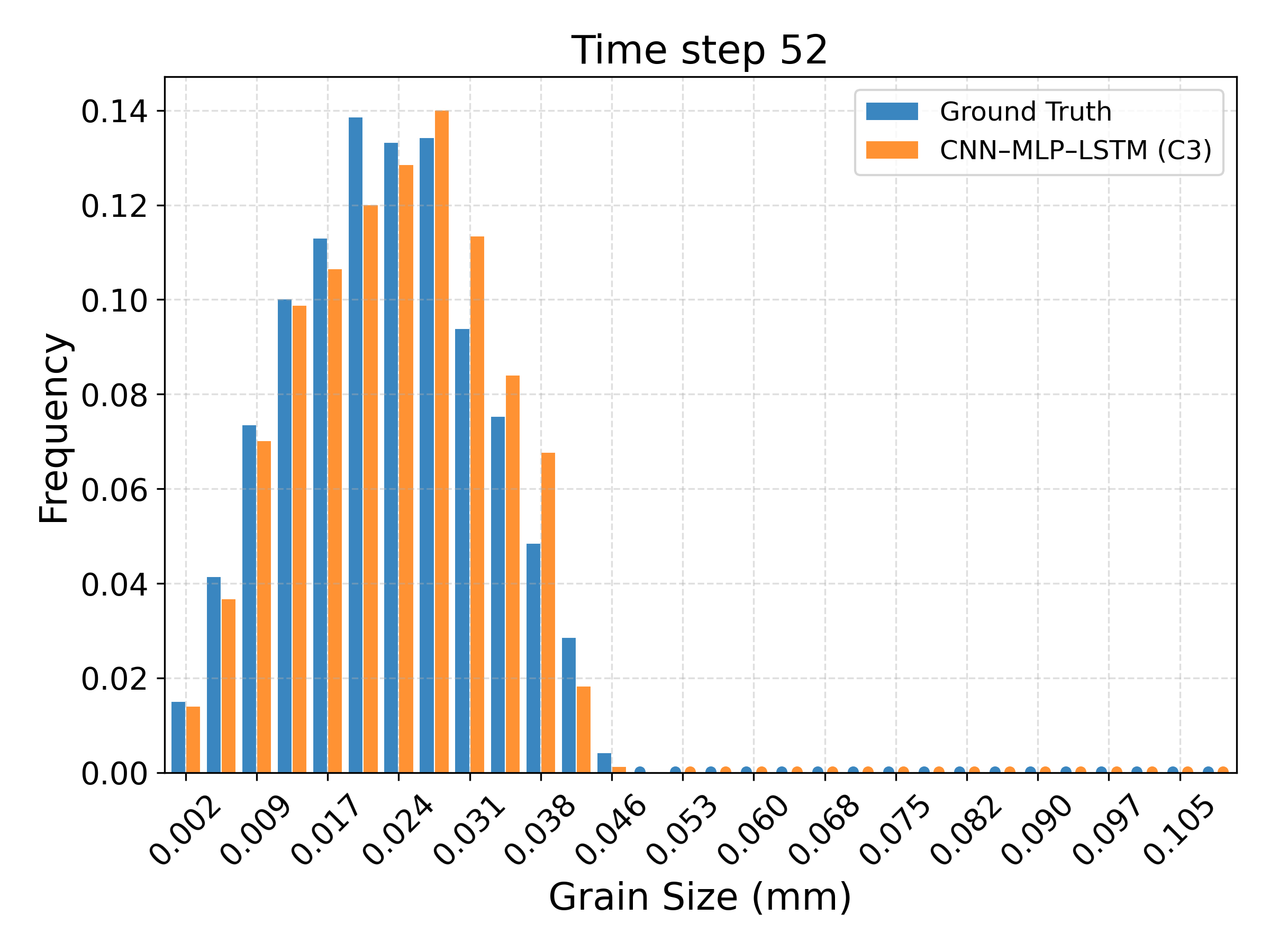}
\caption{}
\end{subfigure}

\vspace{0.2cm}

\begin{subfigure}[t]{0.40\textwidth}
\centering
\includegraphics[width=\linewidth]{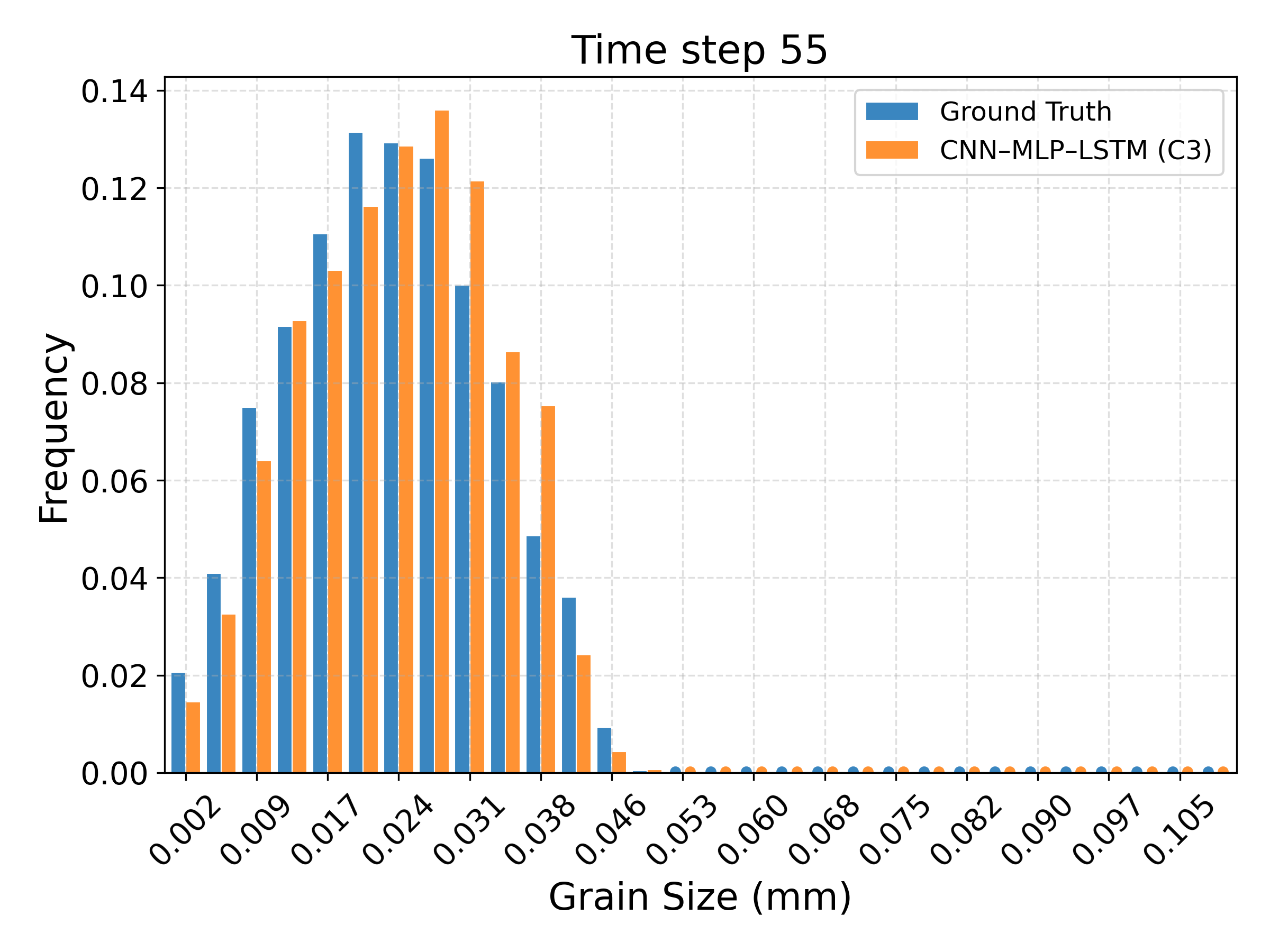}
\caption{}
\end{subfigure}
\hspace{0.01\textwidth}
\begin{subfigure}[t]{0.40\textwidth}
\centering
\includegraphics[width=\linewidth]{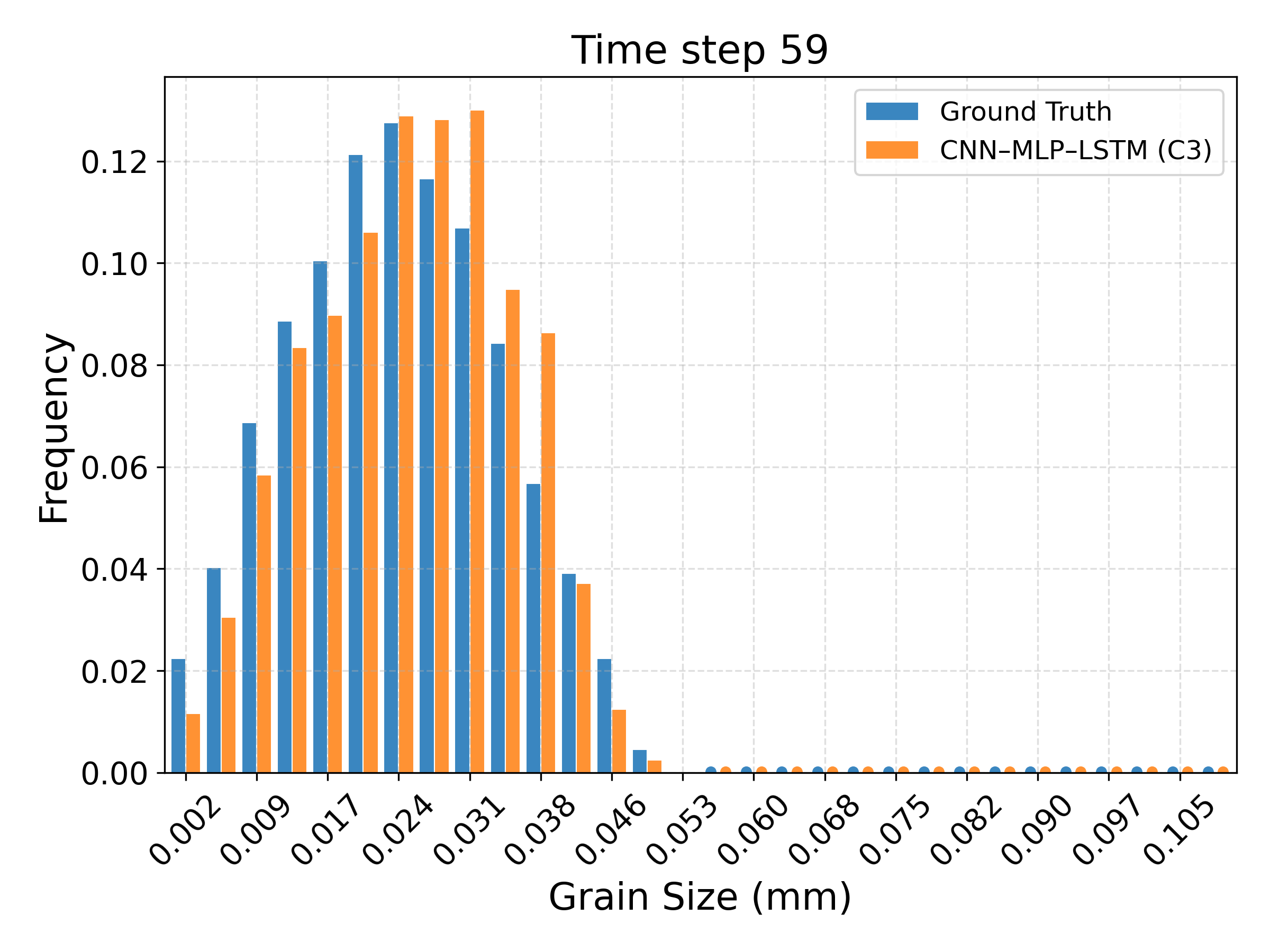}
\caption{}
\end{subfigure}

\caption{Case~4: Comparison of reference (TRM) and predicted (CNN--MLP--LSTM) normalized grain-size distributions for a representative multi-stage non-monotonic thermal cycle at (a) the initial state, (b) the first predicted time step, (c) the end of the first heating stage, (d) after the first holding stage, (e) after reheating and holding at high temperature, (f) following the subsequent holding stage, (g) at the end of the final heating stage, and (h) after the final holding stage.}

\label{fig:gen_case4}
\end{figure}

Case 3 considers a multi-stage thermal cycle involving successive heating, holding, partial cooling, and reheating stages. The predicted distributions follow the reference distributions across these successive thermal transitions (Fig.~\ref{fig:gen_case3}). In particular, the changes in the position and shape of the distribution after cooling and subsequent reheating are reproduced by the model. An MRE of 6.82\% and an $R^2$ of 0.9916 are obtained for this case; the three other initial distributions of the same cycle give MRE values between 7.36\% and 11.17\%, the largest error corresponding to the widest initial distribution ($\sigma_0 = 32~\mu$m, $R^2 = 0.9404$) (Table~\ref{tab:generalization}).

\subsubsection{Case 4: Multi-stage non-monotonic thermal cycle}

Case 4 considers a more complex non-monotonic thermal cycle with multiple changes in the imposed thermal conditions. The predicted distributions follow the main changes observed in the reference distributions throughout the successive stages (Fig.~\ref{fig:gen_case4}). Larger differences appear toward the end of the cycle, particularly in the largest grain-size classes. An MRE of 6.09\% and an $R^2$ of 0.9930 are obtained. The largest errors occur shortly after the end of the first heating stage and after the last reheating, when the grain growth rate changes abruptly. The variant of this cycle with an intermediate holding stage at 1190.15~K gives an MRE of 5.73\% and an $R^2$ of 0.9922 (Table~\ref{tab:generalization}).

\section{Discussion}
\label{sec:discussion}
This work extends a time-series forecasting framework for grain growth, previously limited to isothermal conditions, to non-isothermal regimes representative of industrial heat treatments. The results clearly demonstrate that explicitly conditioning the surrogate model with thermal information is indispensable to accurately capture thermo-kinetic effects on microstructural evolution.

The poor performance obtained without thermal conditioning shows that the grain-size history alone does not provide sufficient information to describe non-isothermal grain growth (Table~\ref{tab:global_perf}). Since grain boundary mobility depends on temperature, similar grain-size distributions may evolve differently depending on the imposed thermal history. Thermal information is therefore needed to account for these changes in grain growth kinetics.

The improvement obtained when either temperature or thermal rate is introduced shows that both descriptors carry relevant information on the thermal history. Temperature provides information on the current thermal state, whereas the thermal rate describes how this state is changing with time. Their respective contributions therefore differ, which motivates their combined use in C3.

Combining both descriptors provides the model with information on the current temperature and its variation during the thermal cycle. Temperature directly influences grain boundary mobility through the Arrhenius relation, while the thermal rate provides information on the evolution of the imposed thermal conditions. Their combined use therefore allows the model to better account for changes in grain growth kinetics during heating, holding, and cooling, leading to improved prediction performance.

Beyond thermal conditioning, the recurrent formulation also plays an important role in the prediction of grain growth evolution. The clear improvement obtained with the gated recurrent architectures can be related to their ability to retain information over longer temporal sequences. This is particularly important in the present framework, where each prediction depends on the preceding microstructural states and the associated thermal history. The poorer behavior of the Simple RNN suggests that this information is not retained effectively over long auto-regressive sequences. In contrast, the memory mechanisms of the GRU and LSTM are better suited to follow the accumulated effects of the thermal history. The slightly better performance of the LSTM may be attributed to its dedicated cell state, which provides an additional mechanism for preserving relevant information as the prediction progresses (Table~\ref{tab:recurrent_comp}).

The processing of the microstructural and thermal inputs before the recurrent block also influences the prediction performance. Providing both inputs directly to the LSTM results in higher prediction errors. When the grain-size distributions are encoded by the CNN and the thermal descriptors by the MLP before their features are combined and passed to the LSTM, both the mean and the largest errors decrease (Table~\ref{tab:arch_comp}). This separate encoding of the microstructural and thermal information therefore provides a more suitable representation for temporal prediction. 

The prediction accuracy also varies across the different stages of the thermal cycle. The relatively low errors obtained during heating indicate that the conditioned models follow the evolution of grain growth as the temperature and grain boundary mobility increase. Larger errors are observed during the isothermal plateau, where the temperature remains constant while the microstructure continues to evolve, allowing prediction errors to accumulate progressively over successive auto-regressive steps. The errors increase further during cooling, as the decrease in temperature reduces grain boundary mobility and progressively slows down the grain growth kinetics. The larger errors obtained with C2, and to a lesser extent with C1, during cooling show that using either temperature or thermal rate alone is less effective in this regime. In contrast, the lower error maintained by C3 shows the benefit of combining both descriptors when the thermal conditions change and the grain growth kinetics decelerate (Table~\ref{tab:phase_mre}).

The temporal analysis of the selected C3 model also shows how the predictions evolve during the auto-regressive rollout. The preservation of the distribution normalization and the continuous increase in the predicted mean grain size show that the expected grain growth behavior is maintained throughout the prediction. The evolution of MAE, MRE, and $R^2$ also shows a progressive accumulation of error over successive auto-regressive steps, without signs of instability. Despite this accumulation, the predicted distributions continue to follow the displacement of the grain-size population toward larger sizes. These results show that the model reproduces the main trends of grain growth over the full prediction horizon despite the accumulation of forecasting errors (Figs.~\ref{fig:temporal_stability_c3} and~\ref{fig:temporal_distribution_example}).

The generalization tests consider thermal histories that were not included during training or validation. The good agreement obtained for the first case shows that the model can predict grain growth for an unseen thermal cycle, while the second case shows that good predictions are also obtained over a longer prediction period. The multi-stage cases are more difficult, since the model has to follow several changes in the thermal conditions rather than a single heating--holding--cooling sequence. The higher error obtained for the most complex case shows that the prediction becomes more difficult as the thermal history becomes more variable. However, over the ten simulations of the four unseen cycles, the MRE remains between 1.91\% and 14.66\% and the $R^2$ above 0.94 (above 0.99 for six of them), showing that the main evolution of the grain-size distributions is still reproduced. These results show that the proposed framework can be applied to thermal histories outside those used for model training, including longer and non-monotonic thermal cycles (Table~\ref{tab:generalization}).

\section{Conclusion}
\label{sec:conclusion}

This work extends the previously developed time-series forecasting framework for grain growth from isothermal to non-isothermal conditions. Temperature and thermal rate are introduced together with the grain-size distributions to account for the effect of the thermal history on grain growth evolution.

The results show that thermal information is needed to predict grain growth under non-isothermal conditions. When no thermal information is provided, the prediction error remains high. The use of either temperature or thermal rate considerably improves the predictions, while their combined use gives the best performance, with an MRE of 2.99\% and an $R^2$ of 0.9970 over the 86 one-hour simulations, and 4.51\% and 0.9975 over the 208 four-hour simulations (Table~\ref{tab:global_perf}). The phase-wise analysis also shows that this configuration gives the lowest errors during heating, holding, and cooling (Table~\ref{tab:phase_mre}).

We also compare different recurrent formulations within the same architecture. The Simple RNN gives significantly larger prediction errors, while GRU and LSTM perform better. Among them, the LSTM gives the lowest error (Table~\ref{tab:recurrent_comp}). The comparison of the architectural components also shows that the best results are obtained when the CNN and MLP are both used before the LSTM (Table~\ref{tab:arch_comp}). In this configuration, the grain-size distributions and thermal descriptors are processed separately before temporal prediction.

The temporal evolution of the predictions shows that the error progressively increases during the auto-regressive rollout (Fig.~\ref{fig:temporal_stability_c3}). This is expected since each predicted distribution is used again to predict the following state, allowing prediction errors to accumulate with time. The residual formulation, the square-root representation and the pushforward training keep this accumulation limited, so that none of the 86 one-hour and 208 four-hour simulations diverges. Despite this increase, the predicted distributions continue to follow the grain growth evolution over the full prediction period (Fig.~\ref{fig:temporal_distribution_example}).

The model is finally tested on four thermal histories that were not included during training or validation. These cases include longer thermal cycles and multi-stage profiles with successive cooling and reheating stages. Over the ten simulations of these four cycles, the MRE ranges from 1.91\% to 14.66\% and the $R^2$ remains above 0.94 (Table~\ref{tab:generalization}). These results show that the model can predict grain growth for thermal histories different from those used during training.

Overall, the proposed approach extends time-series forecasting of grain growth to non-isothermal conditions while keeping the low computational cost of the surrogate model. The results obtained for different thermal conditions and unseen thermal histories show its potential for predicting grain-size distribution evolution under more realistic heat-treatment conditions.

\section*{Data Availability}
raw data used for the training and trained models (original data) are available at \url{https://github.com/mbernacki/GG_DL_TSM}.

\section*{Acknowledgments}
The authors acknowledge the valuable support of ArcelorMittal, Aperam, Aubert \& Duval, CEA, Constellium, Framatome, and Safran. This work was funded in part by the French National Research Agency (ANR) through the DIGIMU consortium and the RealIMotion Industrial Chair (Grant No. ANR-22-CHIN-0003).

{\small \printbibliography}

\end{document}